\documentclass[useAMS,usegraphicx,usenatbib]{mn2e}

\usepackage{graphicx}
\usepackage{times}
\usepackage{amssymb}
\usepackage{amsmath}
\usepackage{euscript}
\usepackage{longtable,lscape}
\usepackage{color}
\def\210keV{{\rm\thinspace 2--10 keV}}
\usepackage{wasysym}

\title[Optical--to--X-ray emission in low-absorption AGN]{Optical--to--X-ray emission in low-absorption AGN: Results from the Swift-BAT 9 month catalogue}
\author[R.V. Vasudevan, R. F. Mushotzky, L. M. Winter \& A. C. Fabian]
{\parbox[]{6.in} {   R. V. Vasudevan$^1$
    R. F. Mushotzky$^2$,
    L. M. Winter$^3$ and
    A.C. Fabian$^1$\\
    \footnotesize
    $^1$Institute of Astronomy, Madingley Road, Cambridge CB3 0HA\\
    $^{2}$Laboratory for High Energy Astrophysics, NASA/GSFC, Greenbelt, MD 20771, USA\\
    $^{3}$Center for Astrophysics and Space Astronomy, University of Colorado at Boulder, 440 UCB, Boulder, CO 80309-0440, USA\\
  }}

\begin{document}

\maketitle

\begin{abstract}
We present simultaneous optical--to--X-ray spectral energy distributions (SEDs) from \emph{Swift}'s X-ray and UV--optical telescopes (XRT and UVOT) for a well-selected sample of 26 low-redshift ($z<0.1$) active galactic nuclei (AGN) from the Swift/Burst Alert Telescope (BAT) 9-month catalogue, the largest well-studied, hard X-ray selected survey of local AGN to date.  Our subsample consists of AGN with low intrinsic X-ray absorption ($N_{\rm H}<10^{22}\rm cm^{-2}$) and minimal spectral complexity, to more accurately recover the intrinsic accretion luminosity in these sources.  We perform a correction for host galaxy contamination in all available UVOT filter images to recover the intrinsic AGN emission, and estimate intrinsic dust extinction from the resultant nuclear SEDs.  Black hole mass estimates are determined from the host-galaxy 2MASS K-band bulge luminosity.  Accretion rates determined from our SEDs are on average low (Eddington ratios $\lambda_{\rm Edd}\lesssim 0.1$) and hard X-ray bolometric corrections cluster at $\sim$10--20, in contrast with the higher values seen for quasars.  An average SED for the 22 low accretion rate ($\lambda_{\rm Edd}<0.1$) objects is presented, with and without correction for extinction.  Significant dust reddening is found in some objects despite the selection of low $N_{\rm H}$ objects, emphasising the complex relationship between these two types of absorption.  We do not find a correlation of optical--to--X-ray spectral index with Eddington ratio, regardless of the optical reference wavelength chosen for defining the spectral index.  An anti-correlation of bolometric correction with black hole mass may reinforce `cosmic downsizing' scenarios, since the higher bolometric corrections at low mass would boost accretion rates in local, lower mass black holes.  We also perform a basic analysis of the UVOT-derived host galaxy colours for our sample and find hosts cluster near the `green valley' of the colour-magnitude diagram, but better quality images are needed for a more definitive analysis.  The low accretion rates and bolometric corrections found for this representative low-redshift sample are of particular importance for studies of AGN accretion history.

\end{abstract}

\begin{keywords}
black hole physics -- galaxies: active  -- galaxies: Seyfert
\end{keywords}

\section{Introduction}
\label{Intro}

Active Galactic Nuclei (AGN) are known to emit radiation over the whole range of available energies observable using current detectors.  Characterising their spectral energy distribution (SED) is therefore important for understanding the different physical processes at work.  The accepted paradigm behind the radiation output of AGN is accretion onto a supermassive black hole.  The emission directly due to accretion emerges primarily in the optical, UV and X-ray regimes by a combination of thermal emission from an accretion disc and inverse-Compton scattering of UV disc photons by a corona above the disc.  A dusty torus partially absorbs the UV photons from the accretion disc, re-emitting them more isotropically in the infrared.  In this study, we aim to recover the true accretion luminosity in a representative sample of AGN, and use these results to identify trends between different SED parameters.  However, recovering the true accretion luminosity can be complex, and previous studies have found a wide spread in the conversion factors (bolometric corrections) between the X-ray luminosity and the total accretion luminosity.  This is, in a large part, due to intrinsic variation between objects in the whole AGN population, but is also complicated by numerous systematics and biases in the samples used.

The pioneering study of \cite{1994ApJS...95....1E} (hereafter E94) presented radio--to--X-ray SEDs for 47 quasars and has provided a useful template for determining bolometric corrections and identifying trends in the SED shapes across a range of AGN properties.  Recent studies have confirmed the usefulness of the average bolometric corrections and SED parameters from E94, but the sample of quasars used was predominantly X-ray bright and their bolometric luminosities included the re-processed infrared emission.  \cite{2006ApJS..166..470R} present infrared--to--X-ray SEDs for 259 quasars selected by a combination of optical and mid-infrared colour selection criteria, and highlight the spread in SED parameters within the sample.  This large spread in SED parameters was known from E94, prompting refinements such as the template SED of \cite{2004MNRAS.351..169M} who employ the observed correlation between X-ray--to--optical spectral index and X-ray luminosity to construct a luminosity-dependent SED, and importantly, they exclude the reprocessed infrared emission to avoid double-counting of part of the accretion energy budget. A similar template is presented by \cite{2007ApJ...654..731H} in their study on the bolometric quasar luminosity function.  These SED templates allowed the diversity of SED shapes to be taken into account when calculating the supermassive black hole mass density from the X-ray background and AGN luminosity functions.   More recently, \cite{2007MNRAS.381.1235V} (VF07 hereafter) presented optical--to--X-ray SEDs for a sample of AGN observed by the Far Ultraviolet Spectroscopic Explorer (\emph{FUSE}), which yielded interesting trends between SED shape and Eddington ratio, confirmed later by \cite{2009MNRAS.392.1124V} (VF09 hereafter) using simultaneous data and reverberation mapping mass estimates to improve accuracy of the bolometric luminosities and accretion rates.  However, the sample of VF07 is by necessity UV-bright, as it is selected by \emph{FUSE} and the sample of VF09 is restricted to those which are bright enough in the optical/UV for reverberation mapping to be carried out.

These considerations emphasise the pressing need for a representative sample of AGN from which conclusions can be drawn about accretion properties of the wider AGN population.  One major factor influencing the selection of AGN for surveys is the nature of their absorption.  It has been known for some time that X-ray and optical surveys detect different segments of the underlying AGN population \citep{2004ASSL..308...53M}, with X-ray surveys able to probe to higher absorbing column density.  The standard physical mechanisms for the generation of X-ray emission in AGN produce a power-law, but often AGN exhibit X-ray spectra deviating significantly from this, in part due to absorption.  The geometry of the absorption in X-rays is the subject of much debate, and the complex spectra in some sources can be accounted for with either `partial covering' scenarios \citep{2004MNRAS.349L...7G} or strong reflection from the accretion disc due to light bending \citep{2004MNRAS.349.1435M}, and warm absorbers can also complicate our view of the intrinsic emission (\citealt{1997MNRAS.291..403R}, \protect\citealt{2005A&A...431..111B}). Many of the reverberation mapped AGN in VF09 showed signs of such spectral complexity, and the authors identified the difficulties in regaining the true accretion luminosity in these cases.

The Burst Alert Telescope (BAT) on board the \emph{Swift} satellite proves invaluable for addressing many of these considerations. Canonical levels of absorption (column densities $N_{\rm H}<10^{24}\rm cm^{-2}$) imprint signatures on the 0.1--10 keV X-ray band, so the very hard X-ray sensitivity of the instrument (14--195 keV) provides the capability to observe AGN in a bandpass relatively unaffected by such absorption.  The 9-month catalogue of BAT-detected AGN (hereafter the \emph{Swift}/BAT catalogue, \citealt{2008ApJ...681..113T}) therefore provides an unprecedented level of completeness (with respect to absorption) when surveying the AGN population as it is unbiased to all but the most heavily obscured sources.  The study of \cite{2009ApJ...690.1322W} presents a comprehensive overview of the X-ray spectral properties of the 153 sources in the \emph{Swift}/BAT catalogue and allows us to select an appropriate sample from which we can be confident of calculating accurate luminosities and SED shapes.  Most importantly, they present values of absorbing column density $N_{\rm H}$ for each source, determined from fits to the X-ray data from the literature and new fits by the authors themselves, and identify those sources in which significant spectral complexity is present.  In this work, we preferentially select those with lower absorption and minimal spectral complexity, facilitating a more straightforward calculation of the accretion luminosity.  We assume that for such sources, the accretion luminosity excludes the reprocessed IR emission and is principally seen in the optical--to--X-ray regime (as in VF07 and VF09).  Additionally, AGN with more X-ray absorption are expected to display higher levels of optical--UV reddening, the calculation of which depends on the precise form of the extinction curve and is difficult to account for when calculating the total luminosity.  These considerations motivate our selection of objects with low $N_{\rm H}$.

Variability in AGN can produce large changes in luminosity, over timescales from hours to years.  In order to catch an accurate snapshot of the total energy budget in an AGN at a given time, it is therefore highly desirable to use simultaneous data.  The potential inaccuracies in SED parameter values from non-simultaneous SED data are discussed in detail in VF07 and VF09, with the latter following the approach of \cite{2006MNRAS.366..953B} by using contemporaneous optical--to--X-ray data from XMM-Newton's PN and Optical Monitor instruments.  In this work, we make use of \emph{Swift}'s own co-aligned UV--optical telescope (UVOT) and X-ray telescope (XRT) to determine simultaneous SEDs.  Many of our target subsample from the \emph{Swift}/BAT catalogue have simultaneous UVOT and XRT observations available in the archives; typically the UVOT data span six filters between 5468$\rm \AA$ and 1928$\rm \AA$, providing a window onto a substantial fraction of the disc emission.  The central frequencies of the UVOT filters are given in Table \ref{uvotfilters}.

\begin{table}
\begin{tabular}{|l|l|}
\hline
Filter&Central wavelength ($\rm \AA$)\\
\hline
V&5468\\
B&4392\\
U&3465\\
UVW1&2600\\
UVM2&2246\\
UVW2&1928\\
\hline
\end{tabular}
\caption{\label{uvotfilters}Central frequencies of the UVOT filters.}
\end{table}

The large wavelength coverage, field of view and spatial resolution of the UVOT data also allow a correction for the host galaxy flux to be estimated for each AGN.  We employ GALFIT for this purpose, a program for performing 2D fitting of PSFs and various analytical profile forms to galaxy images \citep{2002AJ....124..266P}.  For a detailed discussion of the requirements of GALFIT, the reader is directed to peruse the accompanying documentation\footnote{http://users.ociw.edu/peng/work/galfit/galfit.html}.  We also present basic results on the host galaxies from the GALFIT fitting process.  The UVOT data used in this study also clarify how the degree of host galaxy contamination varies with wavelength (UVOT filter).

Understanding variations in the AGN SED is important, and one of the main drivers of these variations appears to be the accretion rate (parameterised as the Eddington ratio, $L_{\rm bol}/L_{\rm Edd}$ for an object with bolometric luminosity $L_{\rm bol}$ and Eddington luminosity  $L_{\rm Edd}=1.38\times10^{38}[M_{\rm BH}/M_{\odot}] \rm erg s^{-1}$ for a black hole of mass $M_{\rm BH}$).  The recent Principal Component Analysis of \cite{2008arXiv0810.5714K} on a sample of red 2MASS AGN SEDs identify the Eddington ratio as the `eigenvector 1' for these AGN (being responsible for most of the variation in the SED shapes).  Both VF09 and VF07 highlight substantial variations in the bolometric output (in comparison to the X-ray output) across the observed range of Eddington ratios, and numerous studies explore these themes further (\citealt{2008ApJ...682...81S}, \citealt{2008ApJS..176..355K}).  Eddington ratios require accurate estimates of the black hole mass in AGN, but since the most accurate estimates from reverberation mapping are only available for 35 AGN, here we employ the $M_{\rm BH}-L_{\rm bulge}$ correlation to calculate black hole masses.   The combination of robust determinations of the bolometric luminosity and sensible estimates of the black hole mass provide great opportunities for insight into the accretion process.  In this work, we discuss the analysis of the \emph{Swift} data, the construction of SEDs, the derivation of SED parameters, the identification of correlations and trends, and finally the host galaxy properties.

\section{Sample selection}
\label{sampleselection}

We firstly apply a cut in intrinsic column density $N_{\rm H}^{(int)}$, selecting only those with log($N_{\rm H}^{(int)}$)$<22$ from the list of objects in the 9-month BAT catalogue as presented in \cite{2009ApJ...690.1322W}.  This is the suggested crossover point between `absorbed' and `unabsorbed' classes of AGN in their study, which also classifies the objects in the sample into two broad categories based on X-ray spectral properties: `simple' (S) and 'complex' (C).  The S class have X-ray spectra which are best fit by a simple power law with intrinsic and galactic absorption (sometimes with a soft excess), whereas the C class display more complex X-ray spectra which are better fit by models such as double-power laws or partial covering (these models are illustrative in that they highlight the spectral complexity; more detailed fits including reflection may also be possible).  We also eliminate any objects from the C class, as they will present difficulties when determining the true accretion luminosity (as discussed in VF09).

In order to facilitate profile fits to optical and UV images of the host galaxies, we also impose a redshift cut, requiring $z<0.1$ for our selection. For $z<0.1$, the angular resolution of the UVOT ($\sim 1.7$ arcsec) allows separation of the nucleus from the galaxy to a physical length scale of 3 kpc, thus allowing a reasonable galaxy--AGN separation while still yielding 54 potential objects for study from the sample. We then identify those for which \emph{Swift} XRT and UVOT data are available from the High Energy Astrophysics Science Archive Research Center (\emph{HEASARC}\footnote{http://heasarc.gsfc.nasa.gov/cgi-bin/W3Browse/swift.pl}), yielding 33 objects. For estimating black hole masses, we also obtain K-band magnitudes from the Two-micron All-Sky Survey (2MASS) catalogues.  The key requirement for calculating black hole masses is an estimate of the bulge luminosity.  The K-band is least subject to the effects of Galactic reddening and predominantly traces the older stars in the bulge over the stellar populations of the galaxy disc, motivating its selection over the J and H bands also available in the 2MASS catalogues.  The 2MASS data were gathered with two ground based telescopes: the Whipple Observatory in Arizona, USA and the Cerro Tololo telescope at La Serena, Chile.  The limitations introduced by ground-based observations (most importantly, the level of seeing for each observation) need to be taken into account when attempting to recover the bulge luminosity.  This is discussed at greater length in \S\ref{bhmassesfrom2mass} on black hole mass determination.  We required the centroids reported for the 2MASS data to be within 2.0 arcsec of the NED position in the initial case. For a few objects which lie just outside this limit, the 2MASS images were downloaded and the locations provided were checked against the refined positions obtained from the XRT and UVOT images; if a confident identification of a counterpart to the X-ray AGN was made, the 2MASS data were used in further processing.  At any rate, objects with 2MASS centroids greater than 5.0 arcsec away from their NED positions were rejected altogether. Three potential objects are excluded from our sample by their absence from the catalogues (IGR 21277+5656, IGR J07597-3842 and LEDA 138501).

Lastly, we check for blazars in the sample.  The SED properties of blazars make them unsuitable for performing a bolometric luminosity calculation based on the `disk + power-law' model employed here since their UV and X-ray luminosity is dominated by jet emission.  Out of the 30 objects which satisfy all the above criteria, four objects are exlcuded based on their NED classifaction as blazars: Mrk 501, 2MASX J19595975+6508547, ESO 362-G021 (PKS 0521-36) and PKS 0548-322.  The final selection of 26 objects is presented in Table~\ref{obsdetails}, along with their redshifts and details of the \emph{Swift} observations used.

\begin{table*}
\begin{tabular}{|l|l|l|l|l}
\hline
AGN&redshift&\emph{Swift} observation ID&Total UVOT exposure time (ks)&XRT exposure time (ks)\\\hline
UGC 06728&0.006518&00035266001&2.1&6.33\\
MCG-06-30-15&0.007749&00035068003&3.19&8.72\\
NGC 4593&0.009&00037587001&1.83&4.88\\
Mrk 766&0.012929&00030846039&1.34&3.71\\
ESO 548-G081&0.01448&00035250002&2.8&6.33\\
Mrk 352&0.014864&00035243002&6.38&15.91\\
NGC 7469&0.016317&00031245005&1.38&4.27\\
NGC 5548&0.017175&00030022062&1.84&5.13\\
WKK 1263&0.02443&00035268002&1.45&8.94\\
ESO 490-G026&0.02485&00035256001&2.87&8.51\\
Mrk 590&0.026385&00037590001&1.43&4.46\\
1RXS J045205.00+493248&0.029&00035281002&0.65&1.99\\
SBS 1301+540&0.0299&00035269001&2.3&7.85\\
Mrk 279&0.030451&00037591001&1.75&5.22\\
MCG +04-22-042&0.032349&00035263001&3.0&9.11\\
Ark 120&0.032713&00037593003&1.44&4.11\\
IRAS 05589+2828&0.033&00035255001&2.68&5.91\\
3C 120&0.03301&00036369001&2.1&6.37\\
Mrk 509&0.034397&00035469003&2.5&6.8\\
Mrk 841&0.036422&00035468002&3.33&8.4\\
Mrk 1018&0.042436&00035166001&1.41&4.53\\
NGC 985&0.043143&00036530005&3.16&8.65\\
3C 390.3&0.0561&00037596001&2.25&6.46\\
IRAS 09149-6206&0.0573&00035233002&1.87&5.04\\
SBS 1136+594&0.0601&00035265001&3.22&9.17\\
2MASX J21140128+8204483&0.084&00035624002&1.72&5.09\\
\hline
\end{tabular}
\caption{\label{obsdetails}Details of UVOT observations used for the objects in our sample.}
\end{table*}

\section{Processing of multiband data}

We download the optical--UV and X-ray data for our selection from HEASARC.  When multiple observations were available, those with the maximum UVOT exposure time and maximum number of UVOT filters were preferentially selected.  Pipeline-processed `level 2' \textsc{fits} files are readily available from HEASARC.  The data for the 26 sources identified in section \ref{sampleselection} were then processed according to the procedure outlined in the following sections.

\subsection{Optical--UV photometry: Correcting for the host galaxy}

We employ the custom-built software tools designed specifically for processing UVOT data where possible.  Each individual UVOT filter data file in general contains a number of exposures which were summed using the tool \textsc{uvotimsum}.  The \textsc{uvotsource} tool was then used to extract magnitudes from simple aperture photometry.  Source and background regions were created for this purpose, with the position of the source region being obtained from the NASA Extragalactic Database (NED) in the first instance, followed by fine adjustment of the source region position if necessary.  The required source region size for \textsc{uvotsource} is 5 arcsec, and the background region can be any size (at the time of writing): we used background regions with radii from 5 arcsec to $\sim$30 arcsec depending on the frequency of other foreground sources in the image.  The magnitudes from \textsc{uvotsource} provide a useful first-order estimate of the nuclear flux, and were saved for a later comparison with the more carefully determined nuclear fluxes using point spread function (PSF) fitting.

The images were then prepared for use within GALFIT.  One of the requirements for accurately determining the nuclear flux is to have suitable PSFs available for each image.  Preliminary studies of the UVOT PSF in the six filters show that the PSF shape may depend on a variety of factors including position on the detector, countrate of the source and the filter being used.  Average PSF full-width half-maxima range between 1.7--2.5 arcsec, when considering all six filters.  As discussed by \cite{2008arXiv0807.1334K} in the context of Hubble Space Telescope (\emph{HST}) images, the choice of PSF can significantly affect PSF fitting results, since a poorly chosen PSF can give profile parameters that deviate dramatically from their `true' values (also see \citealt{2008ApJ...683..644S} for another discussion of the factors affecting AGN--host-galaxy decomposition).   To mitigate these problems and minimize PSF mismatch, we adopted the following approaches.  The first and preferred approach was to generate a unique PSF for each filter image from a few known guide stars in the image, requiring the stars selected to be within a certain range of the count rate of the source of interest (as determined from simple aperture photometry with \textsc{uvotsource}).  Firstly, the UVOT utility \textsc{uvotdetect} was used to provide a list of detectable sources in each image (\textsc{uvotsource} essentially uses the \textsc{Sextractor} package for this purpose).  These were then cross-checked against published catalogues of known guide star positions.  The \textsc{xbrowse} tool in the \textsc{heasoft} suite of utilities provides access to many such catalogues, such as the \emph{US Naval Observatory} (USNO) and \emph{HST} guide star catalogues.  Typically, the USNO catalogue was used to obtain a list of such stars, requiring them to be within 13 arcmin of the AGN of interest to ensure they were within the UVOT field of view.  The guide stars were selected such that between 3-20 stars were used to generate each PSF image.  For the lowest energy V-band, this translated into a requirement that the stars were typically within 0.1--0.2 dex of the target AGN count rate; for the highest energy UVW2 band, the limits were typically within 1.5-2.0 dex of the target AGN count rate to obtain a similar number of guide stars.  Finally, the regions of the images containing these stars were then summed to form a final PSF image, using IDL code by A. A. Breeveld\footnote{Mullard Space Science Laboratory - http://www.mssl.ucl.ac.uk/} designed specifically for generating PSFs from UVOT images.  The resulting PSFs were used in GALFIT to model the central AGN in the galaxy profiles.

When this approach was not possible due to lack of guide star detections within reasonable count-rate limits (or lack of point sources which were also identified in the guide star catalogue), the filter image was viewed using the \textsc{ximage} package and a `King' profile model was fit to an identifiable point source which looked similar in form and intensity to the AGN nucleus.  The parameters from the fit were then used as a model for the PSF in GALFIT (via the `Moffat' model option), in lieu of a real PSF image.  In the vast majority of cases, a real PSF image could be extracted from the images and a `King/Moffat' model PSF was rarely required.  In only one filter image for one object, a Gaussian model was found to fit the nucleus better than other available models, with the Gaussian width fixed at the average FWHM of the UVOT PSF in that filter as reported in the CALDB documentation\footnote{http://heasarc.nasa.gov/docs/heasarc/caldb/swift/docs/uvot/}.

The PSF or PSF model thus generated was then fit along with other galaxy profile components in GALFIT.  We performed tests to ascertain the level of detail discernible within the UVOT images, starting initially with a four-component model consisting of a constant sky background, a central nuclear point source, a Sersic profile for the galaxy bulge and an outer exponential disk.  Such tests overwhelmingly indicated that four components incorporated too high a degree of degeneracy, due to the large UVOT PSF size. The PSF of the UVOT instrument is typically the same size as the galaxy bulge would be in higher resolution images of galaxies near redshift $\sim$0.05, and indeed for some of the more distant sources, the UVOT PSF is as large as the host galaxy itself.  Often the Sersic profile component would shrink to a size comparable to that of the PSF, and would `absorb' some of the flux which would otherwise been reported as coming from the PSF component.  Therefore, three components were adopted as sufficient to give sensible model reconstructions of the profiles seen in the images (sky + PSF + exponential disk).  To get as accurate a representation as possible of the nuclear flux, the following iterative algorthim was employed using GALFIT:

\begin{enumerate}
\item All bright foreground stars were excluded from the fit by creating a bad pixel mask which covered these unwanted objects.
\item The sky background was fit independently to a blank region of the sky close to the AGN of interest.
\item The central point source and sky were fit together, keeping the sky parameters constant from the previous step.  The central point source magnitude was seeded with that obtained from \textsc{uvotsource} aperture photometry, to provide a sensible initial guess of the nuclear flux.
\item The sky, central source and exponential disk components were fitted together, keeping the sky and central source parameters fixed at the values from the previous step.
\end{enumerate}

After these steps, the resultant model fits were compared with the original images and the residual images obtained from subtracting the model from the data were inspected.  In the first instance, the positions of the different components were left free to fit.  If any obvious signs of PSF mismatch or other problems with the fit were evident, these positions were frozen manually by inspection and the fit was re-evaluated.  In some cases, such as for point-like AGN at the higher end of our redshift range, a simpler model consisting of just a PSF and a constant background was adopted, if the first-pass attempt at fitting an exponential disk did not show a significant galaxy disk component.  Where the PSF image displayed obvious PSF mismatch, a `King' profile fit was also attempted.  The best fit obtained from all these approaches, determined by visual inspection of the residual image, was chosen to obtain the final nuclear magnitudes.  An example of the sky-nucleus-disk decomposition is shown in Fig.~\ref{galfitimages_example}.  The choices of profiles used for each filter in each object are given in table~\ref{galfit_modelfits_table}.

\begin{figure*}
\includegraphics[width=3.7cm]{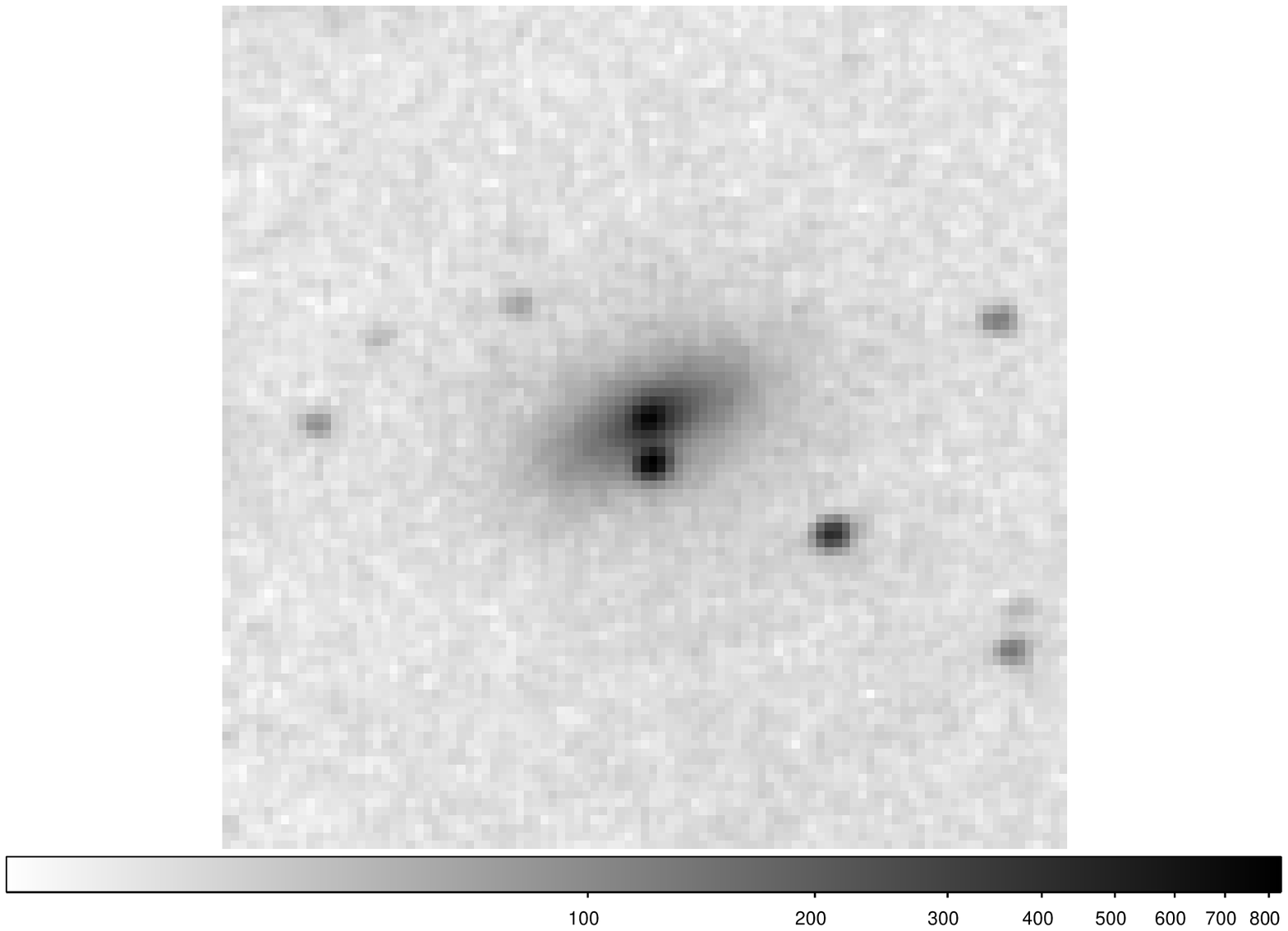}
\includegraphics[width=3.7cm]{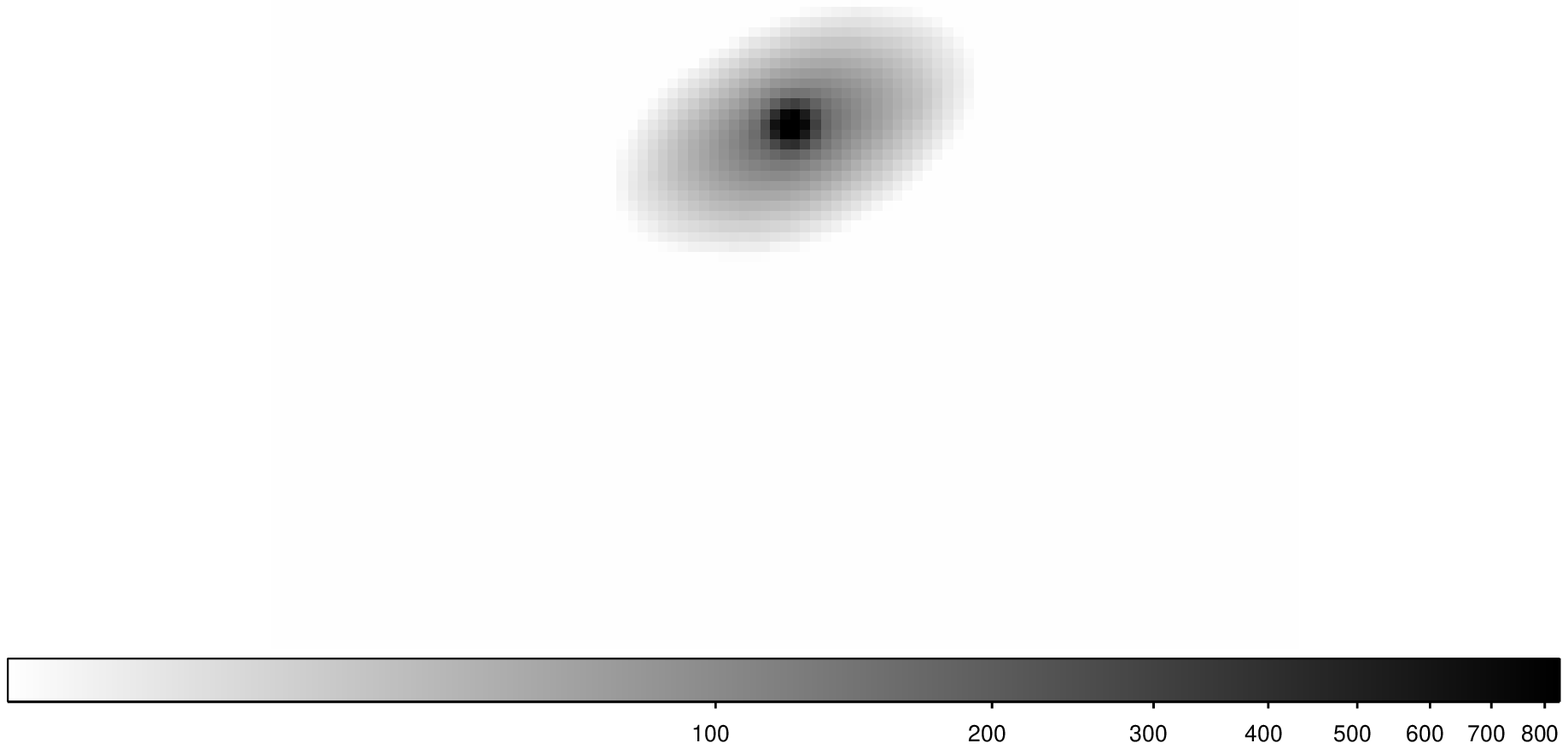}
\includegraphics[width=3.7cm]{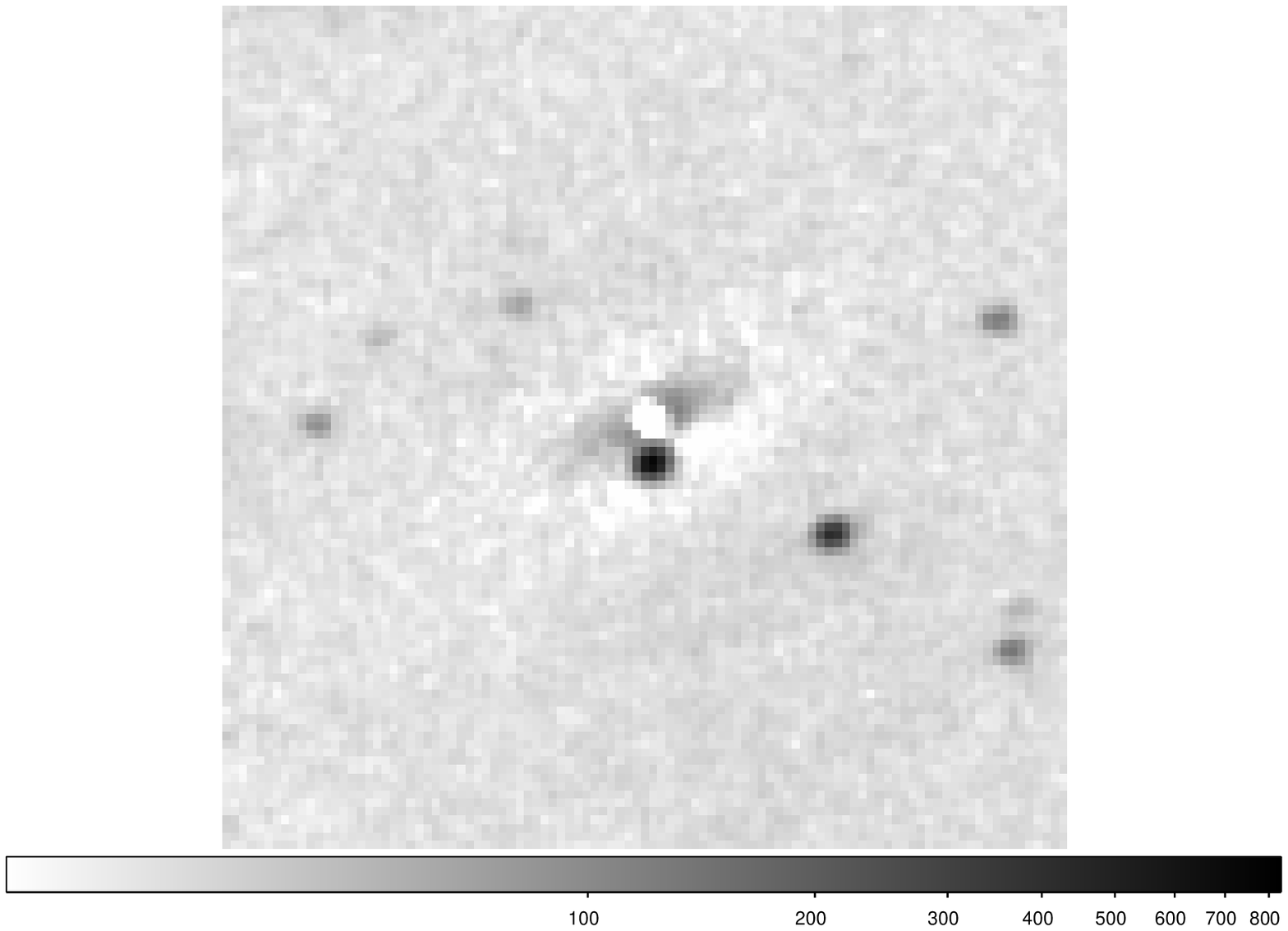}

\includegraphics[width=3.7cm]{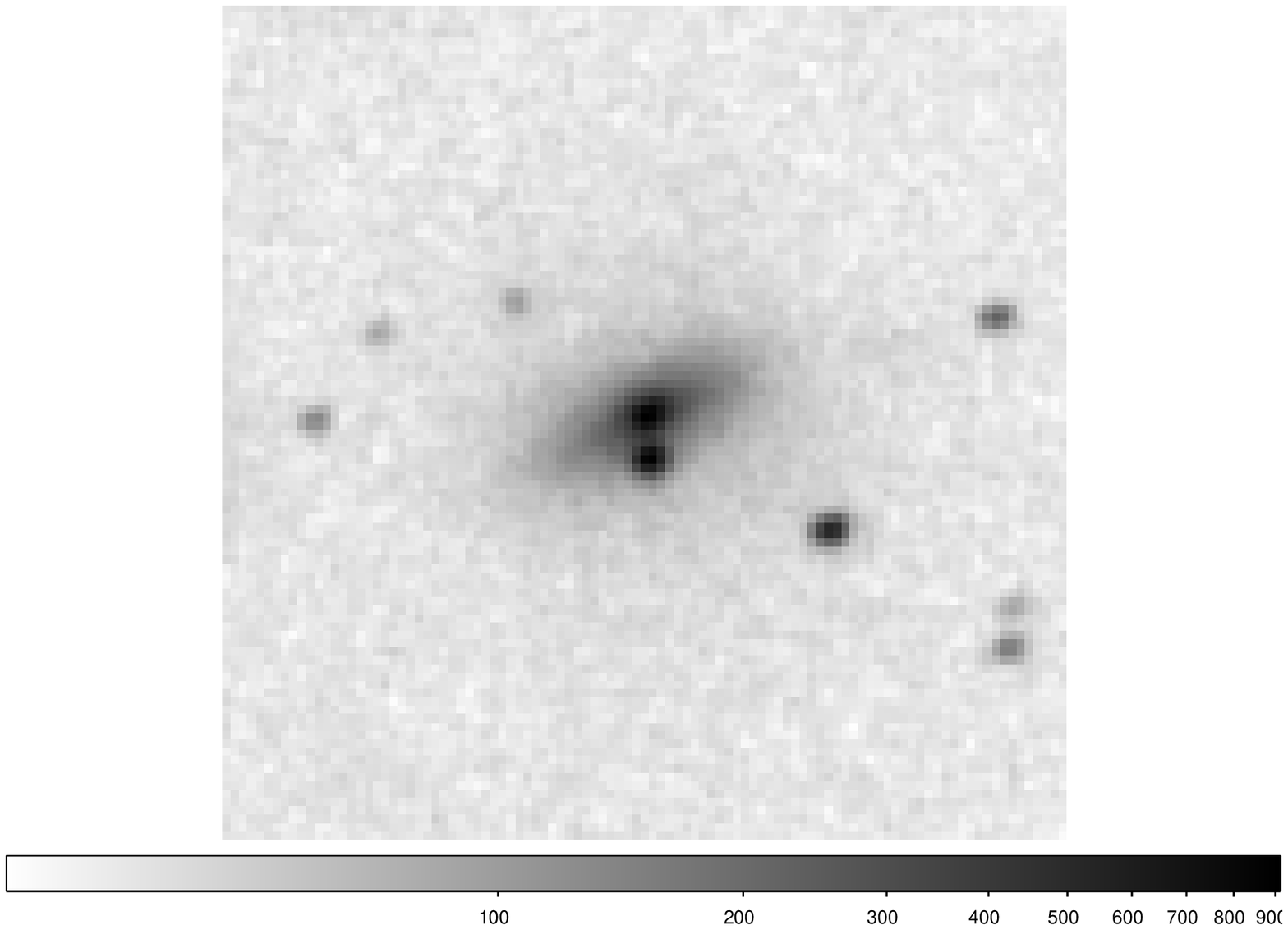}
\includegraphics[width=3.7cm]{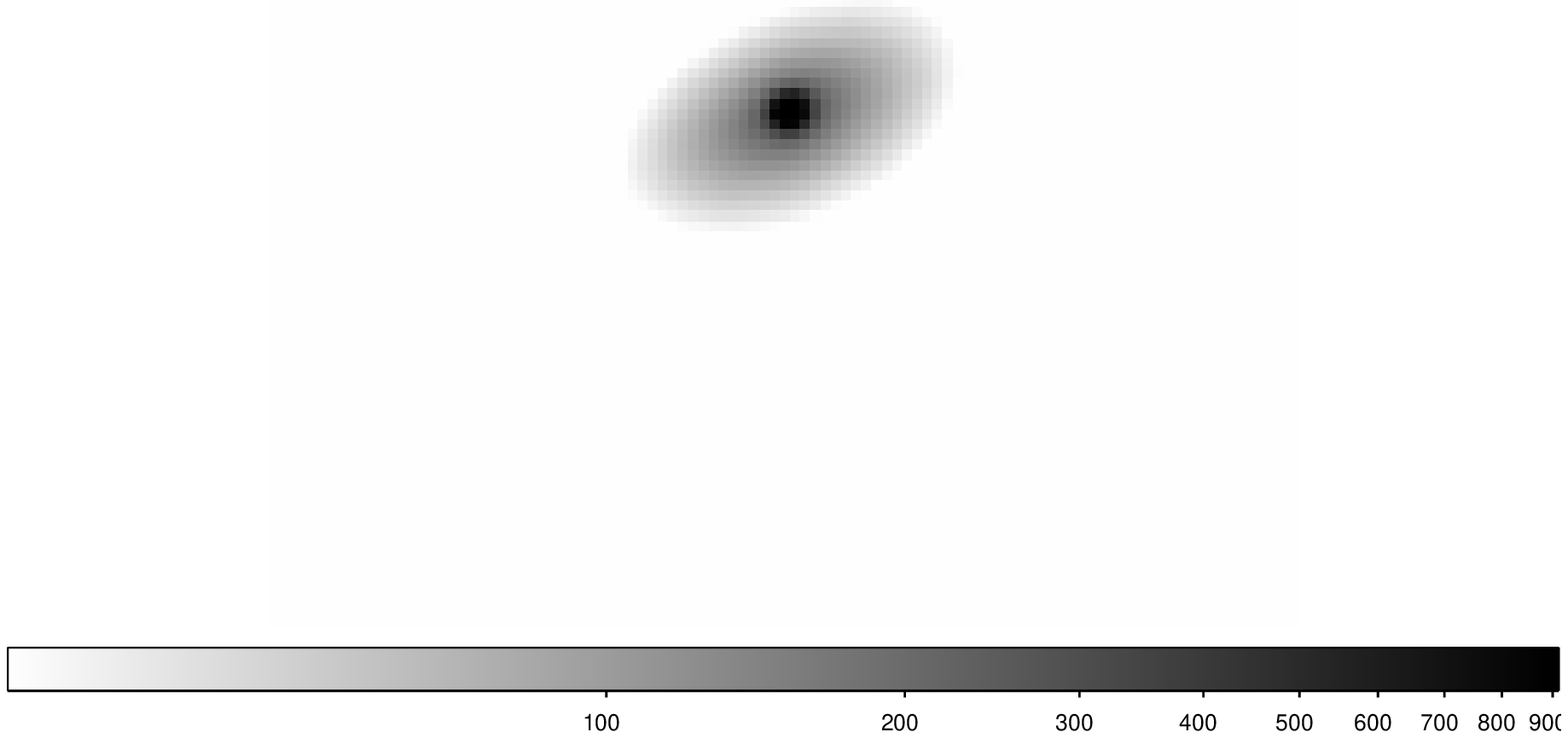}
\includegraphics[width=3.7cm]{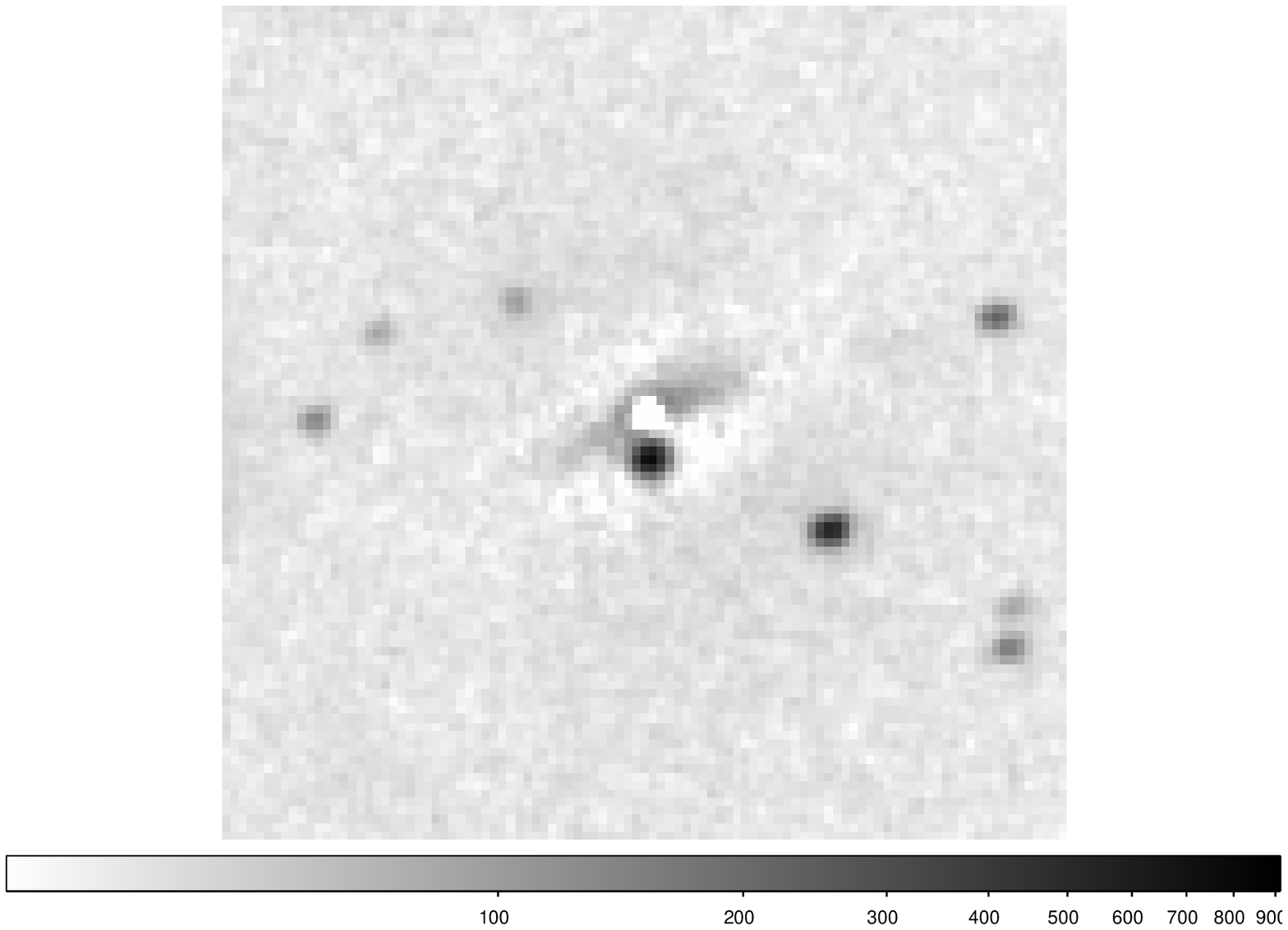}

\includegraphics[width=3.7cm]{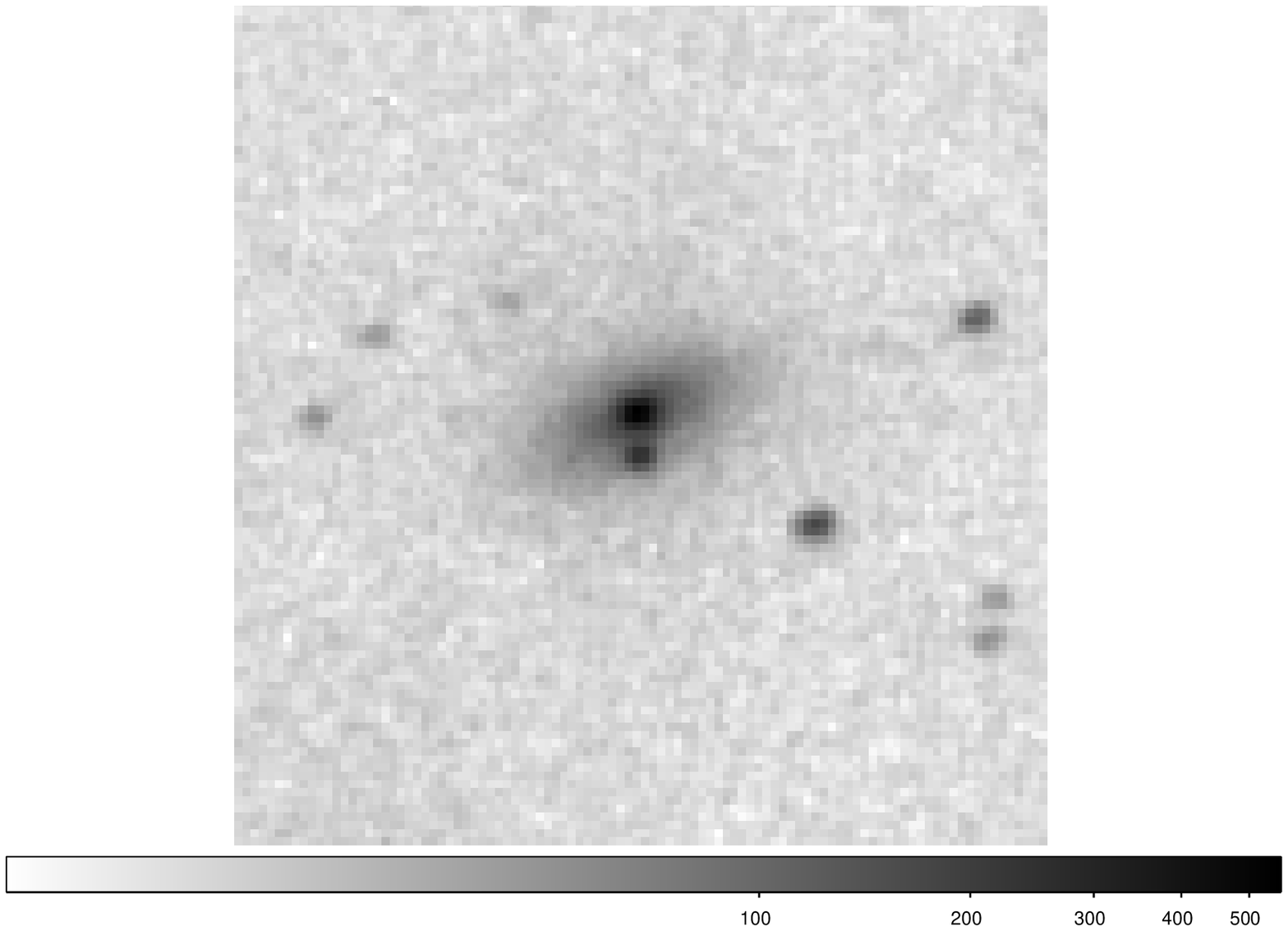}
\includegraphics[width=3.7cm]{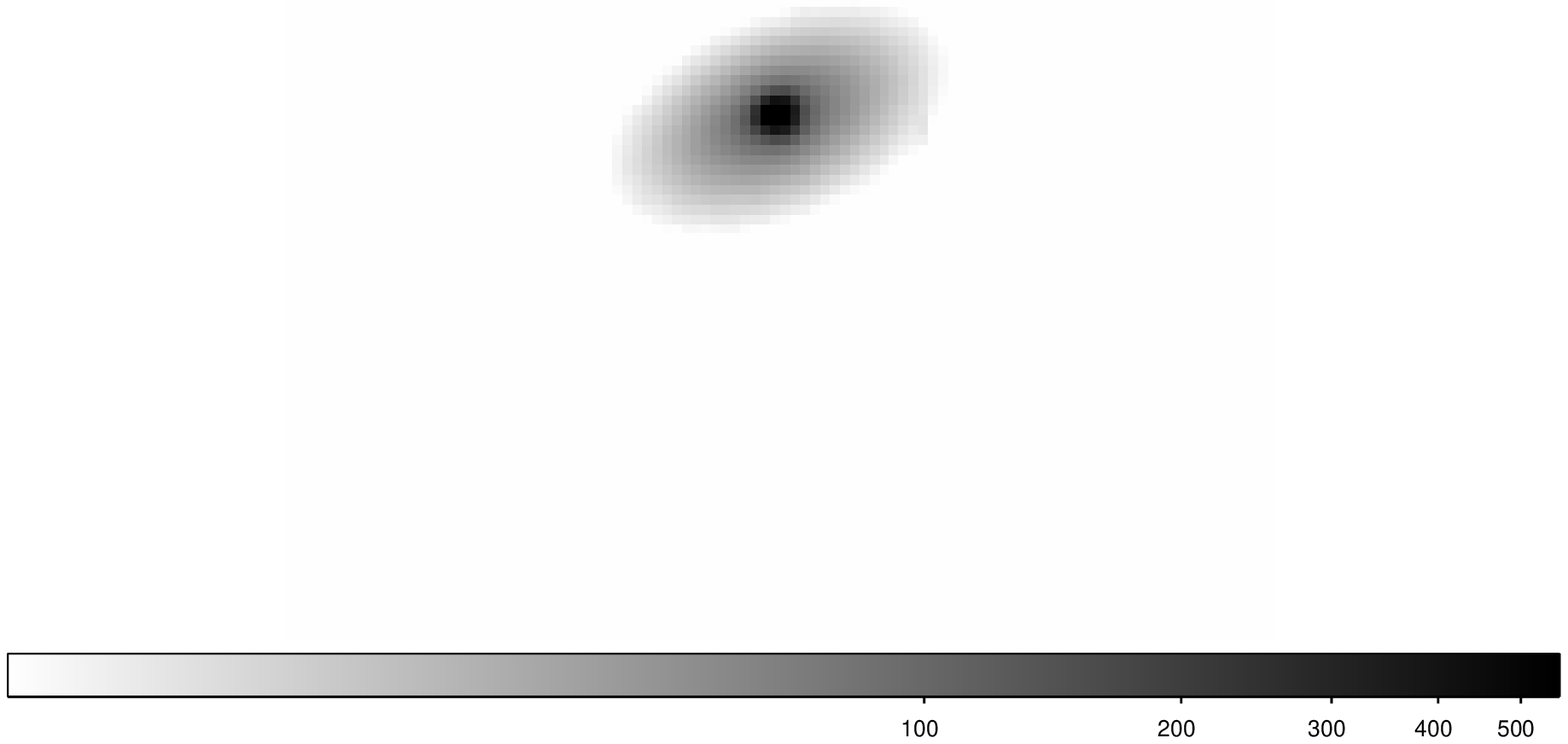}
\includegraphics[width=3.7cm]{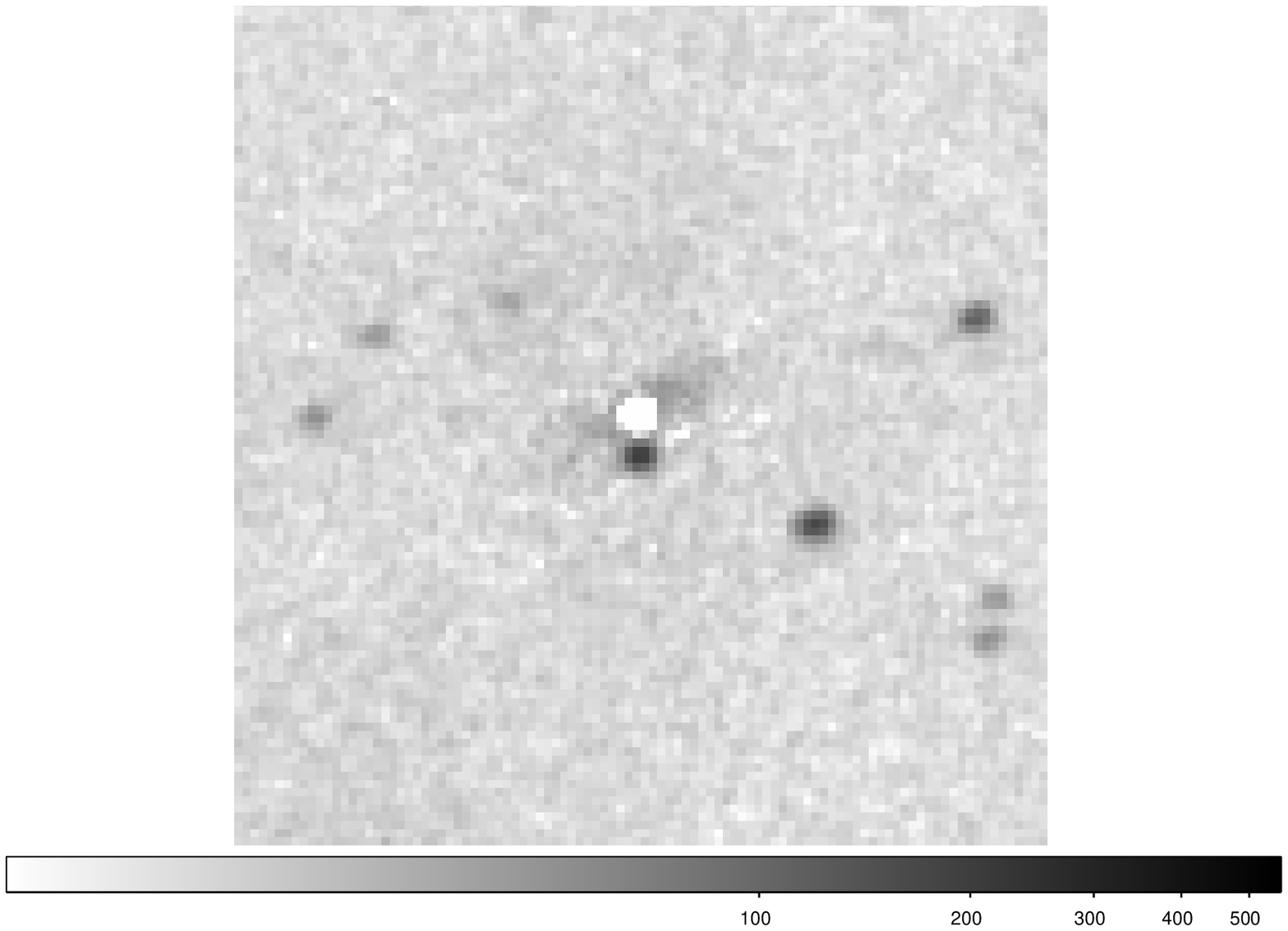}

\includegraphics[width=3.7cm]{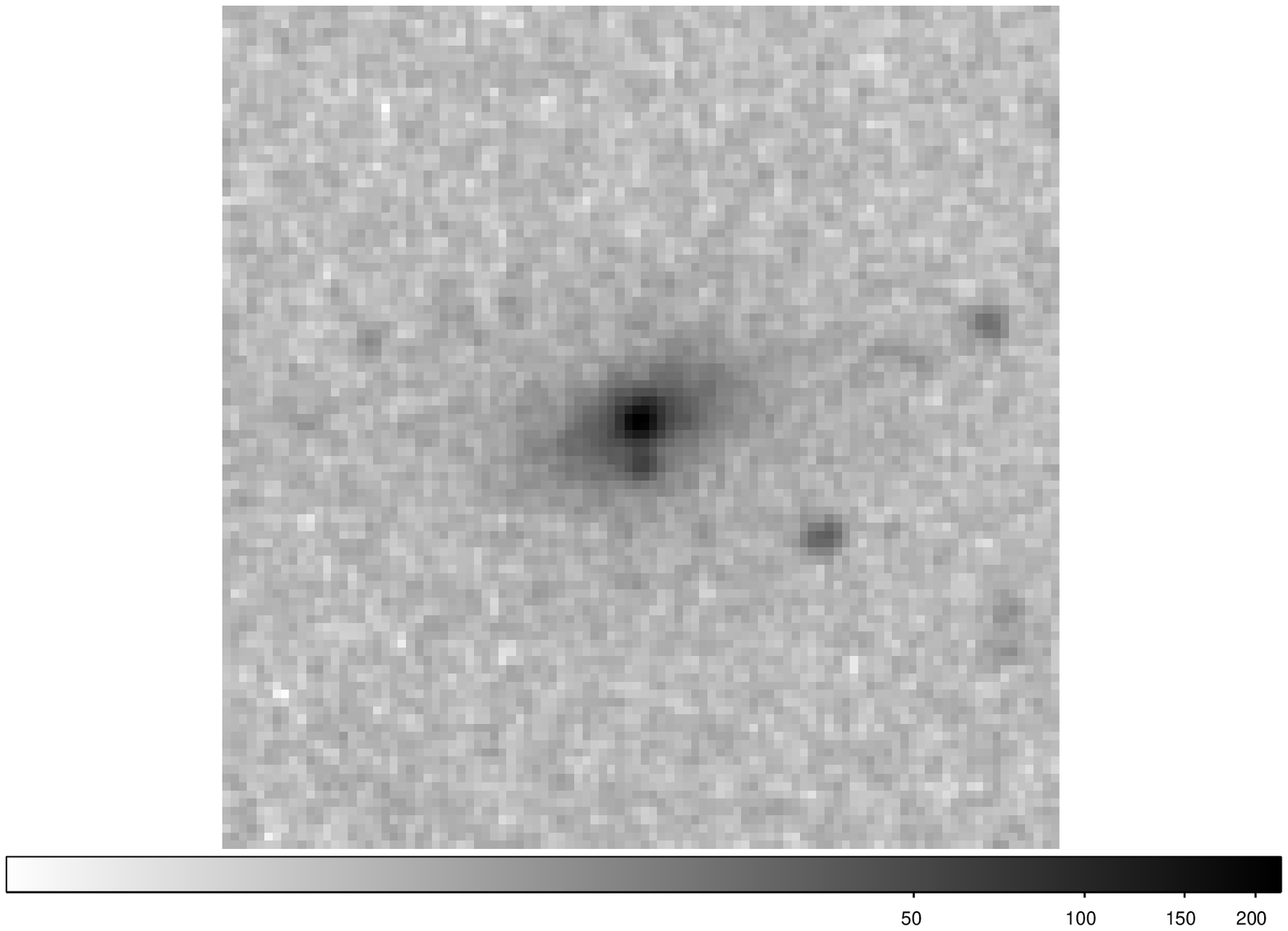}
\includegraphics[width=3.7cm]{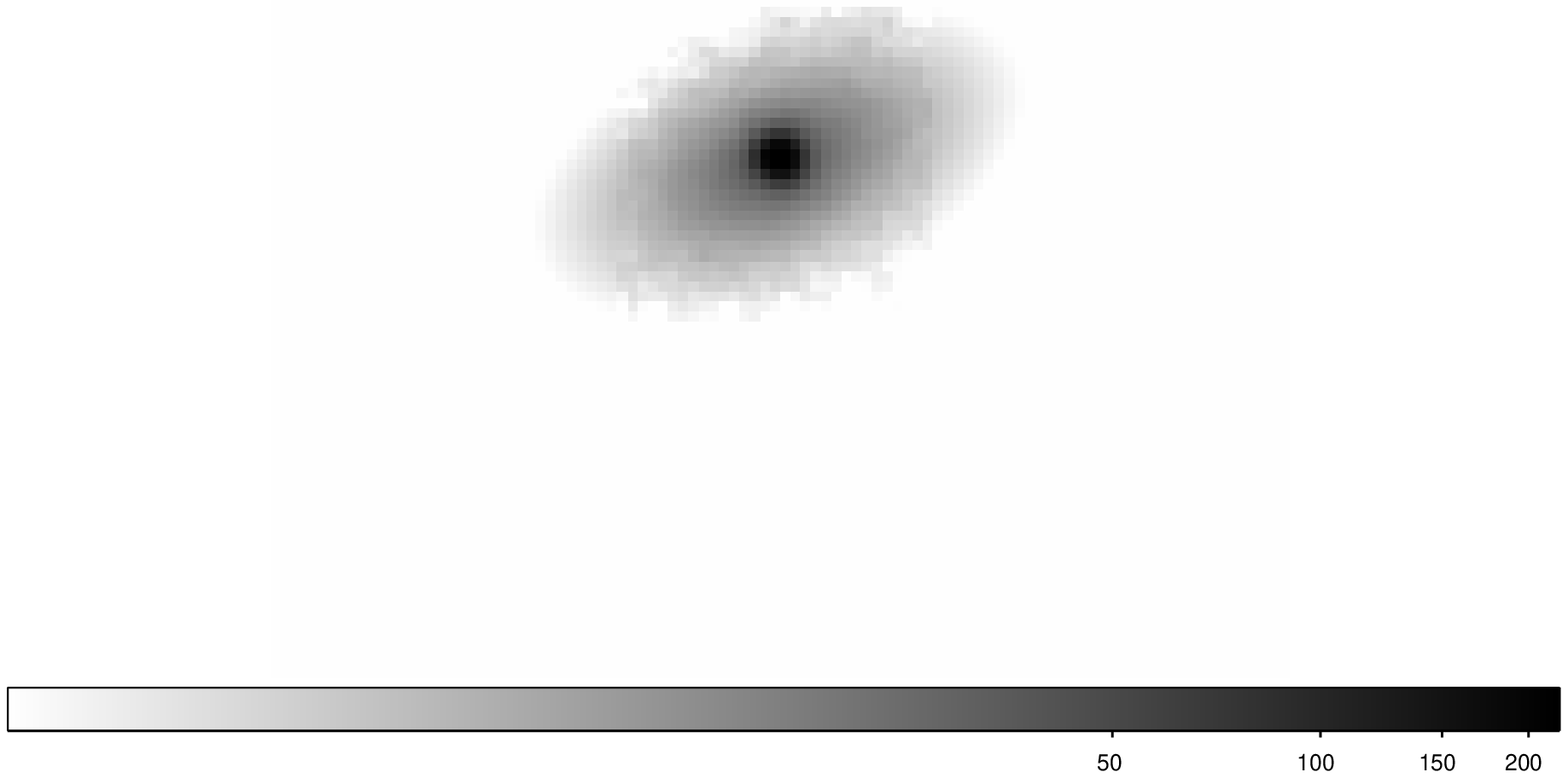}
\includegraphics[width=3.7cm]{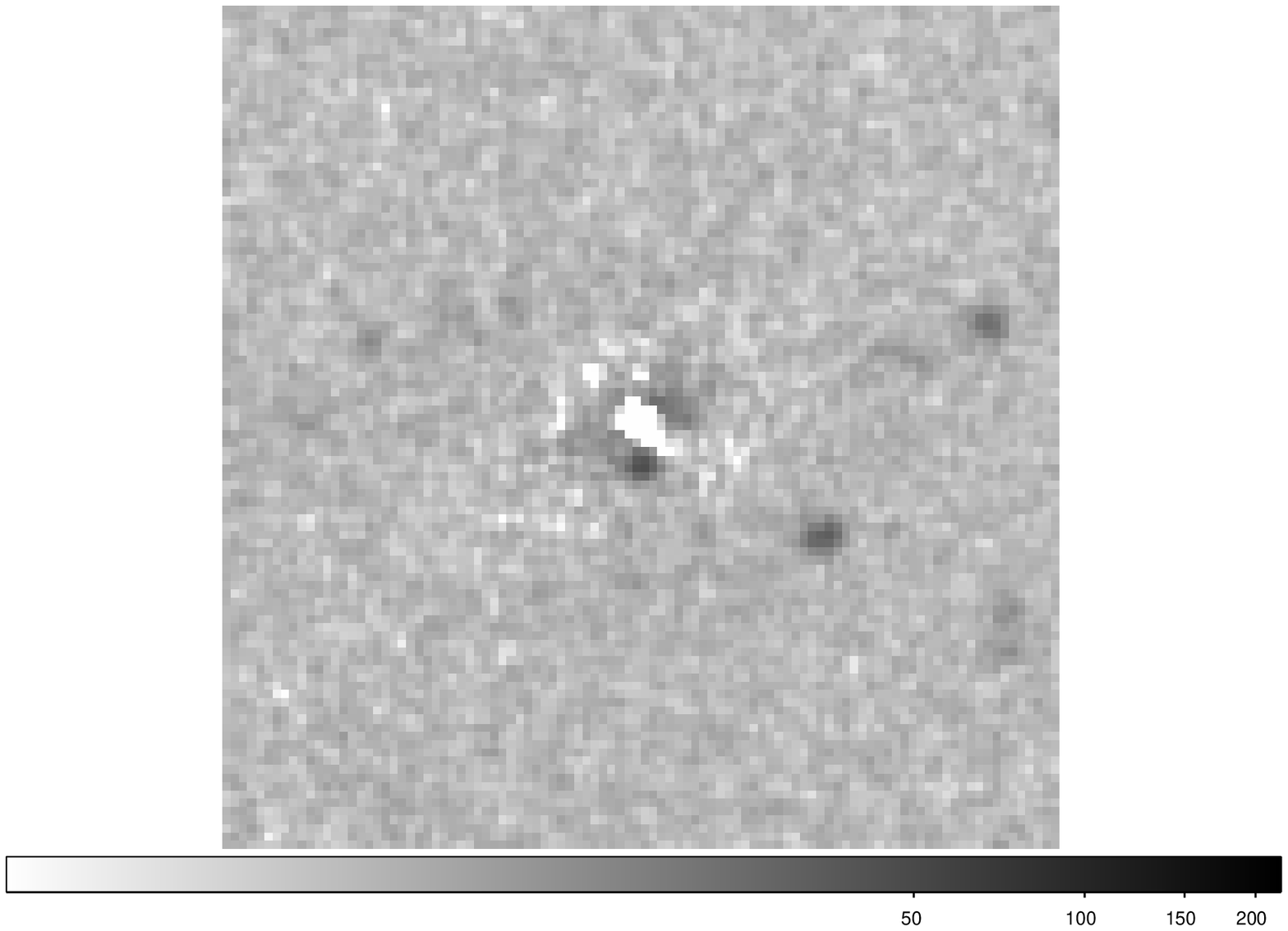}

\includegraphics[width=3.7cm]{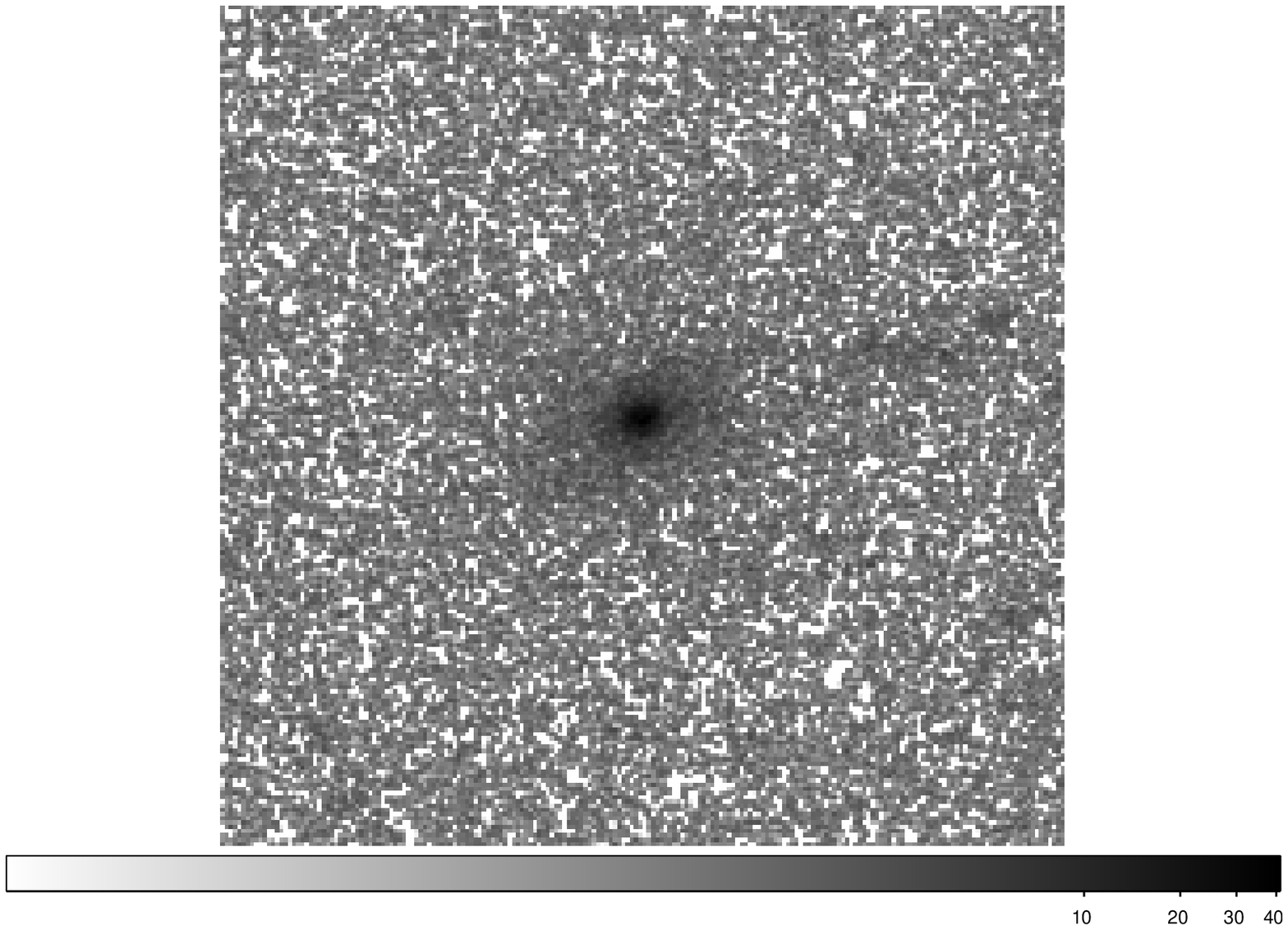}
\includegraphics[width=3.7cm]{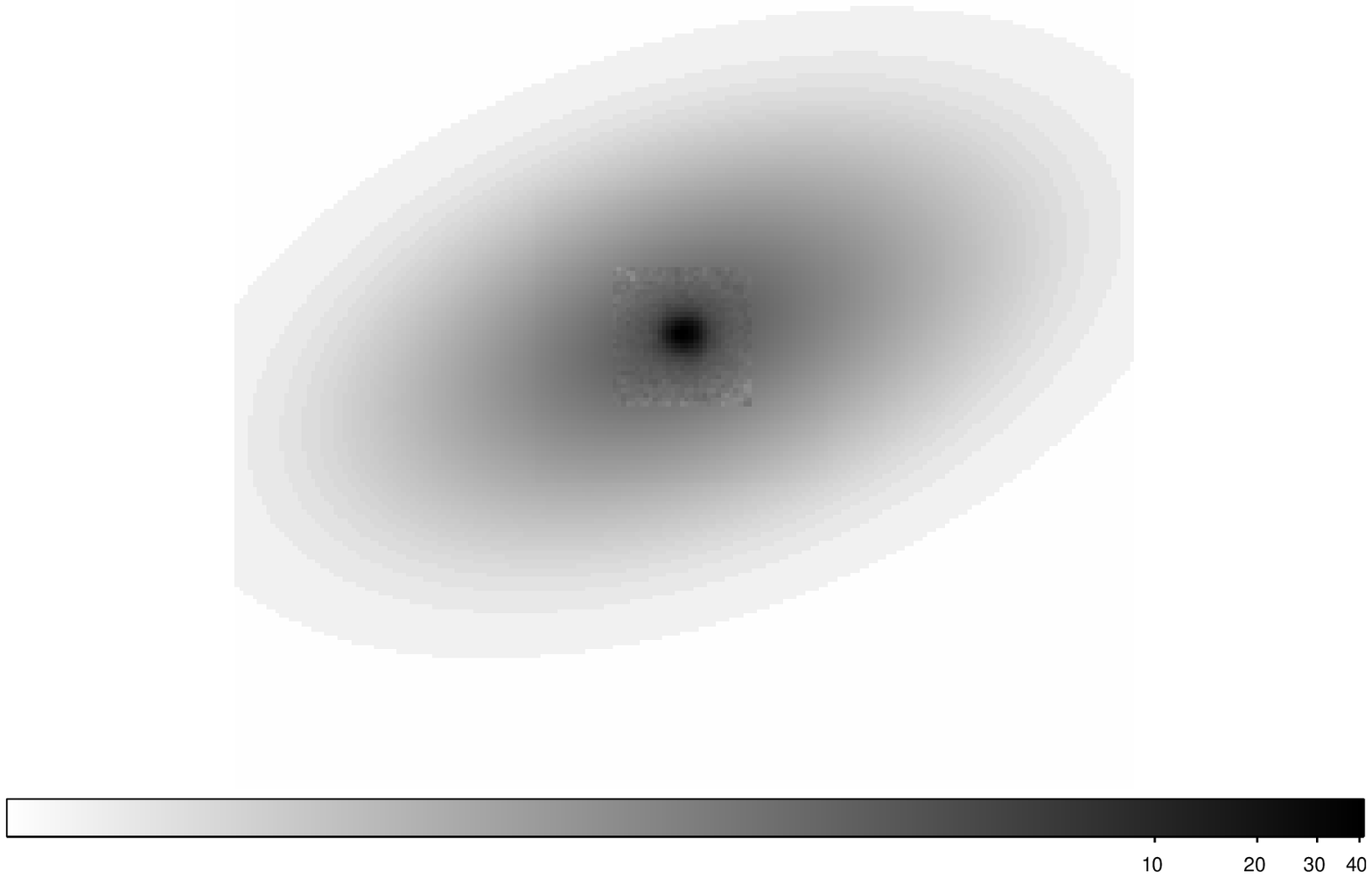}
\includegraphics[width=3.7cm]{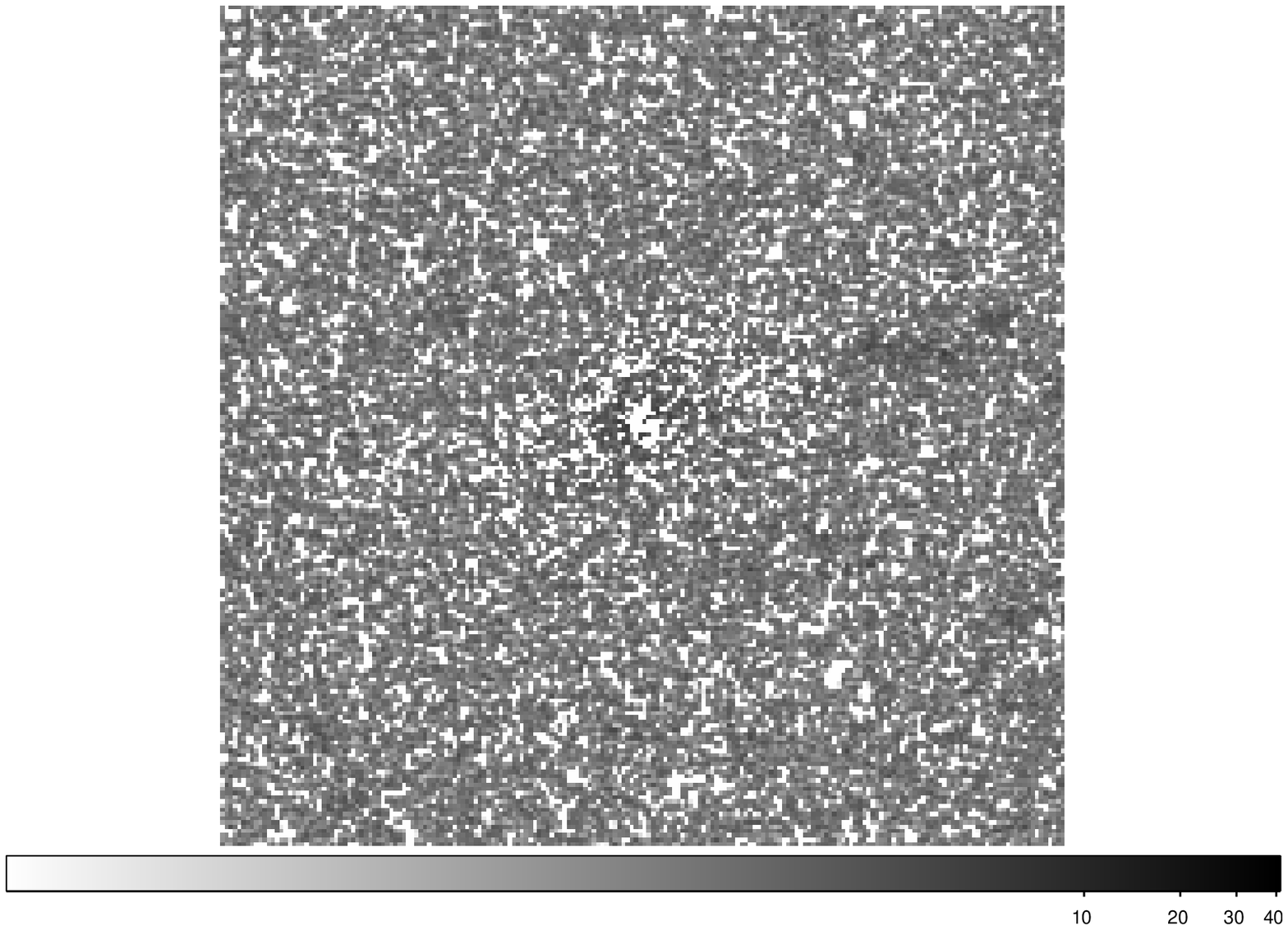}

\includegraphics[width=3.7cm]{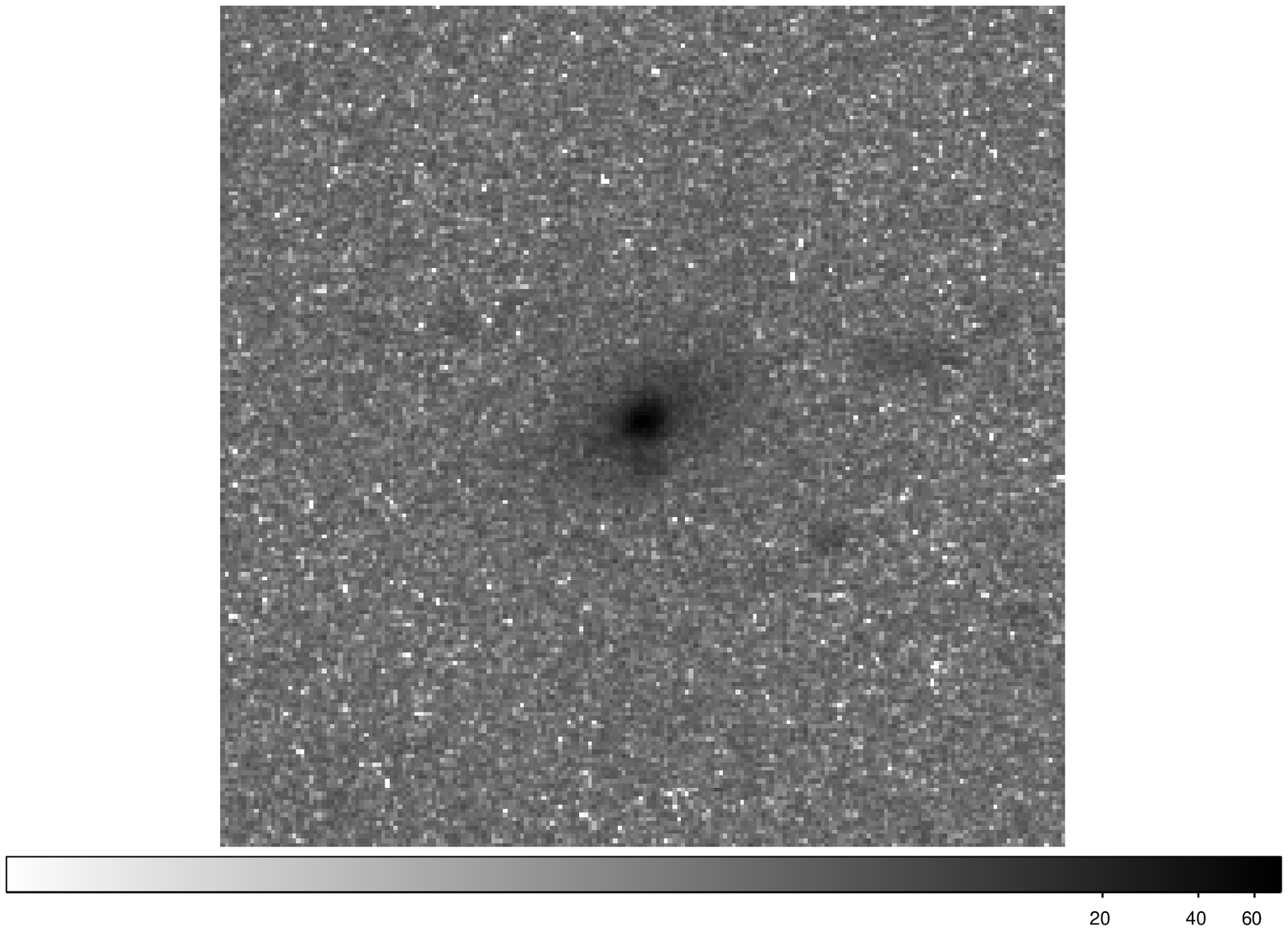}
\includegraphics[width=3.7cm]{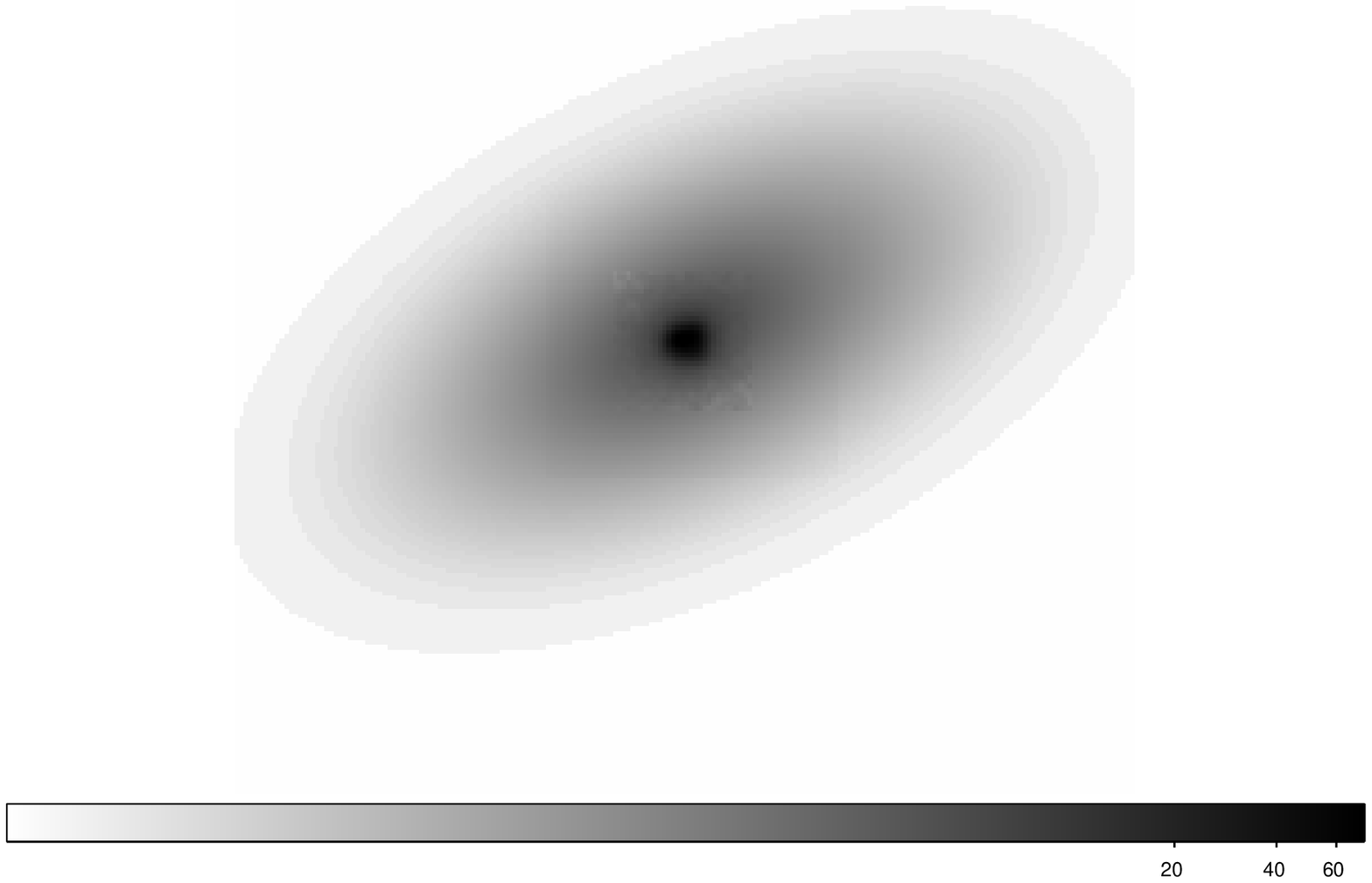}
\includegraphics[width=3.7cm]{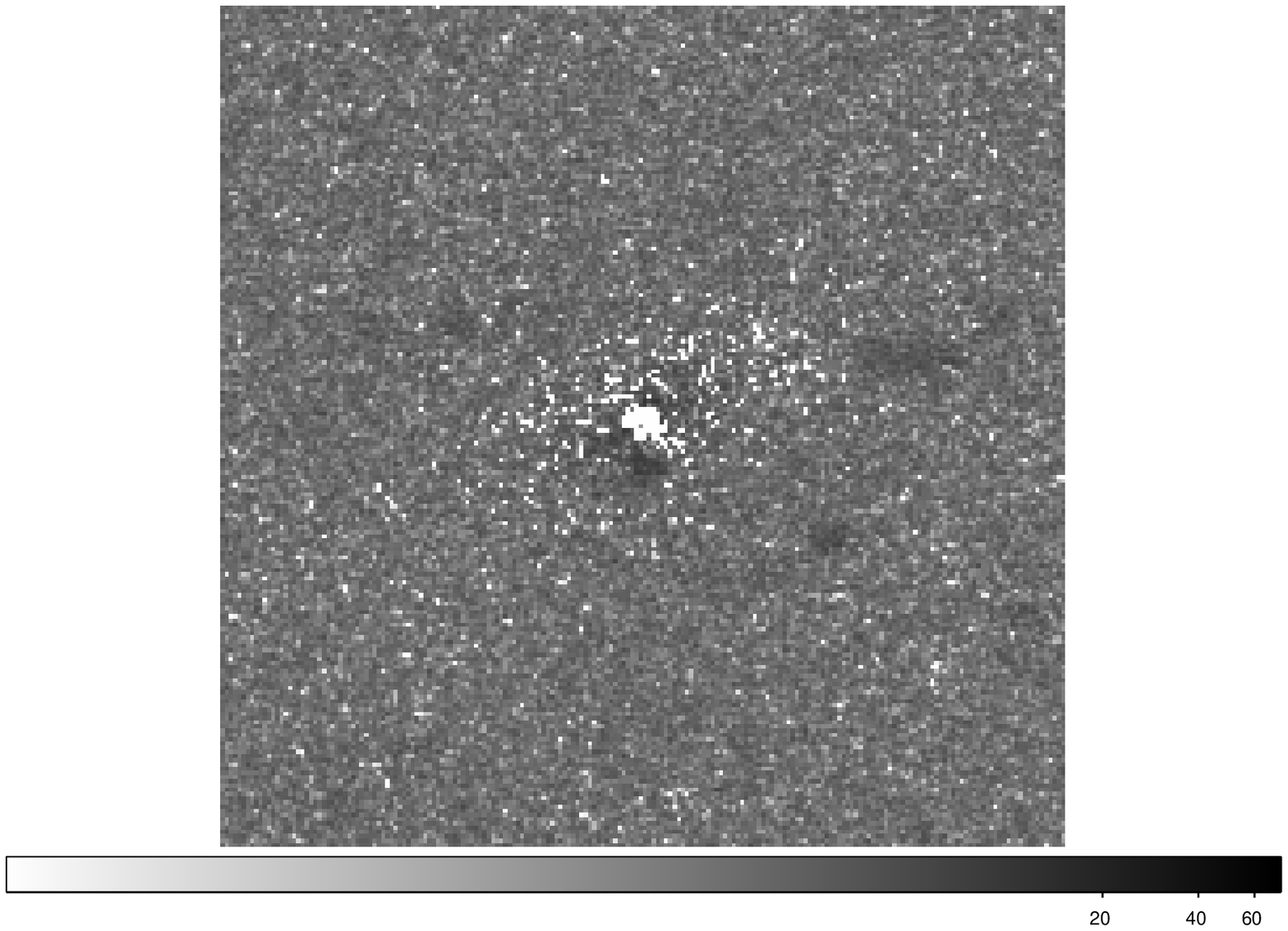}

    \caption{Example GALFIT profile fitting results for MCG-06-30-15.  The first column in each row shows the original data image, the second shows the model fit and the third shows the residual obtained from subtracting the model from the data.  The rows, from top to bottom, show the results for the V, B, U, UVW1, UVM2 and UVW2 bands, in that order.  In the V and B band, the residuals imply that a bulge component may also be discernible in this object, but the UVOT data did not allow robust fits including a bulge component for many of our objects.  A simple analysis using \textsc{ds9} regions indicate that the putative bulge may account for up to 10 per cent of the total galaxy flux in the V and B band images.}
\label{galfitimages_example}
\end{figure*}

\begin{table*}
\begin{tabular}{|l|l|l|l|l|l|l}
\hline
AGN&V&B&U&UVW1&UVM2&UVW2\\\hline
1RXS J045205.00+493248&P&P&P&P&--&--\\
2MASX J21140128+8204483&P+E&P&P+E&P&P+E&P+E\\
3C 120&P+E&P+E&P+E&P+E&M+E&P+E\\
3C 390.3&P+E&P+E&P&P&P+E&P+E\\
Ark 120&P+E&P+E&P+E&P+E&P+E&P+E\\
ESO 490-G026&P+E&P+E&P+E&P+E&P+E&P+E\\
ESO 548-G081*&P+E&P+E&P+E&P+E&M+E&M+E\\
IRAS 05589+2828&P&P&P&P&M&Gaussian\\
IRAS 09149-6206&P+E&M&P+E&P+E&P+E&P+E\\
MCG +04-22-042&P+E&P+E&P+E&P+E&M+E&P+E\\
MCG-06-30-15&P+E&P+E&P+E&P+E&P+E&P+E\\
Mrk 1018&P+E&P+E&P+E&P+E&M+E&M+E\\
Mrk 279&P+E&P+E&P+E&P+E&P+E&P+E\\
Mrk 352&P+E&P+E&P+E&P+E&P+E&M+E\\
Mrk 509&P+E&P+E&P+E&P+E&P+E&--\\
Mrk 590&P+E&P+E&P+E&P+E&P+E&P+E\\
Mrk 766&P+E&P+E&P+E&P+E&P+E&P+E\\
Mrk 841&P+E&P+E&P+E&P+E&M+E&P+E\\
NGC 4593&P+E&P+E&P+E&P+E&P+E&P+E\\
NGC 5548&P+E&P+E&P+E&P+E&P+E&P+E\\
NGC 7469&P+E&P+E&P+E&P+E&P+E&P+E\\
NGC 985&P+E&P+E&P+E&P+E&P+E&P+E\\
SBS 1136+594&P&P&P+E&P+E&P+E&P+E\\
SBS 1301+540&P+E&P&P+E&P&P&P+E\\
UGC 06728&P+E&P+E&P+E&P+E&--&P+E\\
WKK 1263&P+E&P+E&P&P&P&P\\
\hline
\end{tabular}
\caption{Table showing the model components used in GALFIT with each filter image for each AGN. `P' signifies a PSF generated from guide stars within appropriate count-rate ranges,from each image, `M' signifies a Moffat (King) profile with parameters estimated from a single point source in the image, and `E' signifies an exponential disk model component.\label{galfit_modelfits_table}. *For ESO 548-G081, an extremely bright foreground star was located within $\sim$30 arcsec of the AGN, rendering the photometry obtained susceptible to large errors.}
\end{table*}

\subsubsection{Comparison with HST}
\label{hstcomparison_nucmags}

The detailed study of \cite{2006ApJ...644..133B} highlights the need to take into account host galaxy contamination when calculating nuclear luminosities from optical HST images, specifically in the context of determining the radius--luminosity relationship for the broad line region (BLR) in AGN.  They find that even for an aperture of 1 arcsec, there can be significant host galaxy contamination.  However, the HST Advanced Camera for Surveys (ACS) offers a significant advantage with a PSF of FWHM 0.0575 arcsec in contrast to the typical FWHM of $\sim$2 arcsec for the UVOT PSFs.  This somewhat limits our ability to determine the host galaxy contamination, since the stars used to model the PSF often are as wide in extent on the sky as some of the more distant galaxy disks in our sample.  We attempt to gauge the possible underestimate of host galaxy contribution with UVOT using publically available images from the HST ACS and Wide Field and Planetary Camera 2 (WFPC2) for one of our objects, NGC 4593.  This object is known to have a clearly identifiable spiral galaxy structure surrounding the central AGN so provides a good foundation for comparison between UVOT and HST image quality.  Our UVOT analysis also shows a clear discrepancy between the 5.0 arcsec aperture photometry results and those from PSF fitting in the optical (V, B and U) bands.  Since some level of galaxy contamination is clearly identifiable using UVOT images, it should therefore be more readily identifiable in the HST images.  We perform simple aperture photometry using the \textsc{ds9} package on images from ACS for the F550M and F330W filters (centered on 5580$\rm \AA$ and 3353$\rm \AA$ respectively).  We extract counts from the nucleus with a circular region with radius 0.2 arcsec, which should provide an upper limit on the nuclear flux and contain $>$90 per cent of the counts from the nucleus.  We also determine the total counts from a 5 arcsec region, as used with \textsc{uvotsource} for comparison.

In the F550M filter (closest to the UVOT V-band in wavelength), we identify a difference in fluxes of a factor $\sim$5 between the smaller and larger apertures for HST.  In comparison, the difference in fluxes between the 5.0 arcsec \textsc{uvotsource} magnitude and that from PSF fitting to the UVOT image is a factor $\sim$2.4.  Assuming the 5.0 arcsec fluxes from the two instruments should theoretically be identical, this implies that the flux reported to the nucleus by PSF fitting to UVOT images is a factor $\sim5.0/2.4$ too large.  The problem is less acute in the F330W filter (closest to the UVOT U-band), with an identical analysis yielding that the nuclear flux from the UVOT image is a factor $\sim1.8/1.2$ of that from the HST image.   There is also the issue of optical AGN variability, which could be artificially increasing these ratios (the continuum near $5100\rm \AA$ is known to vary by factors of $\sim7$ in the case of NGC 5548; see \citealt{1999ApJ...510..659P}), but it is difficult to disentangle this from the error intrinsic to the instrument, and the factors we calculate here are therefore probably upper limits.  This reflects a fundamental limitation of the resolution of the UVOT images, but the subsequent analysis shows that in many of the objects seen, attempting to remove the host galaxy does produce an optical--UV SED shape closer to that expected for an AGN despite possible overestimates of the nuclear flux.  A more detailed analysis using observations from the Kitt Peak National Observatory (KPNO) 2.1m telescope should provide a much more detailed galaxy profile decomposition (Koss et al. in prep), which will make use of the smaller PSF available with the KPNO (FWHM $\sim$1.0--1.4 arcsec).

\begin{figure*}
\includegraphics[width=10cm]{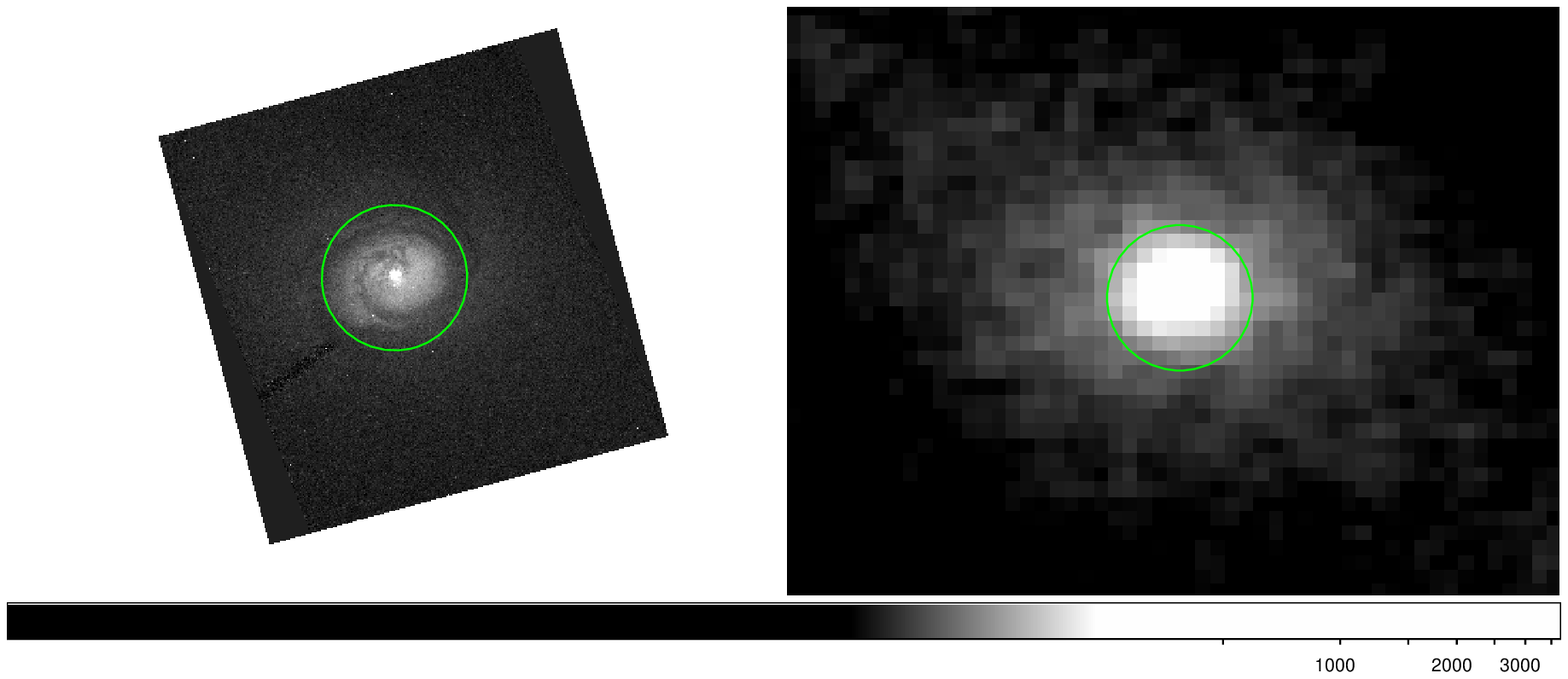}
    \caption{Comparison between images from HST (left) and UVOT (right) for the source NGC 4593.  A 5 arcsec circular region is shown for reference on both images.}
\label{galfitimages_example}
\end{figure*}

\subsection{Optical--UV photometry: Correcting for coincidence loss and Galactic extinction}

For AGN with very high count rates, the fluxes obtained from UVOT images are highly susceptible to coincidence losses and therefore often appear lower than their true values.  Coincidence loss refers to the phenomenon where multiple photons arrive at the same location on the detector during a single frame, and is discussed in detail in \cite{2008MNRAS.383..627P}.  They the theoretical correction to be applied to the measured count rate for individual pixels, but also point out the need to account for the spread of a point source over many pixels.  They also provide an empirical correction which takes the latter effect into account.  We correct all of our magnitudes from GALFIT for coincidence loss manually using these expressions; the \textsc{uvotsource} aperture photometry magnitudes are automatically corrected for coincidence loss.

There is some scope for error in the coincidence loss correction when the PSF image is not quite a point source, as the coincidence loss correction only holds good for point sources.  This problem is especially pronounced for very bright sources.  In these sources, the coincidence-corrected fluxes are sometimes even greater than those obtained from \textsc{uvotsource}, which should theoretically provide an upper limiting nuclear flux.  In cases where this phenomenon is very pronounced, we do not use the magnitudes from GALFIT and instead employ the 5.0 arcsec aperture photometry magnitudes in any further analysis.

After calculation of magnitudes from both GALFIT and \textsc{uvotsource}, these magnitudes were corrected for Galactic extinction using the values for $E(B-V)_{\rm Gal}$ from the NASA/IPAC Infrared Science Archive\footnote{http://irsa.ipac.caltech.edu/applications/DUST/} and the Galactic extinction curve of \cite{1989ApJ...345..245C}.  The final magnitudes were turned into \textsc{xspec} PHA files using the \textsc{flx2xsp} utility, to facilitate fitting along with the contemporaneous XRT data using the \textsc{xspec} analysis software.

\subsection{X-ray spectra from XRT}

The pipeline-processed  event files from the XRT detector were processed using the \textsc{xselect} package, as directed in the Swift XRT user guide.  Source regions of 50 arcsec were used, with larger accompanying background regions (average radius $\sim$150 arcsec).  Background light curves were determined from the event files and inspected for flaring, but this was not found to be a problem in any of our observations.  Source and background spectra were extracted, and the source spectra were grouped with a minimum of 20 counts per bin.

\subsection{Black hole mass estimates using the 2MASS All-Sky Survey Catalogues}
\label{bhmassesfrom2mass}

The $M_{\rm BH}-L_{\rm bulge}$ correlation for galaxies provides a useful way of estimating their central black hole masses.  The key challenge is to obtain an accurate estimate of the bulge luminosity.  The study of \cite{2003ApJ...589L..21M} presents a decomposition of images from 2MASS using GALFIT to obtain the bulge luminosity and correlate it with the black hole mass from direct determination methods (stellar or gas kinematics, maser kinematics, etc.).   However, the redshifts of the objects under scrutiny here are generally significantly higher than those in \cite{2003ApJ...589L..21M}, presenting problems for performing a full nucleus-bulge-disc decomposition.  \cite{2008arXiv0807.4695M} present a method for calculation of black hole mass for the Swift-BAT catalogue AGN from 2MASS K-band total source magnitudes, which involves subtracting the central nuclear component from the extended source flux to estimate the bulge luminosity.  We attempt to refine this method here by incorporating information on the expected angular size of the bulge on the sky and the resolution limitations of the 2MASS data. The crucial factor is the degree of seeing affecting the 2MASS observations; typically the seeing takes a value of 2.6 arcsec.  For bulges with typical size $\sim0.5-1$ kpc \citep{2008MNRAS.388.1708G}, the bulge becomes unresolvable at redshifts of about $z\approx0.01$ (assuming $H_{\rm 0}=71 \thinspace \rm km \thinspace s^{-1} \thinspace Mpc^{-1}$, $\Omega_{\rm M}=0.27$ and a flat Universe).  Therefore, for the overwhelming majority of objects in our sample, the bulge should be unresolved.  Inspection of the atlas images from 2MASS for the lowest redshift objects in our sample confirms that the visually identifiable bulge component is not resolved.  The bulge light is therefore mixed in with the nuclear light in these sources and identified as a `point source'.  The point sources identified by 2MASS have been collated, along with their magnitudes, in the \emph{2MASS Point Source Catalogue (PSC)}; similarly extended sources have been catalogued in the \emph{2MASS Extended Source Catalogue (XSC)}.  Based on our assumption that the bulge is unresolved, we initially download only the magnitudes from the PSC for further use, as they will contain the bulk of the bulge light we wish to recover.

By way of calibrating our attempt to obtain black hole masses, we first turn to the reverberation mapping (RM) sample of AGN presented by \cite{2004ApJ...613..682P}, representing the most secure mass determinations for AGN to date (augmented with the refined result from \citealt{2006ApJ...653..152D} for NGC 4593).  There are uncertainties in the RM method, connected to assumptions about the geometry of the broad-line region and whether it is gravitationally bound, but the $\sim35$ AGN for which RM mass estimates exist nevertheless constitute the sample of AGN with the most carefully determined black hole masses.  Initially, we download the 2MASS atlas images for those RM AGN below a redshift of 0.01, in which the bulge should be theoretically resolved and perform a GALFIT 3-component decomposition using a gaussian PSF for the nuclear point source, a Sersic bulge and an exponential disc.  Our GALFIT analysis yields that there is a significant degree of uncertainty in the identification of the bulge component in these AGN, and the magnitude ascribed to the bulge is heavily dependent on the prior constraints imposed.  We explore a number of different constraints on the model profiles; for example, the Sersic index $n$ is often constrained to an upper limit to avoid the fitting routine confusing it with the nuclear point source, and often requires a lower limit to avoid confusion with the disc component.  Despite such strategies, in some sources the bulge and disc profiles acquire very similar or identical radii and `share' the disc luminosity, with the Sersic profile partially tracing the disc instead of the bulge.  In such sources, three-component model fits do not yeild meaningful bulge parameters. Our attempts at constraining the bulge using GALFIT result in bulge luminosities that vary by up to an order of magnitude depending on the choice of model and constraints imposed, and show that even for the close by sources, the bulges are probably not resolved.

\begin{figure}
\includegraphics[width=9cm]{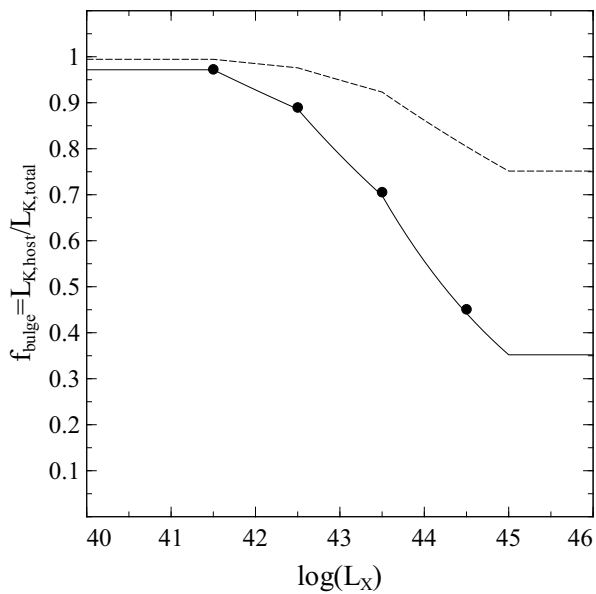}
    \caption{Variation of $L_{\rm K,host}/L_{\rm K,total}$ with X-ray luminosity, extrapolated from the host and nuclear IR SED templates of \protect\cite{2004MNRAS.355..973S} (black filled points - determined using the four fiducial luminosities provided in their paper; solid line - extrapolation between points as detailed in the text).  These fractions are employed to calculate the contribution of the bulge to the 2MASS point source magnitudes, for determining black hole masses.  The curve for an AGN with intrinsic X-ray absorption $N_{\rm H}=10^{23.5} \rm cm^{-2}$ is shown (dashed line) for comparison (the curves for other absorptions above $N_{\rm H}=10^{22} \rm cm^{-2}$ are very similar; the significant difference is between obscured and unobscured AGN).}
\label{hostoverLX_Kband}
\end{figure}

In light of this, we then employ the 2MASS PSC catalogue magnitudes to provide a simple estimate of the total bulge and nuclear flux.  In order to estimate the relative contributions of the nucleus and bulge in the PSC magnitudes, we employ the infrared SED templates presented in \cite{2004MNRAS.355..973S}.  In their study, they present nuclear IR SED templates constructed using available near-to-mid IR data on 33 Seyferts, using a radiative transfer models for dust heating to interpolate between the wavelengths covered by the data.  They also present host galaxy SED templates by subtracting the nuclear template contribution from the total photometry at each available wavelength.  They present a number of different nuclear SED templates appropriate for different levels of X-ray absorption, and host SED templates are presented for different intrinsic 2--10 keV AGN luminosity ($\rm L_{X}$) regimes.  Using the appropriate nuclear SED along with the corresponding host galaxy SED, it is therefore possible to estimate the fraction of the total luminosity which can be accounted for by the host for an AGN with a particular $L_{\rm X}$.  In the process we make the assumption that, in the case of the K-band, the host flux should be dominated by the bulge as discussed in \S\ref{Intro}.  We calculate the ratio of host to total luminosity (nuclear plus host luminosity) from these SED templates in the K-band for the four different luminosity regimes ($\rm 41.0< log(L_{X})<42.0$, $\rm 42.0< log(L_{X})<43.0$, $\rm 43.0< log(L_{X})<44.0$, $\rm 44.0< log(L_{X})<45.0$).  The fractions are plotted in Fig.~\ref{hostoverLX_Kband} against the central luminosity of each bin.  We also plot a logarithmic interpolation between the points, to allow determination of the ratio of $L_{\rm K,host}/L_{\rm K,total}$ at a general value of $\rm L_{X}$, assuming the ratio varies continuously.  For values of $\rm L_{X}$ greater than those spanned by the \cite{2004MNRAS.355..973S} study, we simply employ the fraction calculated for $\rm L_{X}=10^{45} erg \thinspace s^{-1}$, extrapolating from the direction of the trend shown between the bins centred on $\rm log(L_{X})=43.5$ and $44.5$.  For luminosities $\rm log(L_{X})<41.5$, we use the fraction evaluated at $\rm log(L_{X})=41.5$ (extrapolating from the trend would yield fractions larger than Unity). The high luminosity extrapolation is clearly not likely to be accurate for extremely powerful objects ($\rm log \thinspace L_{X}>46$), but none of the reverberation mapped AGN or the AGN in our Swift-BAT subsample are in this regime.  In the absence of information on IR SEDs outside this luminosity range we do not attempt a more complex extrapolation.  

Before using these fractions to calculate bulge luminosities from the 2MASS PSC magnitudes, we consider potential sources of bias.  Since \cite{2004MNRAS.355..973S} adopt a reprocessing scenario to account for the K-band continuum, it is possible that the model nuclear K-band luminosity is a function of the Eddington ratio (since it is ultimately linked to the UV accretion disc luminosity, the dominant component of the total accretion luminosity).  The X-ray luminosity can be recast as the product of the Eddington ratio, bolometric correction and black hole mass (via the Eddington ratio), so the trend seen in Fig.~\ref{hostoverLX_Kband} could be the result of two functions of Eddington ratio being plotted against each other.  However, we again note the low X-ray luminosities probed by the Swift/BAT 9-month catalogue (up to a few times $10^{44} \rm erg \thinspace s^{-1}$), and find that the host-to-total fraction varies at most by a factor of two in this regime.  Therefore, the dependence is not strong and will not introduce significant biases compared to other sources of error, such as the intrinsic spread in RM masses and the intrinsic dispersion of the $M_{\rm BH}-L_{\rm K,bulge}$ relation.  Regardless of the particular model used to predict the nuclear K-band continuum, it is clear that the fraction of an unresolved bulge that is attributable to the nucleus will decrease for lower luminosity AGN and vice versa; the simple relationship in Fig.~\ref{hostoverLX_Kband} provides an observationally-rooted estimate of this effect.

\begin{figure}
\includegraphics[width=9cm]{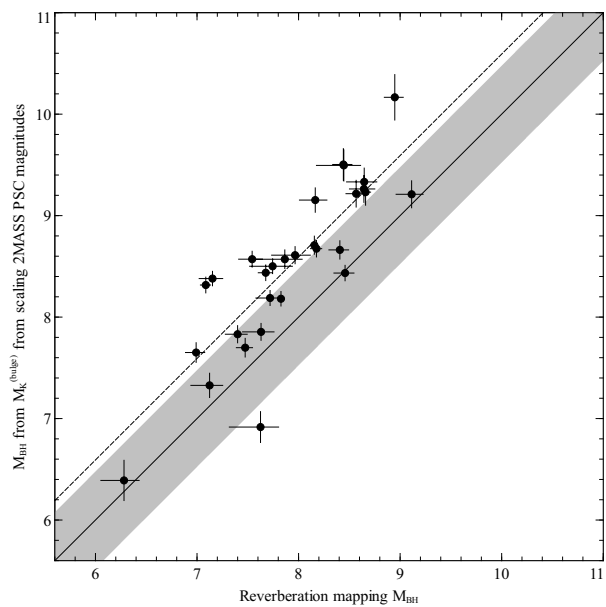}
    \caption{Comparison between masses from reverberation mapping and masses estimated from K-band bulge luminosities.  The solid line represents the desired one-to-one correspondence between the two methods and the shaded area depicts the intrinsic uncertainties in reverberation mapping.  The dashed line shows the best fit of the form ${\rm log}(M_{\rm BH,L_{K,bulge}})=A + {\rm log}(M_{\rm BH,revamp})$.}
\label{revmap_vs_kband_mass}
\end{figure}

We employ the values of $\rm L_{X}$ from XMM-Newton reported in \cite{2009MNRAS.392.1124V} to calculate the K-band host-to-total ratio for each of the RM objects, and assume the \cite{2004MNRAS.355..973S} Seyfert 1 IR SED template for all objects in the sample. In the absence of a detailed, uniform study on the absorbing columns of the RM AGN sample this is a reasonable assumption, since the reverberation mapping sample is heavily biased towards Seyfert 1 AGN; none of the RM AGN are classified as Seyfert 2s in NED.  We scale the PSC magnitude by the appropriate fraction, using the expression

\begin{equation}
\label{bulgemag}
M_{\rm K, bulge}=M_{\rm K, PSC} - 2.5 {\rm log_{10}}(f_{\rm bulge}(L_{\rm X})),
\\
\end{equation}
where the fraction $f_{\rm bulge} = \rm L_{\rm K,host}/L_{\rm K,total}$ is calculated as above.  The black hole masses are then calculated using the $M_{\rm BH}-L_{\rm K,bulge}$ relation of \cite{2003ApJ...589L..21M}.  The resulting mass estimates are compared with RM mass estimates in Fig.~\ref{revmap_vs_kband_mass}.

A reasonable agreement between the two methods is seen, albeit with some significant deviation at higher masses.  The grey shaded area highlights the uncertainties in RM masses due to the lack of precise knowledge of the BLR geometry.  The largest discrepancy is for the AGN with the highest mass, 3C 273, for which the K-band bulge luminosity estimate gives a black hole mass an order of magnitude too high.  This is known to be a very powerful object, and it is possible that our simple approach to estimating the bulge fraction in the point source light has still underestimated the powerful nucleus.  There is also likely to be a significant synchrotron (non-accretion) contribution to fluxes at all wavelengths in this source, which further increases uncertainties. This AGN is far more powerful than the majority of the Swift-BAT catalogue AGN, so such problems are not likely to affect the lower-power objects we consider here.

\begin{figure}
\includegraphics[width=9cm]{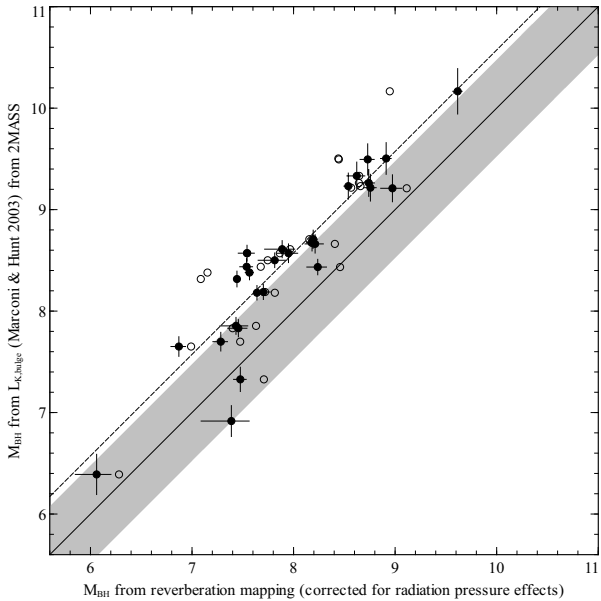}
    \caption{Comparison between masses from RM (corrected for radiation pressure effects as discussed by \protect\citealt{2008ApJ...678..693M}) and masses estimated from K-band bulge luminosities.  The empty white circles show the results using the reverberation masses before correcting for radiation pressure.  Other key conventions as in Fig.~\ref{revmap_vs_kband_mass}}
\label{revmap_RADPRESSCORR_vs_kband_mass}
\end{figure}

\cite{2008ApJ...678..693M} point out that the previous reverberation mass estimates have been calculated with the assumption that the BLR is gravitationally bound, and introduce a correction based on an estimator of the accretion rate (the $5100 \rm \AA$ monochromatic luminosity) to provide revised mass estimates for the RM AGN catalogue.  We also provide a comparison with their revised mass estimates, in Fig.~\ref{revmap_RADPRESSCORR_vs_kband_mass}.  The correlation between the two mass estimators is strengthened significantly if these first-order corrections to the reverberation masses are applied, and particularly reduces the scatter seen for those objects with high accretion rates (for which the correction is maximal and works to increase the reverberation mass estimate).  The offset is still present however, and a simple fit of the form ${\rm log}(M_{\rm BH,L_{K,bulge}})=A + {\rm log}(M_{\rm BH,revamp})$ yields $A\approx0.57$ in both cases (using reverberation masses with or without correction for radiation pressure), implying an offset of a factor of up to $\sim3.7$ in black hole mass, just outside the typical quoted tolerance for reverberation masses (a factor of $\sim 3$).  We now suggest a possible contributor to this offset and present a correction for it.

The \cite{2004MNRAS.355..973S} host galaxy SEDs have been calculated using large-aperture and extended-source data (such as the 2MASS XSC catalogue magnitudes) which could include more than the bulge light.  If the galaxy disc is significant in some cases, the \cite{2004MNRAS.355..973S} host-to-total ratios actually provide the fraction of bulge \emph{plus galaxy disc} to the total luminosity rather than just the bulge, and therefore using them to scale the `nucleus plus bulge' luminosities from the 2MASS PSC is not strictly correct.  It is therefore necessary to gauge the effect of neglecting the galaxy disc in this calculation.  We do this by obtaining the 2MASS Extended Source Catalogue magnitudes (matching the centroids appropriately as before), using these fluxes as estimates of the total nuclear, bulge and disc light ($L_{\rm total}=L_{\rm nuc}+L_{\rm bulge}+L_{\rm disc}$).  Employing the PSC magnitudes as the bulge and nuclear light as before ($L_{\rm bulge}+L_{\rm nuc}$), we then can utilise the fraction of nuclear-to-total luminosity predicted from the \cite{2004MNRAS.355..973S} IR SED templates ($f_{\rm nuc}=L_{\rm nuc}/[L_{\rm nuc}+L_{\rm bulge}+L_{\rm disc}]$) to obtain a theoretically more accurate estimate of the bulge magnitude:

\begin{equation}
\label{bulgemag_withxsc}
M_{\rm K, bulge}^{(2)}=M_{\rm K, PSC} - 2.5 {\rm log_{10}}(1-f_{\rm nuc} 10^{\frac{M_{\rm K,PSC}-M_{\rm K,XSC}}{2.5}})),
\\
\end{equation}
This expression is obtained by solving for $L_{\rm bulge}$ from the available quantities.  Put simply, it arises from using the nuclear fraction from the IR SED templates to estimate the nuclear flux from the XSC flux; this nuclear component is then subtracted from the PSC flux to obtain the bulge flux.  This estimate is not available for all of the objects available previously, as some objects do not have 2MASS XSC magnitudes.  For a couple of sources, the predicted fraction $f_{\rm nuc}=L_{\rm nuc}/(L_{\rm nuc}+L_{\rm bulge}+L_{\rm disc})$ from the \cite{2004MNRAS.355..973S} templates was larger than the fraction $(L_{\rm nuc}+L_{\rm bulge})/(L_{\rm nuc}+L_{\rm bulge}+L_{\rm disc})$ found from the ratio of the PSC and XSC fluxes, leading to an impossible solution for $L_{\rm bulge}$.  This indicates that the nuclear-to-total fraction obtained from the \cite{2004MNRAS.355..973S} templates is not appropriate for that object, or that the bulge is not completely contained in the PSC flux.  These problems limit the use of this method, but for those objects where a calculation is possible, we find that the new K-band mass estimates are systematically shifted down by 0.14 dex (a factor $0.72$) from the values from the simpler approach of using the PSC magnitude alone. We therefore apply this correction to all masses obtained using the PSC alone, which brings the best-fit line for the new mass estimates within the tolerance of the reverberation masses.  The strength of the correlation is good, as indicated by the Pearson correlation coefficients of 0.83 and 0.93 obtained (correlating with reverberation masses before and after radiation pressure correction).  This analysis also indicates that the galaxy disc contribution in the host-to-total ratios is of the order $\sim30$ per cent, and a similar value applies to all AGN in the RM sample. There still remains some offset above reverberation and within the tolerance region, but any further attempt to account for this is somewhat redundant in light of the intrinsic uncertainties in reverberation masses, so we do not make any other corrections to our approach.  The offset of virial mass estimates below the $M_{\rm BH}-L_{\rm bulge}$ relation has been noted and discussed in detail by \cite{2008ApJ...687..767K}; they find that it is a complex function of a variety of parameters including the black hole accretion rate and host galaxy morphology.

\begin{figure}
\includegraphics[width=9cm]{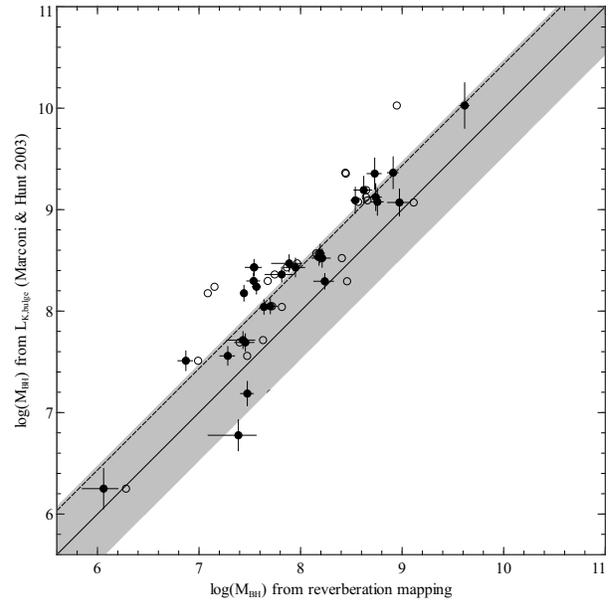}
    \caption{Comparison between masses from reverberation mapping and masses estimated from K-band bulge luminosities (with offset due to underestimation of the galaxy disc contribution taken into account).  Key as in Fig.~\ref{revmap_RADPRESSCORR_vs_kband_mass}.}
\label{revmap_RADPRESSCORR_vs_kbandOFFSET_mass}
\end{figure}

\begin{figure}
\includegraphics[width=9cm]{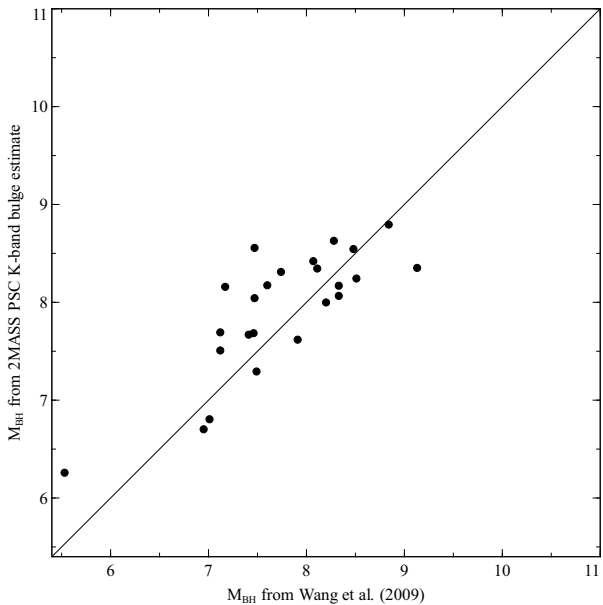}
    \caption{Comparison between masses from $\rm {H}-\beta$ linewidths \protect\citep{2009AJ....137.3388W} and masses estimated from K-band bulge luminosities.  The solid line represents the desired one-to-one correspondence between the two methods. \protect\cite{2009AJ....137.3388W} do not provide error estimates on their black hole masses, and the typical (random) error in the K-band estimates can be gauged from Figs~\ref{revmap_vs_kband_mass} and \ref{revmap_RADPRESSCORR_vs_kband_mass} (it is typically dominated by uncertainties intrinsic to the $M_{\rm BH}-L_{\rm bulge}$ relation), although the discussion in the text highlights that systematic errors are likely to dominate.}
\label{wangetal09_vs_kband_mass}
\end{figure}

We finally calculate the masses using this approach for all of the $z<0.1$ objects in the Swift-BAT 9-month catalogue for which 2MASS PSC magnitudes are available.  The K-band host-to-total ratios are calculated using the $L_{\rm X}$ values reported in \cite{2009ApJ...690.1322W}, and the $N_{\rm H}$ values in the same paper are used to select between the two host-to-total ratio curves for obscured or unobscured AGN (Fig.~\ref{hostoverLX_Kband}) as required.  By way of another check on the accuracy of our mass estimates, the recent study of \cite{2009AJ....137.3388W} presents virial mass estimates for a sample of Seyferts which partially overlaps with Swift-BAT 9-month catalogue AGN, using the single-epoch $\rm {H}\beta$ linewidth-based black hole mass estimator (as given in \citealt{2005ApJ...630..122G}).  We plot a comparison between their values and ours in Fig.~\ref{wangetal09_vs_kband_mass} for the 24 overlapping objects.  While the relation shows some scatter, we caution that the linewidth-based black hole mass estimation methods also show significant scatter themselves when compared to RM.  The two estimators still appear to trace each other (with a correlation coefficient of 0.80); additionally, preliminary calculations of black hole masses from \emph{Kitt Peak National Observatory} (KPNO) and \emph{Sloan Digital Sky Survey} (SDSS) spectra for the Swift/BAT 9-month catalogue AGN also seem to match our estimates well, with a scatter of a factor of $\sim2$ (Winter et al., Koss et al. in prep).  We see some indications that the local Swift-BAT 9-month catalogue AGN have predominantly low black hole masses, as expected in `anti-hierarchical' structure formation scenarios.  The masses thus determined are used in all subsequent analyses for calculation of accretion rates and for constraining accretion disc models in SED construction.

\section{Generation of Spectral Energy Distributions}

We fit the optical, UV and X-ray SED data points with a simple multicolour disc \textsc{diskpn} and absorbed power-law model combination using XSPEC, following the approach outlined in VF09.  In this work, we extend the approach of VF09 by inclusion of a multiplicative \textsc{zdust} component on the \textsc{diskpn} model used for the UV bump, to estimate intrinsic reddening where possible.  The presence of data in at least four UVOT filters for all of the objects in the sample makes this possible, and indeed since most of the objects have UVOT data in all six filters, we are able to constrain the shape of the SED over a wider energy range than was generally possible in VF09.  Signs of intrinsic reddening (such as an optical--UV spectrum steeply declining with energy) can then be more unambiguously identified.

The full model used in XSPEC consisted of the optical--UV reddening component multiplied by a thermal accretion disk model, and a broken power law in X-rays with both Galactic and intrinsic absorption components (\textsc{zdust(diskpn) + wabs(zwabs(bknpower))}).  The normalisation of the \textsc{diskpn} component was determined as in VF09, assuming an inclination angle of zero and a colour-to-temperature ratio of Unity.  Unification scenarios for AGN \citep{1993ARA&A..31..473A} suggest that AGN with lower gas column densities such as the ones in this sample should be preferentially oriented with their accretion disks almost face-on, and therefore small inclination angles are appropriate (see VF09 for a discussion of the effect of varying inclination angle in the \textsc{diskpn} component on luminosities calculated from SEDs).  We also freeze the inner radius of the disc at 6.0 gravitational radii, which is appropriate for radiatively efficient accretion.  The local Galactic absorption was determined using the \textsc{nh} utility from the \textsc{ftools} suite and frozen in the fit.  The low energy branch of the \textsc{bknpower} model was set to have negligible values within the energy range of the UV bump (as in VF09).  We fit the XRT data in the range 0.3--10.0 keV. While there is an interesting debate as to the precise nature of the intrinsic reddening in AGN \citep{2004ApJ...616..147G}, we adopt the Small Magellanic Cloud (SMC) reddening curve for our `first-order' estimation of intrinsic extinction $E(B-V)$ and freeze the ratio of total to selective extinction, $R_{V}$, at 3.0.  The intrinsic extinction $E(B-V)$, the intrinsic column density $N_{\rm H}^{(int)}$, the maximum temperature $T_{\rm max}$ for the \textsc{diskpn} model and the parameters for the high-energy branch of the \textsc{bknpower} component were all left free in the fit.  Occasionally, if the combined optical, UV and X-ray data resulted in a photon index in the X-ray regime which was influenced by the slope between optical and X-ray, the photon index was frozen at the value obtained from just fitting the X-ray data alone.

\begin{figure*}
    \includegraphics[width=4.5cm]{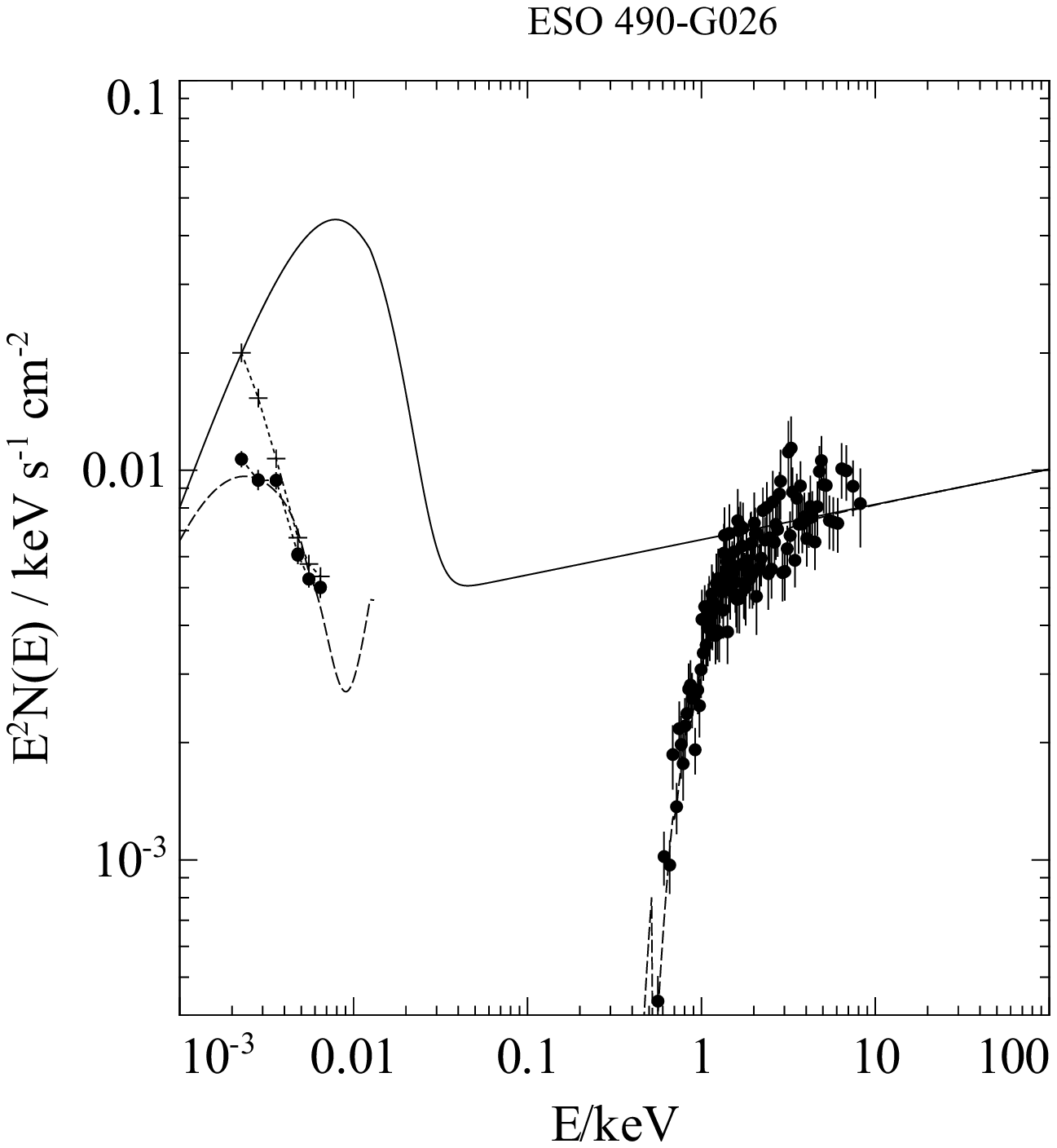}
    \includegraphics[width=4.5cm]{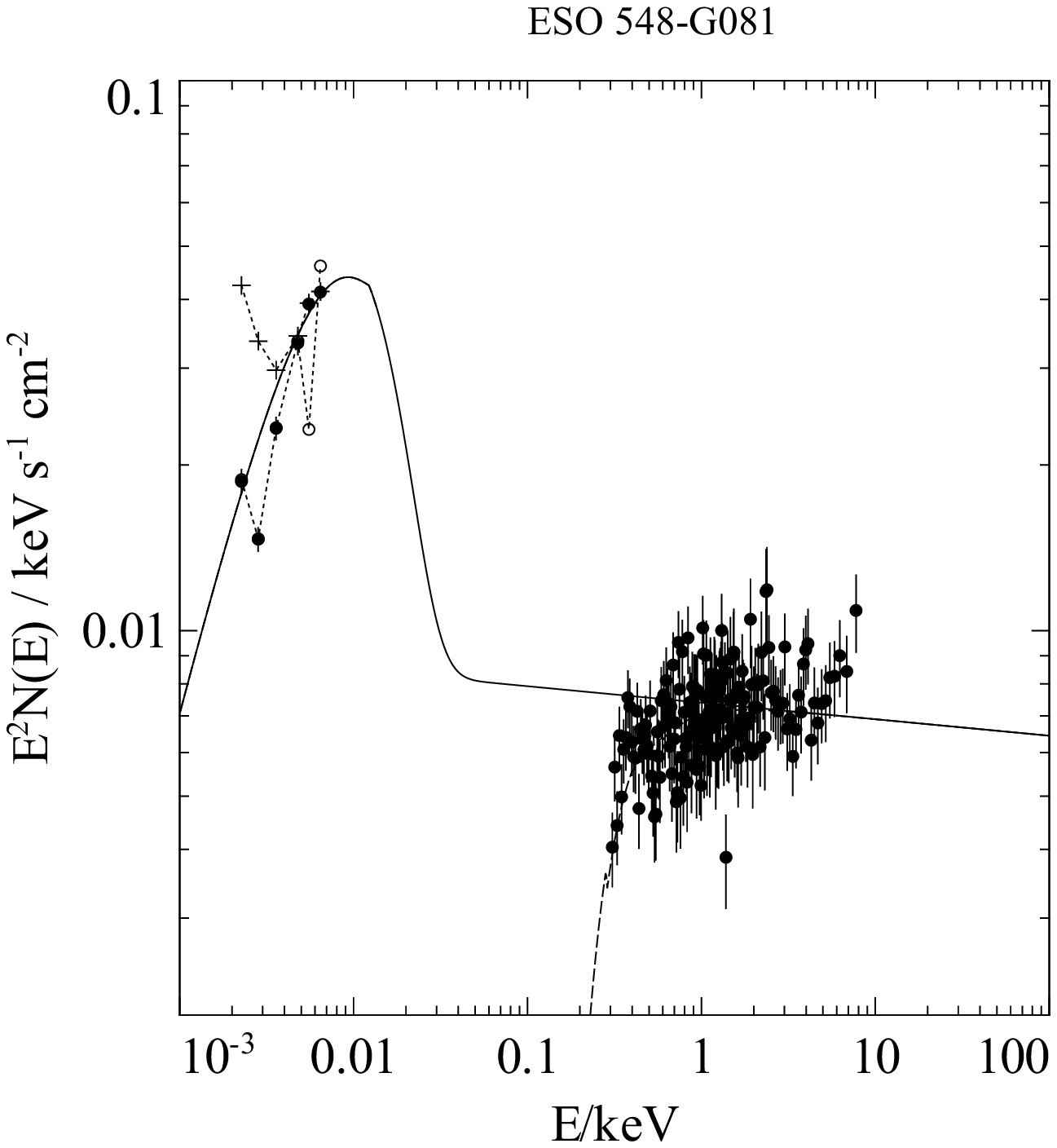}
    \includegraphics[width=4.5cm]{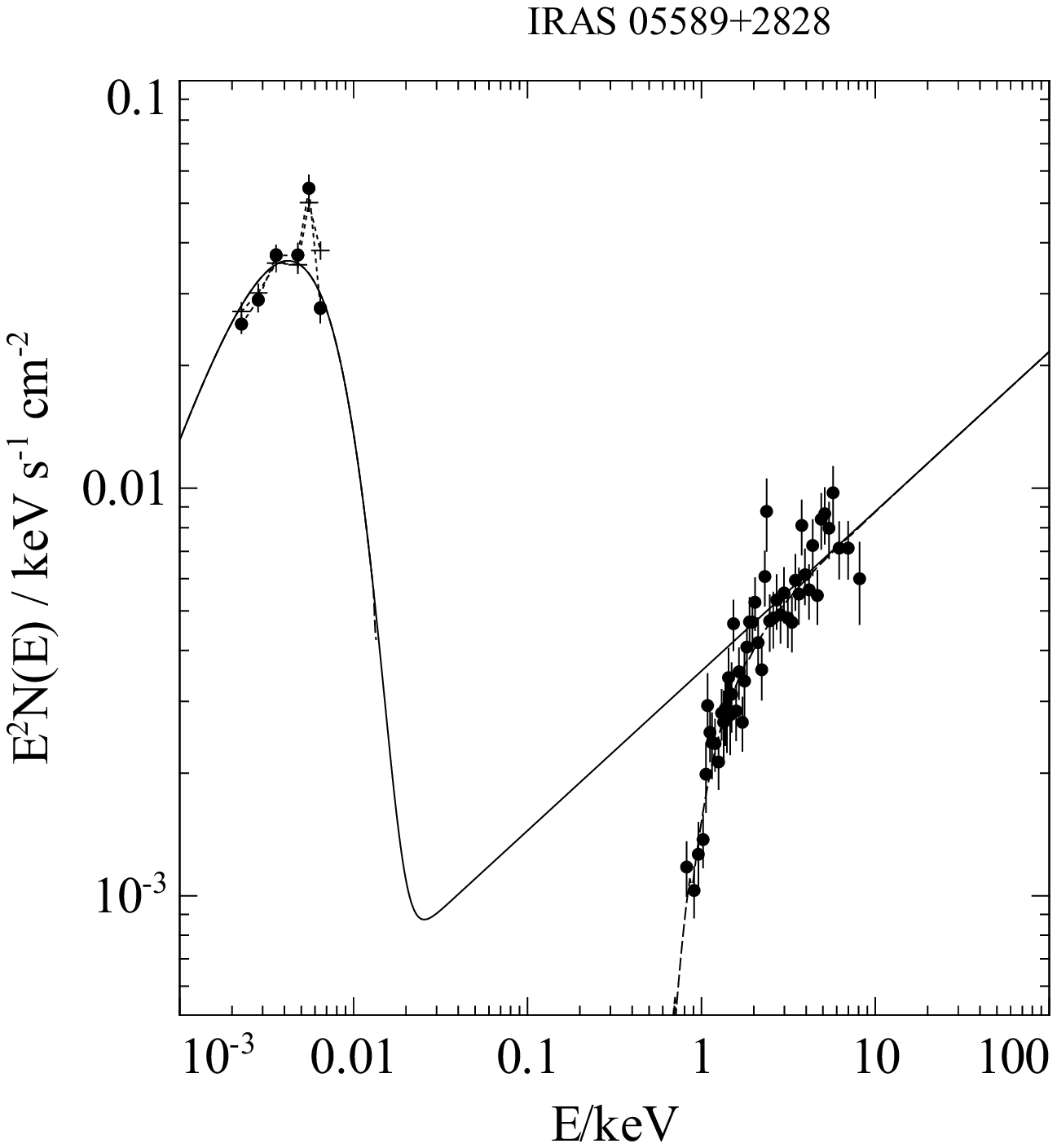}
    \includegraphics[width=4.5cm]{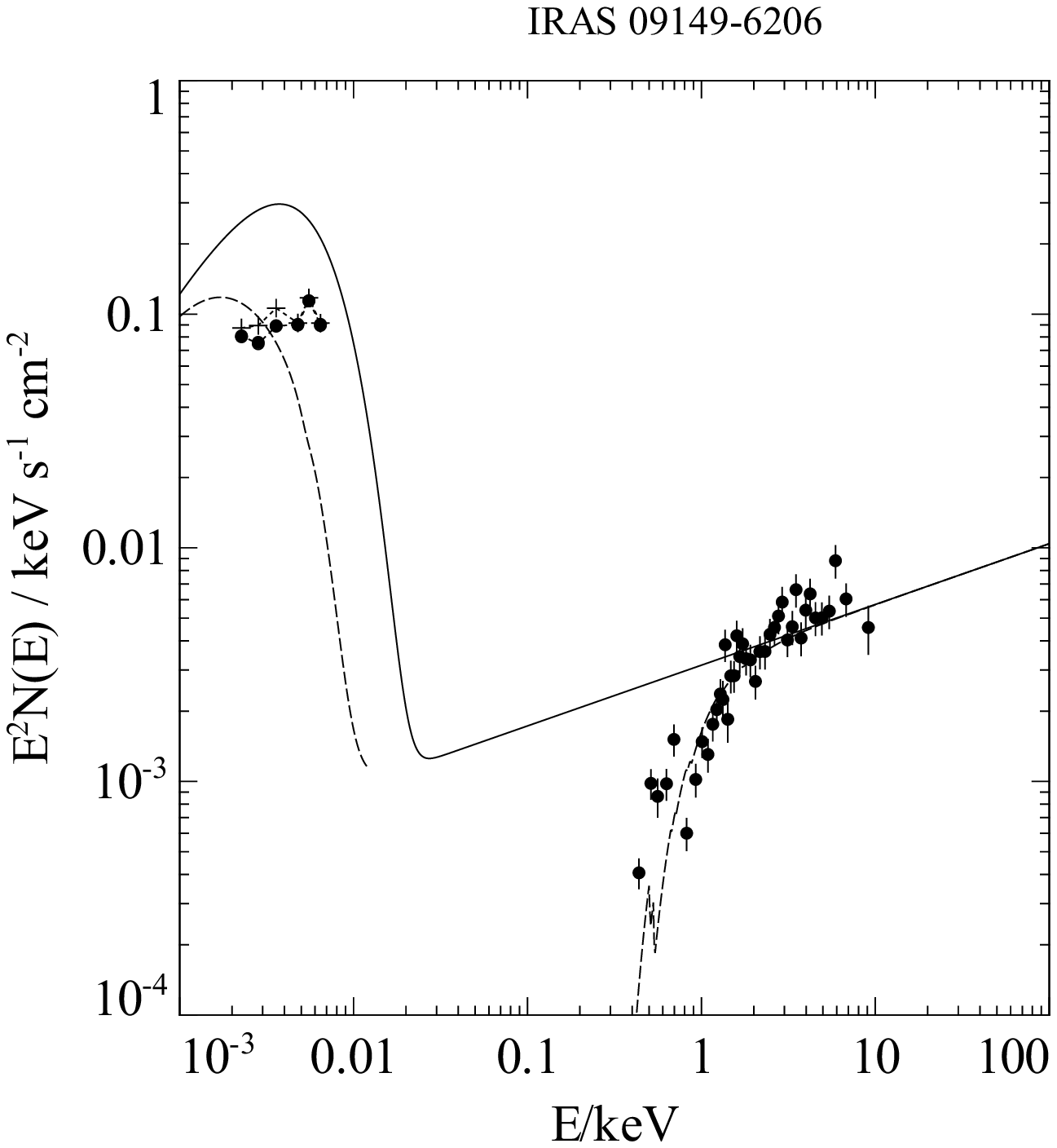}
    \includegraphics[width=4.5cm]{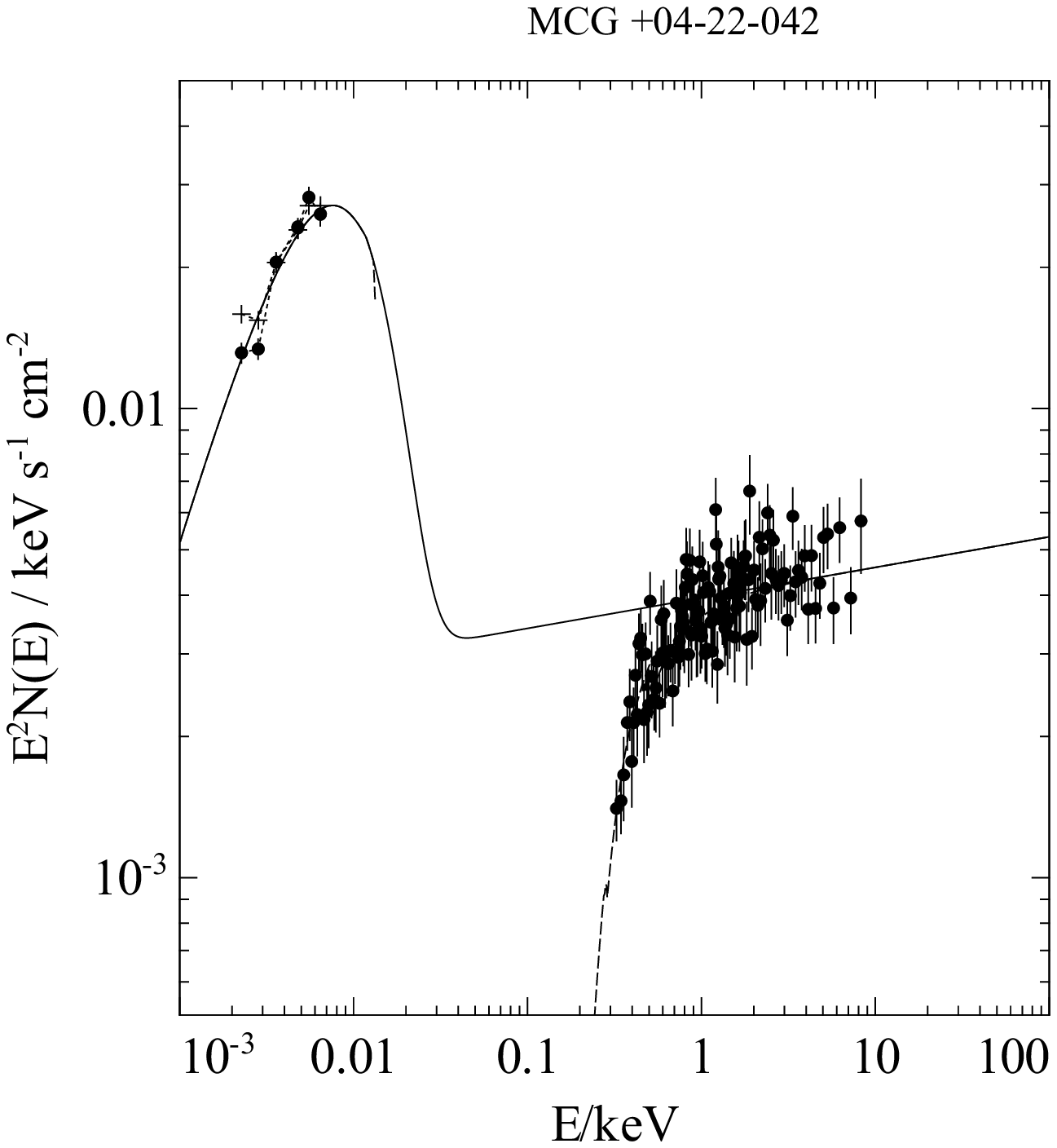}
    \includegraphics[width=4.5cm]{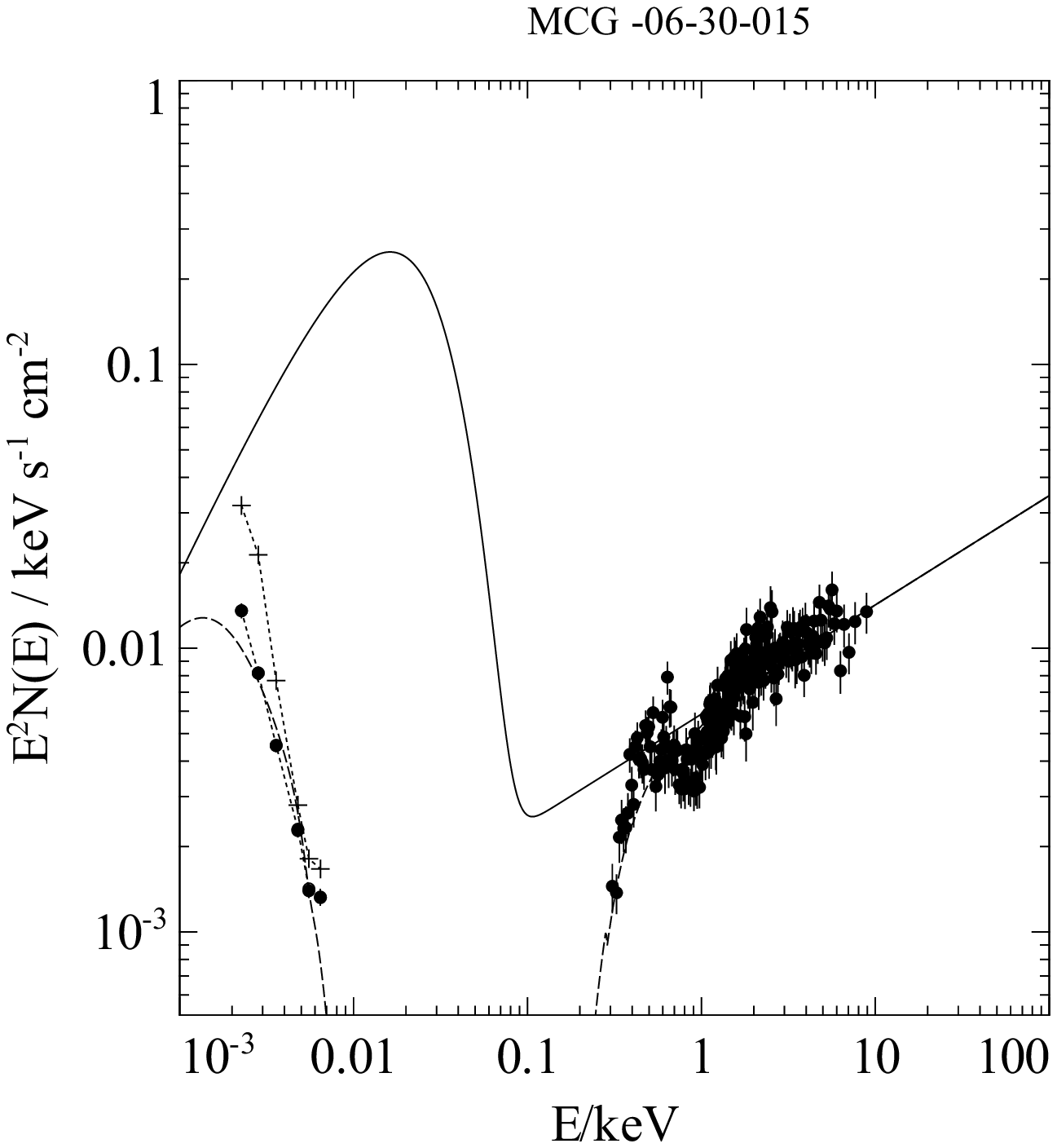}
    \includegraphics[width=4.5cm]{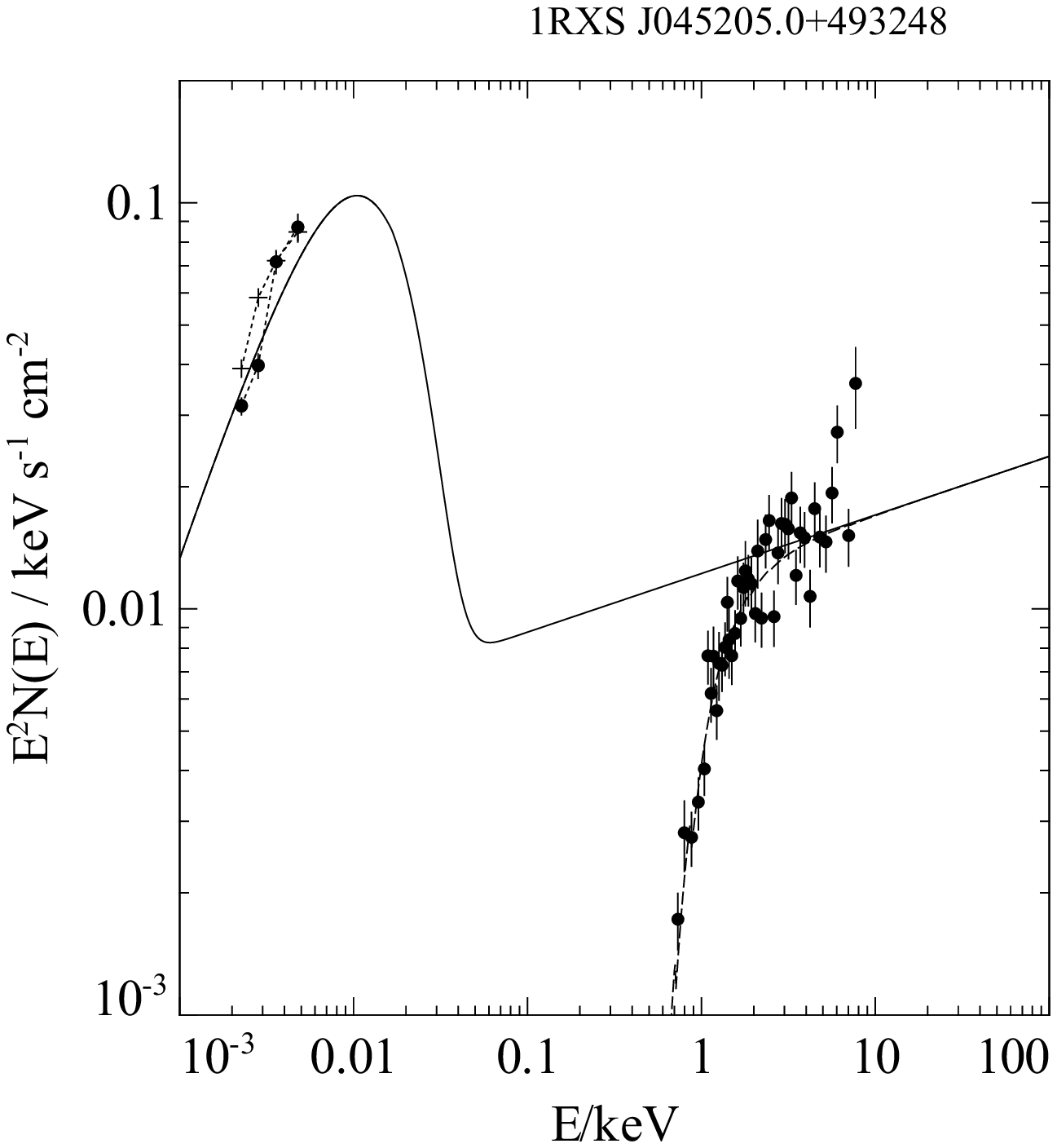}
    \includegraphics[width=4.5cm]{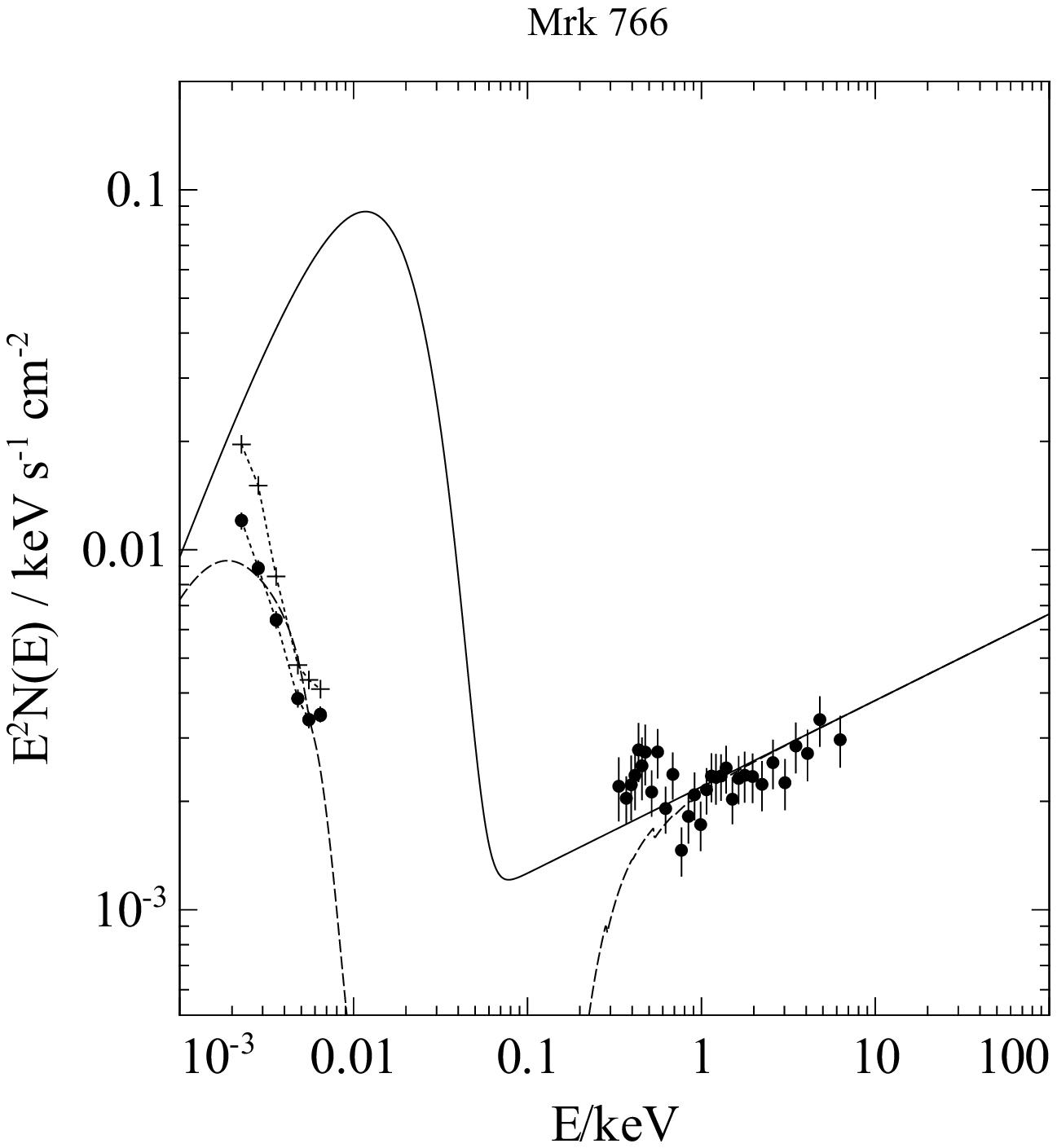}
    \includegraphics[width=4.5cm]{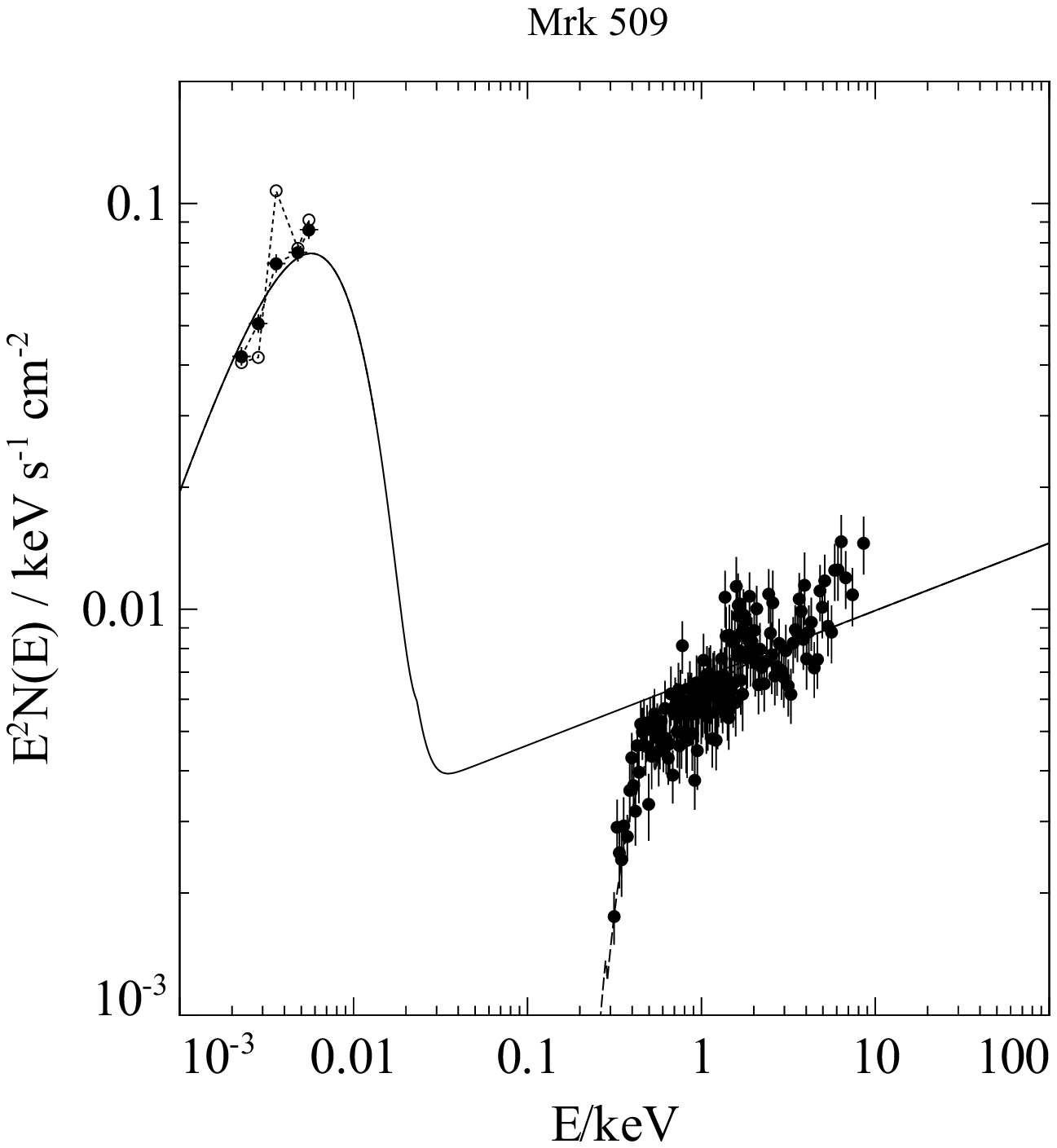}
    \caption{A selection of AGN SEDs. The crosses represent the 5 arcsec aperture photometry results from \textsc{uvotsource}, and the filled circles represent the data points used for the fit.  In the UV and optical regime, the black filled points represent the fluxes with the host galaxy component removed using GALFIT as detailed in the text.  Empty circles were not used in the fit in a few cases where significant coincidence losses risk affecting the accuracy of the fluxes.}
\label{allseds_mass2MASS}
\end{figure*}

We freeze the normalisation of the UV bump \textsc{diskpn} component using the black hole mass estimate from the 2MASS K-band bulge luminosity estimate (normalisation $K = M_{\rm{BH}}^2 cos(i)/D_{\rm{L}}^2\beta^4$ for black hole mass $M_{\rm{BH}}$, inclination angle $i$, luminosity distance $D_{\rm{L}}$ and colour-to-effective temperature ratio $\beta$). We present the SEDs determined using this approach in Fig.~\ref{allseds_mass2MASS}.  For a few objects, the \textsc{diskpn} model thus constrained fits poorly to the data despite their relatively `blue' optical--UV spectral shapes, thought to be typical for the lower-energy part of an un-reddened accretion disc spectrum (i.e. Ark 120, Mrk 1018, Mrk 509 and NGC 985), which is probably attributable to a slightly too large black hole mass.  This is confirmed later in the case of Ark 120 for which a reverberation mass estimate is available (see \S\ref{discussion:bolcoredd}).  In the case of Mrk 590, the distinctly reddened optical--UV spectral shape is poorly fit using the 2MASS black hole mass-derived \textsc{diskpn} normalisation, but the reverberation mass estimate produces a more plausible fit.  The \textsc{diskpn} model fit for IRAS 09149-6206 also looks poor, although this object possesses a rather unusual, flat optical--UV SED (noted initially by \citealt{1991AJ....102.1933K}) unlike the other AGN in the sample, which may be due to some other process.  These five cases might point to a genuine offset between the reverberation masses and our 2MASS derived estimates, but the degeneracy of the mass with the inclination and other parameters in the \textsc{diskpn} normalisation could indicate, for example, unusual inclination properties in these few AGN.  The remaining 21 objects all exhibit sensible model fits to the optical--UV points.

\begin{figure}
\includegraphics[width=8cm]{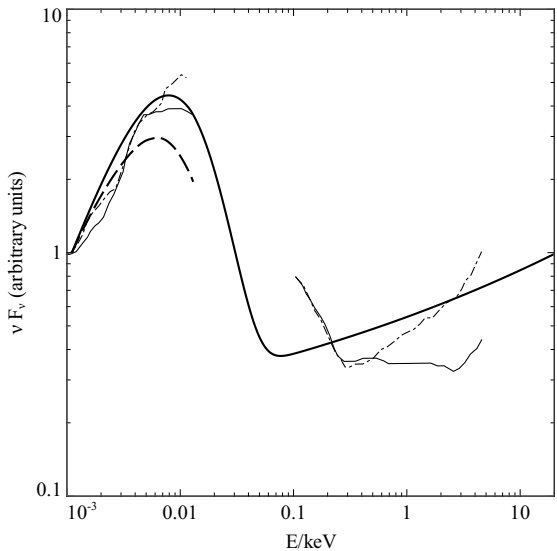}
    \caption{Average, reddening-corrected optical--to--X-ray SED for the 22 low Eddington ratio ($\lambda_{\rm Edd}<0.1$) AGN, normalised at 1 eV (thick black solid line).  The average fit (in the optical--UV regime) with reddening included is shown using the thick black dashed line.  The mean SEDs for radio-quiet and radio-loud quasars from \protect\cite{1994ApJS...95....1E} are also reproduced (thin black solid and dot-dashed lines, respectively).}
\label{averageSED}
\end{figure}

We present the average model SED (before and after correction for reddening) calculated for the 22 AGN with $\lambda_{\rm Edd}<0.1$ (normalised at 1eV) in Fig.~\ref{averageSED}.  This allows a useful comparison with the average SEDs presented in VF07 and VF09 for high and low Eddington ratio AGN, and the mean quasar SED templates of E94; the latter are also reproduced in Fig.~\ref{averageSED}.  While the E94 template is representative of X-ray bright quasars, ours offers a similar template for local, X-ray unabsorbed and low accretion rate Seyferts, albeit calculated from the model fits instead of the data.  Some differences between our average SED and those of E94 include the slightly reddened shape of the E94 SEDs in the optical--UV (their SEDs trace our extincted SED more closely, at least at longer wavelengths) and the differing X-ray spectral shapes. The radio-quiet quasar X-ray spectrum in particular appears softer in the 0.5--2 keV band which would be expected for a brighter, higher accretion rate sample of AGN (according to the trends seen in e.g. \citealt{2008ApJ...682...81S}); but this could also be in part due to the inclusion of some X-ray absorbed objects in the E94 composite, depressing the flux at this part of the spectrum relative to the flux at higher energies.  The hard shape of the radio-loud template can be accounted for by the expected synchrotron contributions from jets.  The E94 SEDs also show a prominent soft excess; due to the limited spectral resolution of the XRT we do not consider the soft excesses in our SEDs here.  We repeat the cautionary note from E94, that AGN properties determined from the average SED do not necessarily reflect the full degree of variation in the AGN population.

We follow the approach outlined in VF09 for calculating the unabsorbed hard X-ray (2--10 keV) and bolometric (0.001--100 keV) luminosities and errors in \textsc{xspec}, for use in determining accretion rates, bolometric corrections and other SED parameters.  The results are presented in Table~\ref{uvotxrtresults}.  As discussed in \S\ref{Intro}, we define the bolometric luminosity as being the primary emission due to accretion emerging in the optical--to--X-ray regime and assume that the IR is reprocessed emission; it is therefore not included in $L_{\rm bol}$.

\subsection{Extending the SEDs into hard X-rays}

\begin{figure*}
    \includegraphics[width=4.5cm]{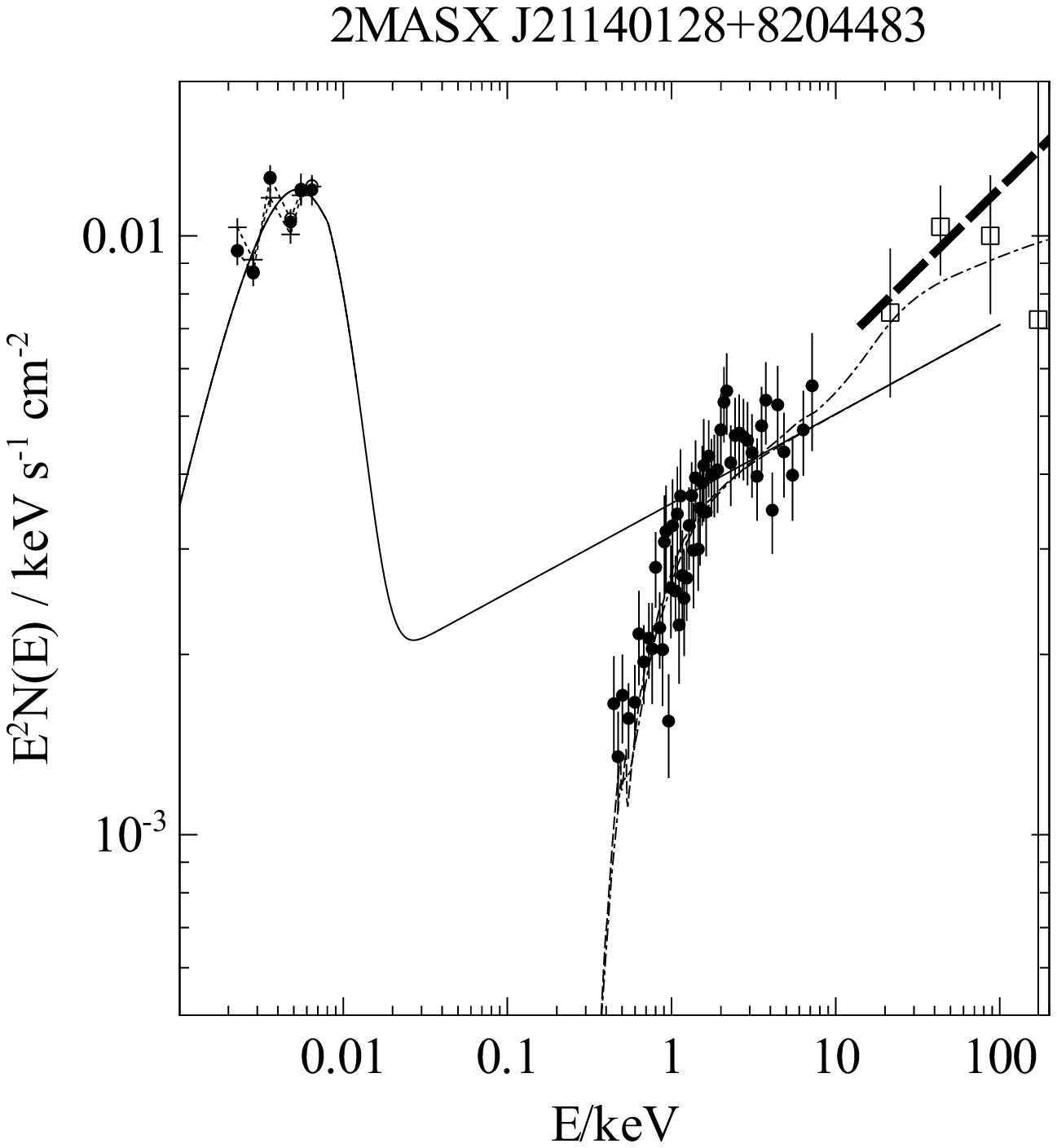}
    \includegraphics[width=4.5cm]{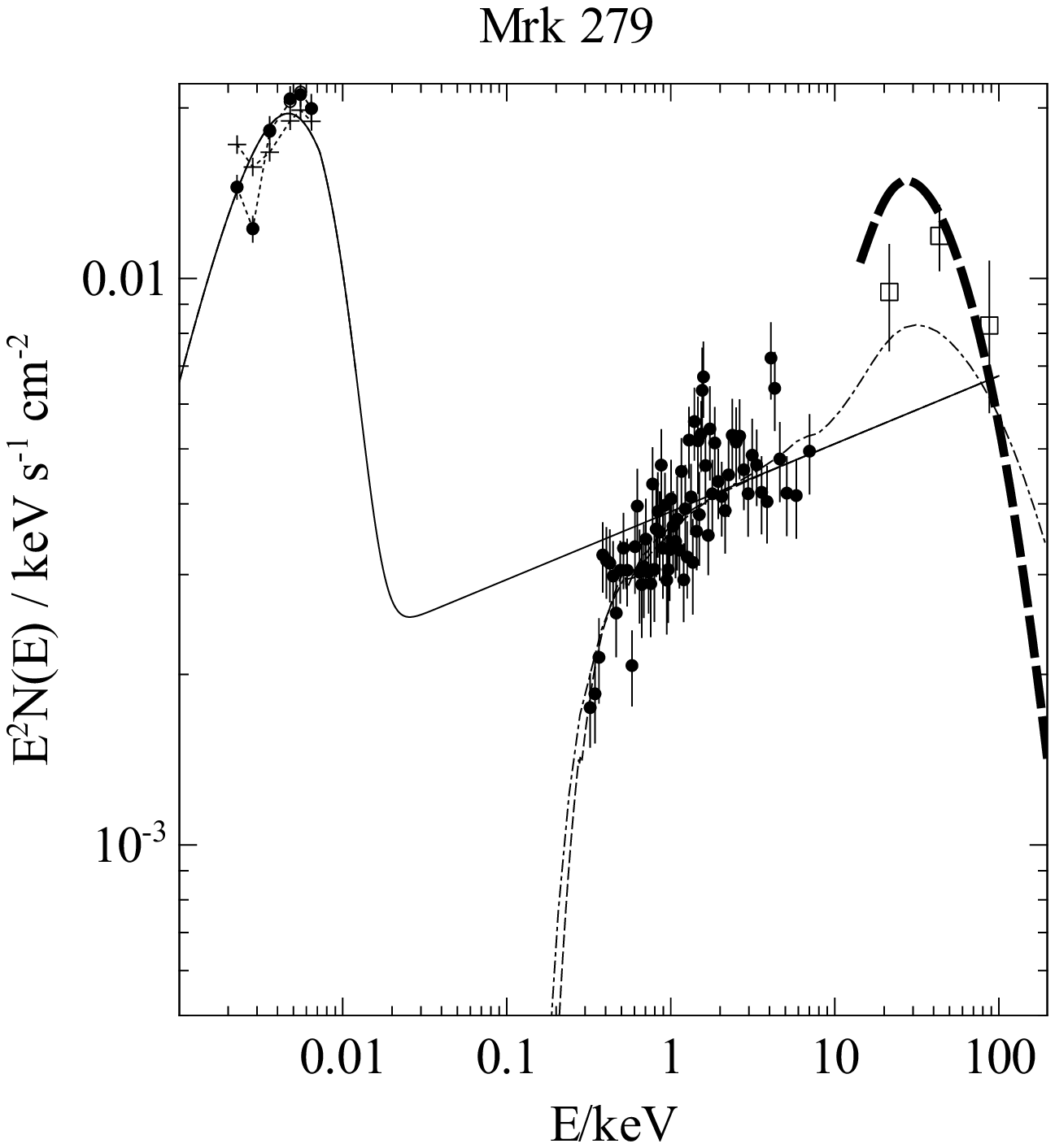}
    \includegraphics[width=4.5cm]{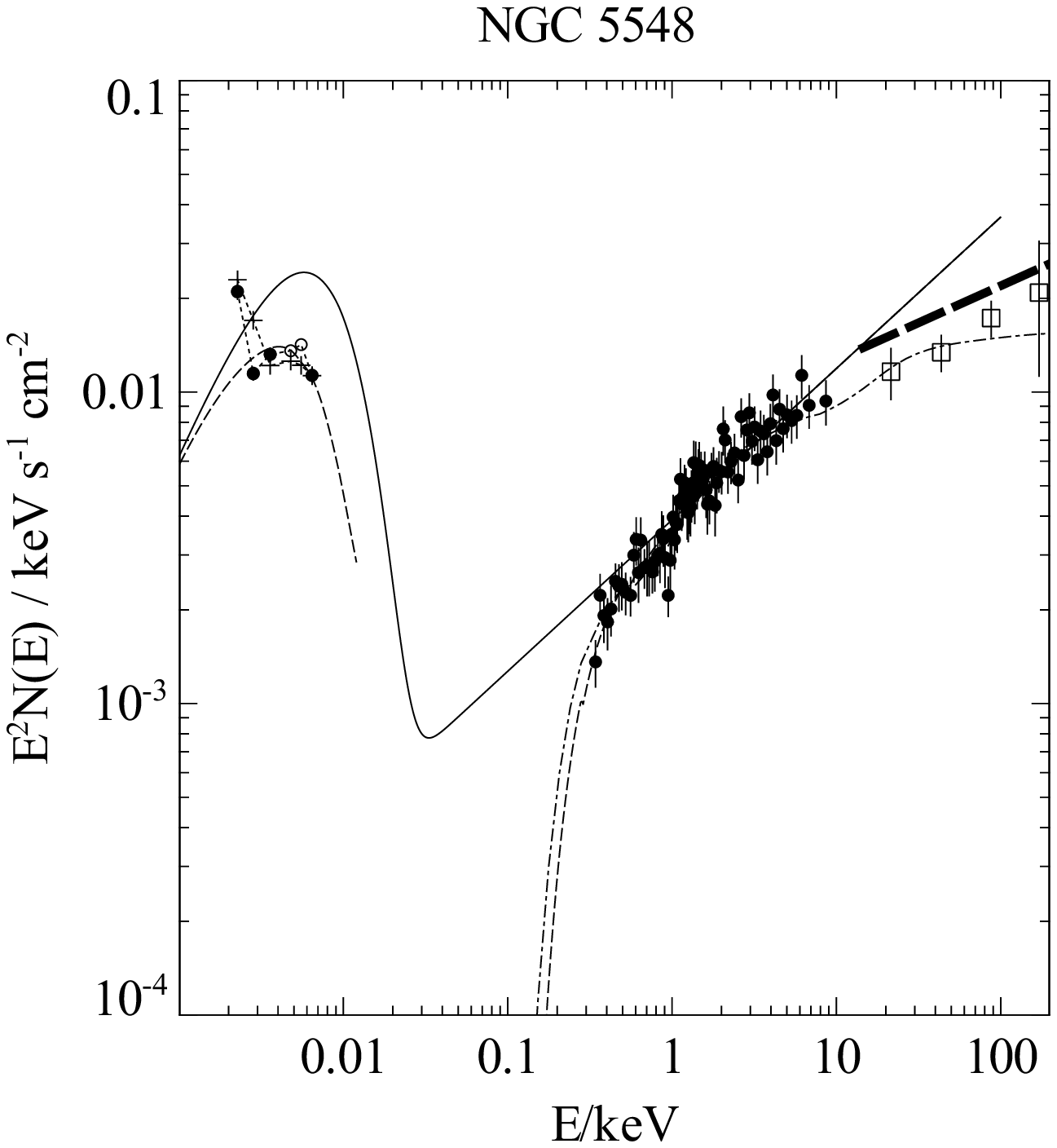}
    \includegraphics[width=4.5cm]{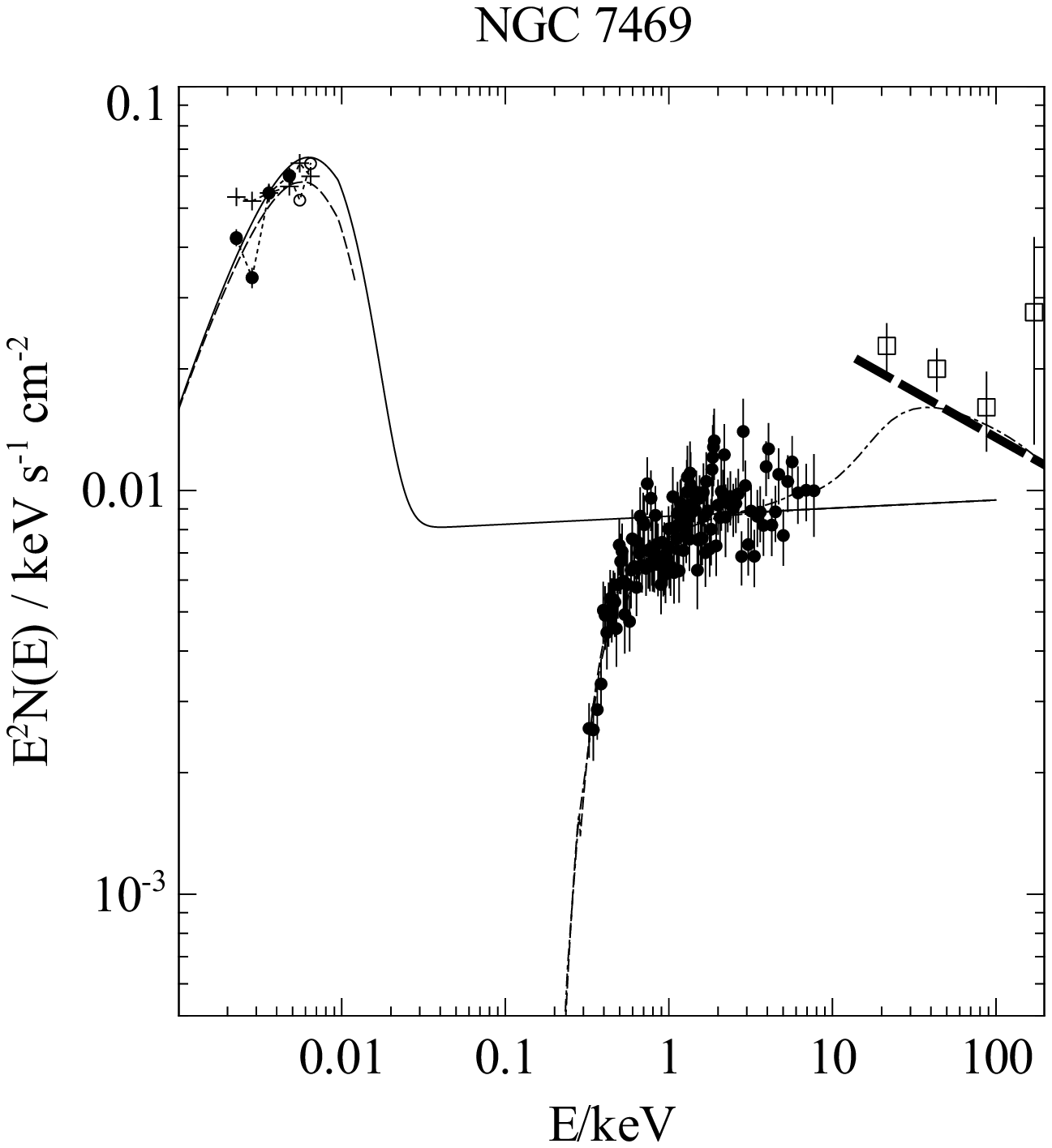}
     \caption{Spectral energy distributions with BAT data included.  Empty squares represent 4-channel BAT data (publicly available).  Thick dashed lines represent fits to the 8-channel data (not public at the time of writing).  The thin dot-dashed line represents a fit to the XRT and BAT data combined, using the model \textsc{[wabs(zwabs{powerlaw})+pexrav]} to illustrate the plausibility of a reflection fit to the data.}
\label{seds_withbat}
\end{figure*}

The BAT data allows us to extend the SEDs into hard X-rays (14--195 keV).  Four-channel BAT spectra are publicly available on the internet\footnote{http://swift.gsfc.nasa.gov/docs/swift/results/bs9mon/}, time-averaged over periods of many months.  We identify the four least variable objects in the 14--195 keV band using measurements of excess variance on the BAT lightcurves, in order to maximize confidence in luminosities extrapolated from fits to the combined X-ray the hard X-ray data.  We present the SEDs for these four sources (NGC 5548, NGC 7469, Mrk 279 and 2MASX J21140128+8204483) including the 4-channel BAT data in Fig.~\ref{seds_withbat}.  We also present the fits to the more robust 8-channel data (the data themselves are not yet public at the time of writing) to illustrate any substantive differences between 4- and 8-channel datasets.  In all cases, the simple model fits (\textsc{powerlaw} or \textsc{pexrav}) do not indicate a signifcant departure from the general shape implied by the 4-channel data.  

In three out of four of these low-variability sources, it appears that the BAT data may be capturing a reflection hump peaking at around $\sim$30 keV, in accordance with standard scenarios for reflection from the accretion disc.  There does not seem to be any pronounced reflection signature in NGC 5548.  We corroborate this suggestion by fitting the model combination \textsc{[wabs(zwabs{powerlaw})+pexrav]} to the XRT and BAT data together in XSPEC, linking the photon index of the illuminating source in \textsc{pexrav} to that for the \textsc{powerlaw} component.   The model fits (thin dot-dashed lines) agree well with the fits to the XRT alone at low energies, and show a plausible reflection hump fit to the BAT data at higher energies for NGC 7469, Mrk 279 and 2MASX J21140128+8204483. There are hints of accompanying Fe K-$\alpha$ lines at 6.4 keV in the XRT data, as expected from reflection models, but the quality of the XRT data does not lend itself to a confident identification of this line.  The superior quality XMM-PN data presented for NGC 7469, Mrk 279 and NGC 5548 in VF09 show a much more convincing Fe K-$\alpha$ line shape at 6.4 keV, although the line signature is less evident in NGC 5548, which would be expected for weaker reflection. The indications of reflection in these sources are confirmed by a more detailed analysis using \emph{XMM} and BAT data (Mushotzky et al. 2009 in prep).

\section{Discussion}

\subsection{Contamination of optical--UV continuum with emission lines}

We briefly comment on the effect of emission lines in optical--UV spectra of AGN which have the potential to introduce systematic effects in our SED measurements.  The study of \cite{1991ApJ...373..465F} presents a high signal-to-noise composite quasar spectrum, allowing the parameters of such emission lines to be accurately determined.  We identify the emission lines with the largest widths and highest peaks from Figs. 2 and 7 of their work and redshift them individually for each AGN in our sample to ascertain whether emission line fluxes are likely to contaminate the UVOT photometry.  

As a first-order attempt to quantify this, we ascertain whether the central wavelength of the redshifted line lies within the band-pass of each of the UVOT filters, as obtained from the CALDB documentation.  The most common contaminant from this analysis would seem to be the Mg II line at 2798$\rm \AA$ (rest-frame), which is most likely to appear in the UVW1 filter, and occasionally in the U filter for two high-redshift objects.  Al III (1958$\rm \AA$) and C III] (1909$\rm \AA$) are also likely to be common contaminants in UVM2 and UVW2 bands.  For the four highest redshift sources (3C 390.3, IRAS 09149-6206, SBS 1136+594 and 2MASX J21140128+8204483), the redshifted O III lines (4959$\rm \AA$ and 5007$\rm \AA$) appear in the UVOT V-band and in the case of 2MASX J21140128+8204483, the $\rm H\beta$ line (4861$\rm \AA$) also.  The presence of these lines may explain, in part, the `angular' behaviour of the optical--UV SEDs in these four sources.  However, the V-band `bumps' in more local sources (ESO 548-G081, Mrk 279, NGC 4593 and NGC 7469 for example) are not accounted for by the presence of such lines and must have some other cause.  In any case, for the vast majority of objects in the sample, the SED points in most filters are highly unlikely to be affected by the presence of any prominent emission lines.

\subsection{Intrinsic reddening due to dust extinction}

\begin{figure}
\includegraphics[width=8cm]{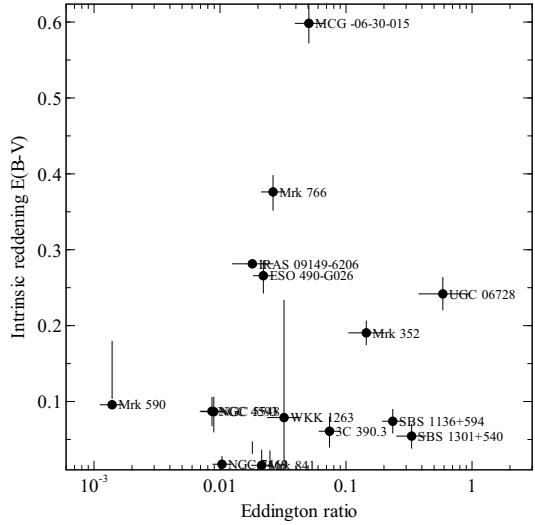}
    \caption{Intrinsic extinction $E(B-V)$ against Eddington ratio $\lambda_{\rm{Edd}}$, for those objects for which reddening was included in the model fit via the \textsc{zdust} XSPEC model.}
\label{EBminusVvsEdd}
\end{figure}

We present the values for intinsic extinction $E(B-V)$ determined from the SEDs against Eddington ratio in Fig.~\ref{EBminusVvsEdd}.  This allows a comparison between the amount of estimated dust extinction in the source and the accretion power available from the central AGN.  One notable feature of the distribution of $E(B-V)$ observed is that despite the selection of objects with low $N_{\rm H}$, there appears to be significant dust reddening in many of these sources, based on the shape of their optical--UV spectrum. This was noted previously by \cite{2000ApJ...535...53K} and provided the basis for their `lukewarm absorber' scenario to account for this discrepancy.  Our results contrast with those of \protect\cite{2001A&A...365...28M}, who find that dust-to-gas ratios -- $E(B-V)/N_{\rm H}$ -- are systematically lower in their sample AGN than in the Galaxy.  However, their sample is selected specifically to include higher $N_{\rm H}$ objects with relatively un-absorbed optical/UV/IR broad lines, and the data used are not simultaneous.  The results of both of these studies results imply that objects which are significantly absorbed in X-rays can show little sign of dust extinction in the optical--UV and vice-versa, and in general that classifications of AGN into `obscured' or `unobscured' based on their optical properties can differ significantly from X-ray classifications.   Our sample of objects confirms that optical and X-ray obscuration scenarios can differ significantly in an object, influenced as they are by different physical processes (as summarised in \citealt{2001ApJ...562L..29C}). We caution of a degeneracy between the mass and $E(B-V)$ in the model fits (noting in the case of Mrk 590 that using its RM mass gives a best-fit $E(B-V)$ of $0.33$ compared to $\sim0.1$ using the 2MASS mass estimate), although the SEDs for ESO 490-G026, MCG-06-30-15, Mrk 590 and Mrk 766 show more unambiguous signs of significant intrinsic extinction.  In support of this, the aforementioned study on reddened Seyfert 1 galaxies of \cite{2001ApJ...562L..29C} confirms that observations from the \emph{International Ultraviolet Explorer} (IUE) also strongly suggest intrinsic reddening in MCG-06-30-15, Mrk 590 and Mrk 766. The high optical polarisation for Mrk 766 discussed by \cite{1995AJ....109...81U} may also be an indicator of the dust responsible for reddening in this source. Encouragingly, the more detailed study of MCG-06-30-15 by \cite{1997MNRAS.291..403R} reports a lower limit on $E(B-V)$ very similar to the value of $\sim$0.6 that we determine for this source, implying that our estimation of intrinsic extinction with the \textsc{zdust} model can provide reddening estimates in line with more detailed studies.

The studies of \cite{2008MNRAS.385L..43F} and \cite{2009arXiv0901.0250F} highlight the connection between the radiation pressure exerted by the AGN and the configuration of the surrounding absorbing material.  They identify the effective Eddington limit for dusty gas in the $N_{\rm H}-\lambda_{\rm Edd}$ plane; above this limit is a `forbidden region' in $N_{\rm H}-\lambda_{\rm Edd}$ space within which absorbing dusty gas clouds are unstable to radiation. They verify that most local and deep samples of AGN avoid the forbidden region, and the latter work in particular verifies this in the case of the unbiased \emph{Swift}/BAT catalogue.  Objects near the effective Eddington limit are known to exhibit AGN winds/warm absorbers, which would be expected when radiation pressure locally exceeds gravity.  Since the dust and gas are expected to be coupled, the dust extinction parameterised by $E(B-V)$ may also exhibit a similar distribution in the $E(B-V)-\lambda_{\rm Edd}$ plane.  We see from Fig.~\ref{EBminusVvsEdd} that the majority of objects occupy the lower-left part of the plot, and the only object with particularly high reddening -- MCG-06-30-15 -- is also known from \cite{2009ApJ...690.1322W} to have a relatively high $N_{\rm H}$ ($\sim1.9\times10^{21}\rm cm^{-2}$) and $\lambda_{\rm Edd}$, placing it close to (but not within) the forbidden region.  This object showcases a warm absorber, as expected for an object near the effective Eddington limit.  The absorbing column reported from analyses of \emph{BeppoSAX} and \emph{ASCA} data is strongly model-dependent (see for example, \citealt{1997MNRAS.291..403R} who find $N_{\rm H}$ to be an order of magnitude lower than the Winter et al. result; or \citealt{2000MNRAS.315..149M}, whose detailed multi-zone warm absorber scenario gives a column density of $3\times10^{22}\rm cm^{-2}$).     

  These hints may suggest that an analogous `forbidden region' could be proposed in the upper-right portion of the $E(B-V)-\lambda_{\rm Edd}$ plane. A more complete study of intrinsic reddening properties for a larger sample of sources spanning a wider range of X-ray column densities would be valuable for exploring this possibility further.

\subsection{Bolometric corrections and Eddington ratios}
\label{discussion:bolcoredd}

\begin{figure*}
\includegraphics[width=12cm]{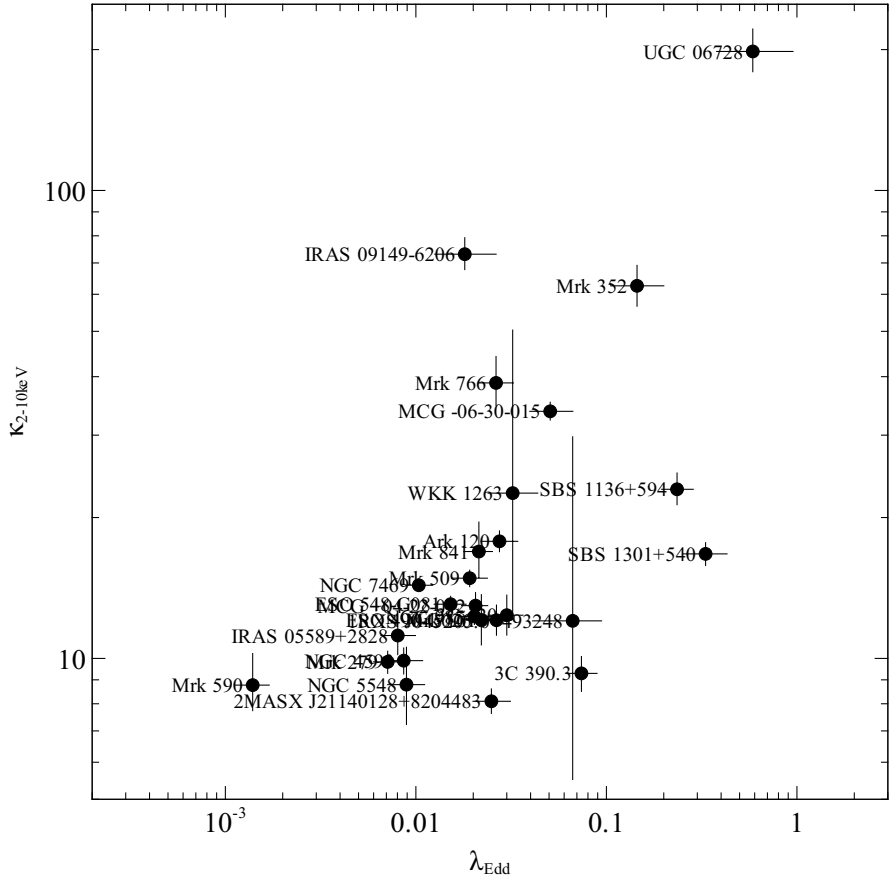}
    \caption{Hard X-ray (2--10keV) bolometric correction against Eddington ratio.  The mass estimates used for model fitting and calculation of Eddington ratios are determined from the 2MASS bulge luminosity.}
\label{bcvsedd_2mass}
\end{figure*}

We present the bolometric corrections $\kappa_{\rm 2-10keV}=L_{\rm bol}/L_{\rm 2-10keV}$ against Eddington ratio in Fig.~\ref{bcvsedd_2mass}.  The bolometric corrections cluster between 10--20 with a low fraction of objects possessing higher bolometric corrections, and Eddington ratios are overwhelmingly below $<0.1$.  The low bolometric corrections obtained are expected for low Eddington ratios (as found in VF07 and VF09) and the few objects with high bolometric corrections generally lie at higher Eddington ratios as expected.  Errors in luminosities take into account the errors in model parameters obtained from fitting the data, and the Eddington ratio additionally takes into account the random error in the black hole mass estimate (extrapolated from the uncertainties in the K-band magnitudes).  However, the potential systematic errors due to intrinsic disparities between RM and K-band mass determination methods also contribute an uncertainty on both axes which is not represented in the error bars.  

\begin{figure*}
\includegraphics[width=7cm]{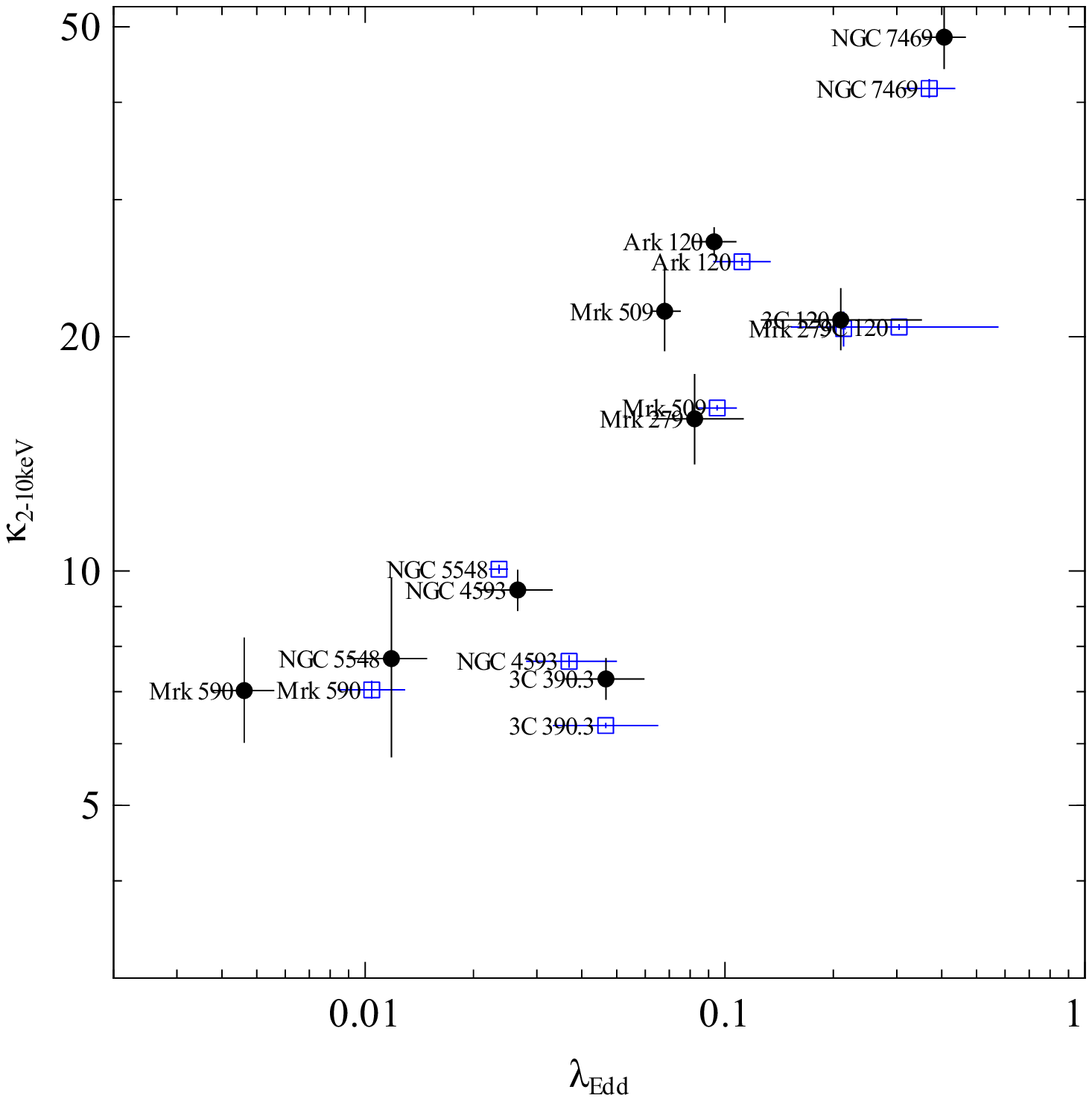}
\includegraphics[width=7cm]{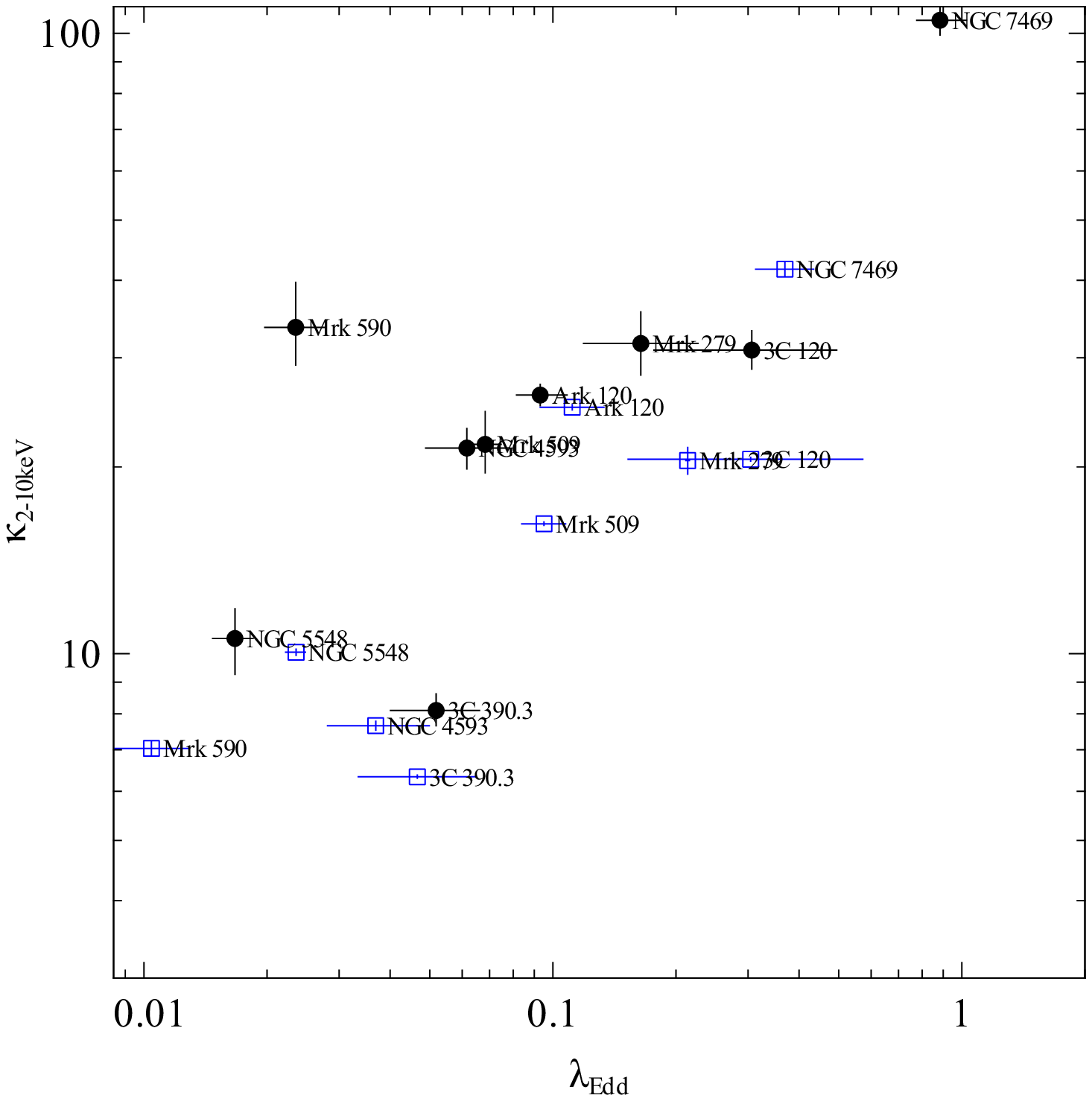}
    \caption{Hard X-ray (2--10keV) bolometric correction against Eddington ratio for the objects with reverberation mapping mass estimates (black filled circles).  \emph{Left panel:} Intrinsic reddening was not included in the fit, as was the case in VF09. \emph{Right panel:} Intrinsic reddening was included in the fit, to contrast with VF09 and assess the role that intrinsic extinction may play in shaping the SED for some of the reverberation mapped AGN.  In both cases, the reverberation mapping masses were used in the fit and to determine the Eddington ratios. The results from XMM-Newton PN and Optical Monitor (VF09) are shown for comparison (blue empty squares).}
\label{bcvsedd_revmap}
\end{figure*}

We therefore invoke RM mass estimates for nine sources common to the objects discussed by \cite{2004ApJ...613..682P} to perform some simple comparisons which provide an estimate of the magnitude of this systematic effect.  If the RM masses are used in the fitting process instead, we obtain bolometric corrections and Eddington ratios as given in Fig.~\ref{bcvsedd_revmap}.  We also present the results from VF09 for comparison, which employed simultaneous data from XMM-Newton.  If reddening is not included, as in VF09, we obtain very similar bolometric corrections and Eddington ratios as seen with XMM (left panel, Fig.~\ref{bcvsedd_revmap}).  However if reddening is taken into account, this generally increases both bolometric corrections and Eddington ratios (right panel, Fig.~\ref{bcvsedd_revmap}) and in the case of objects such as Mrk 590 and NGC 7469, this change is very pronounced.  However, in both scenarios, the general trend for bolometric correction increasing with Eddington ratio identified in VF07 and VF09 is preserved.

\begin{figure}
\includegraphics[width=7cm]{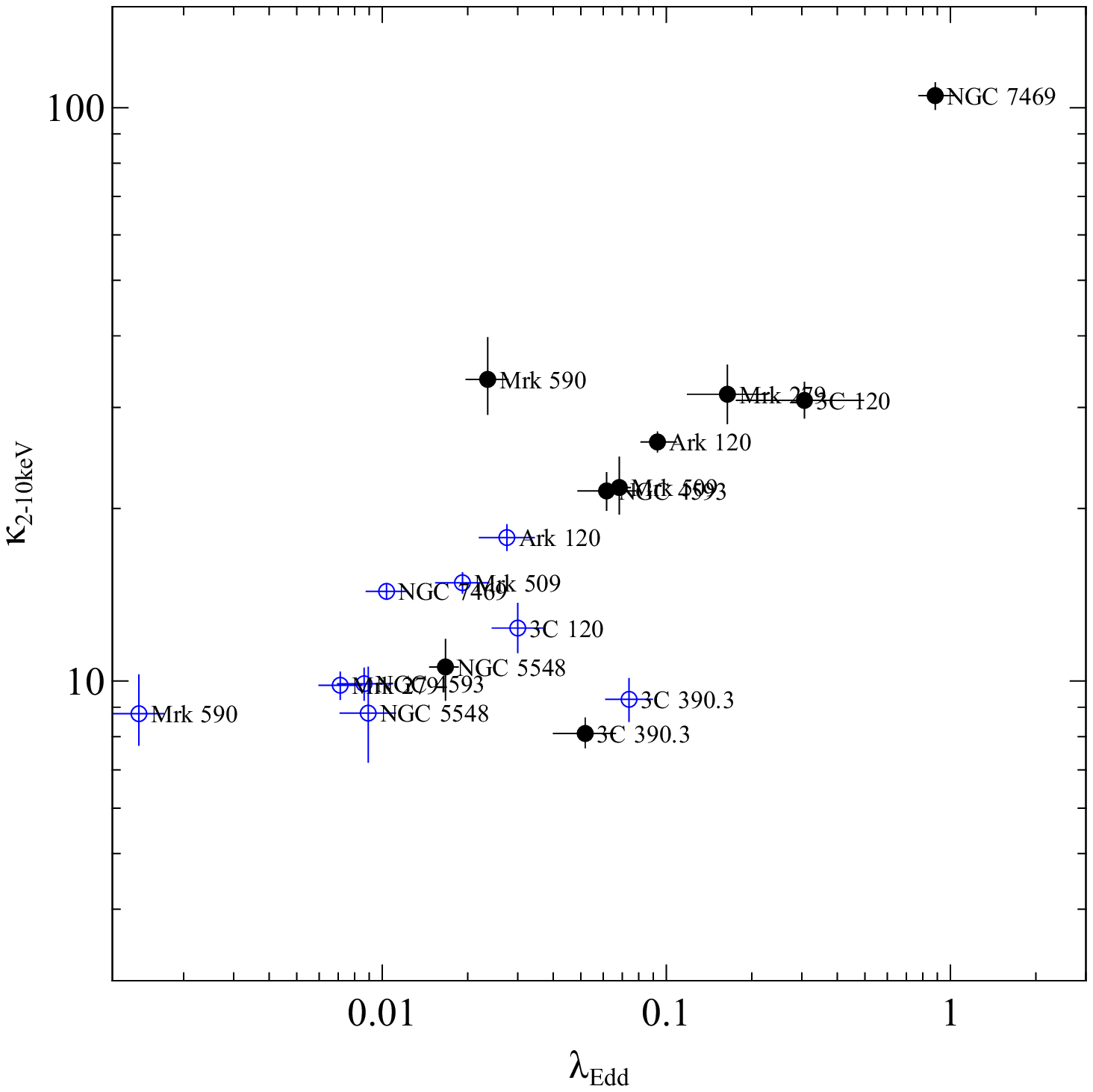}
    \caption{Hard X-ray (2--10keV) bolometric correction against Eddington ratio for the objects with reverberation mapping mass estimates.  Here we present a comparison between the values obtained using mass estimates from 2MASS K-band bulge luminosities (blue empty squares) and reverberation mapping (black), using the UVOT and XRT data in both cases, with intrinsic reddening included in the fit.}
\label{bcvsedd_revmap_2masscompare}
\end{figure}

We also perform a comparison between RM masses and masses from K-band luminosities using only the \emph{Swift} simultaneous data, including intrinsic reddening effects in both cases (Fig.~\ref{bcvsedd_revmap_2masscompare}).  Bolometric corrections and Eddington ratios are clustered at significantly higher values for this subsample when using the RM masses compared to the K-band masses.  There is an increase of around $\sim1$ dex in Eddington ratios when RM masses are used, accompanied by an increase in bolometric correction to an average of $\sim30$.  The most extreme changes are for NGC 7469, Mrk 590 and Mrk 279.  These results illustrate the degree of change which could be seen if the 2MASS K-band mass estimates need to be scaled down to bring them more in line with RM estimates, but given the BLR geometry uncertainty intrinsic to the RM masses, the bolometric corrections and Eddington ratios calculated using better-calibrated RM masses could also be reduced significantly.  As depicted in Fig.~\ref{revmap_RADPRESSCORR_vs_kbandOFFSET_mass}, the two methods are broadly consistent within these uncertainties.  We suggest that the true magnitude of any shift in results may lie in between these two extremes; the 2MASS masses may represent an upper limit whereas the reverberation estimates may represent a lower limit.  Indeed, the radiation pressure corrections may require further calibration as discussed by \cite{2009arXiv0905.0539M}, and could bring these estimates closer together, for example.  More work is required on calibration between different black hole mass estimators for AGN before this issue can be addressed more convincingly, and at this stage we are only able to provide a crude estimate of the limiting uncertainties.

\begin{figure*}
    \includegraphics[width=8cm]{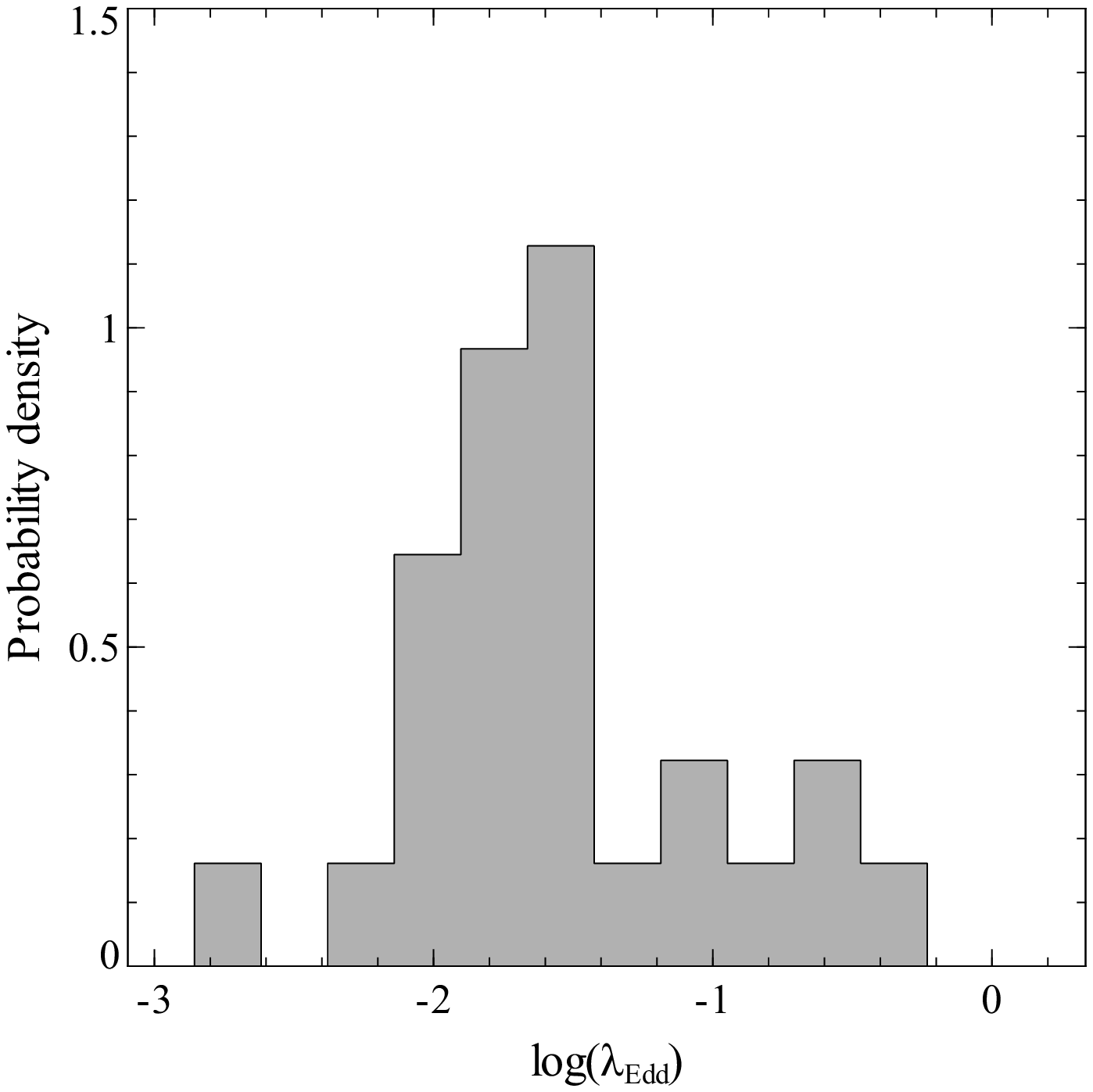}
    \includegraphics[width=8cm]{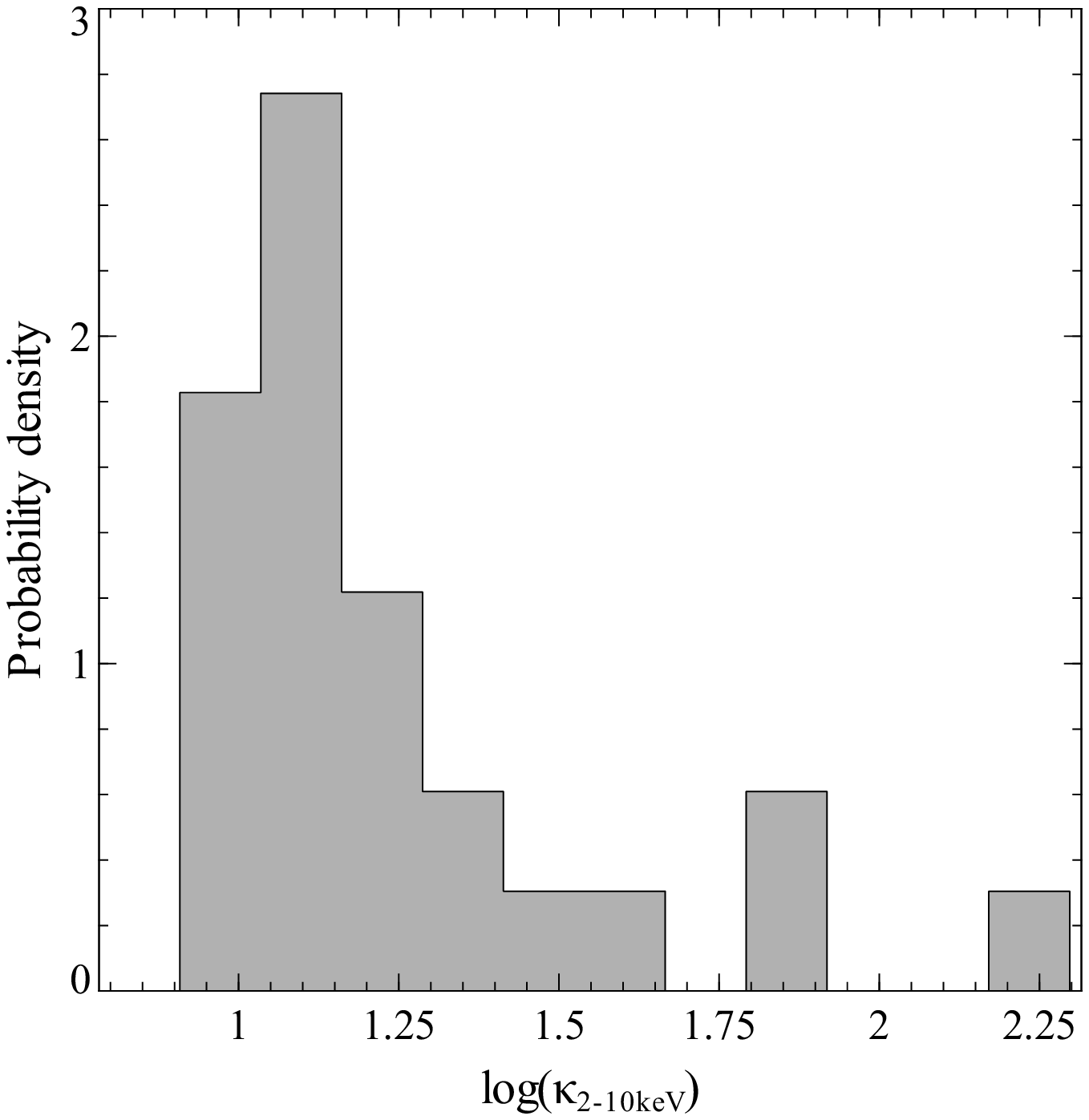}
    \caption{Histograms of Eddington ratios (left panel) and bolometric correction (right panel) for the sample.}
\label{Eddhist}
\end{figure*}

These analyses suggest that the typical bolometric corrections and Eddington ratios for this sample are low, even when uncertainties in the mass estimates are taken into account.  The distribution of Eddington ratios and bolometric corrections are presented in Fig.~\ref{Eddhist}, using the 2MASS K-band mass estimates.  The distribution is similar to that presented for Seyfert 1s in Fig.~7 of \cite{2009ApJ...690.1322W} shifted by approximately an order of magnitude, in line with the relatively low bolometric corrections we find for the majority of the sample.

\subsection{Fraction of ionizing luminosity}

\begin{figure*}
\includegraphics[width=5cm]{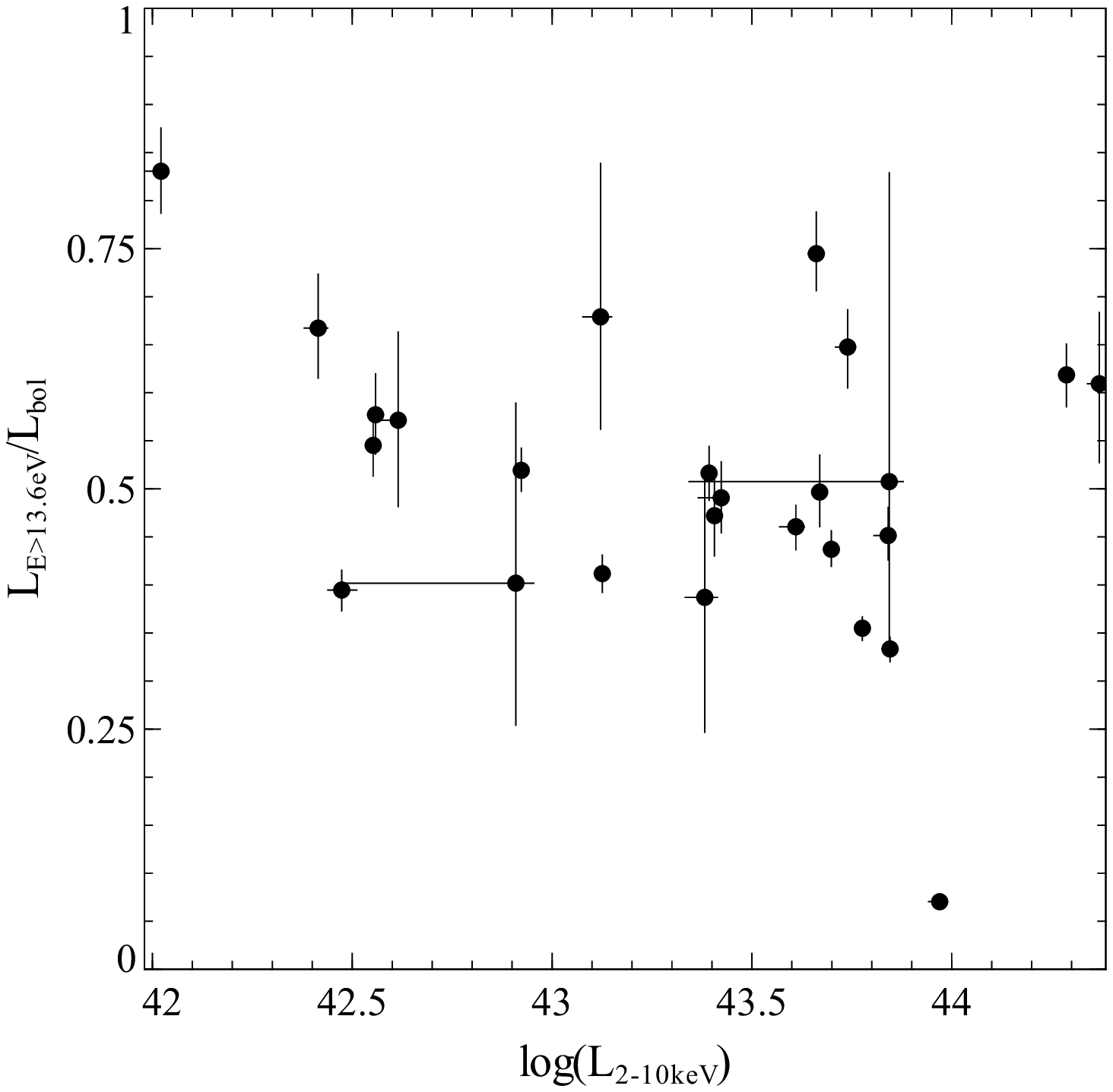}
\includegraphics[width=5cm]{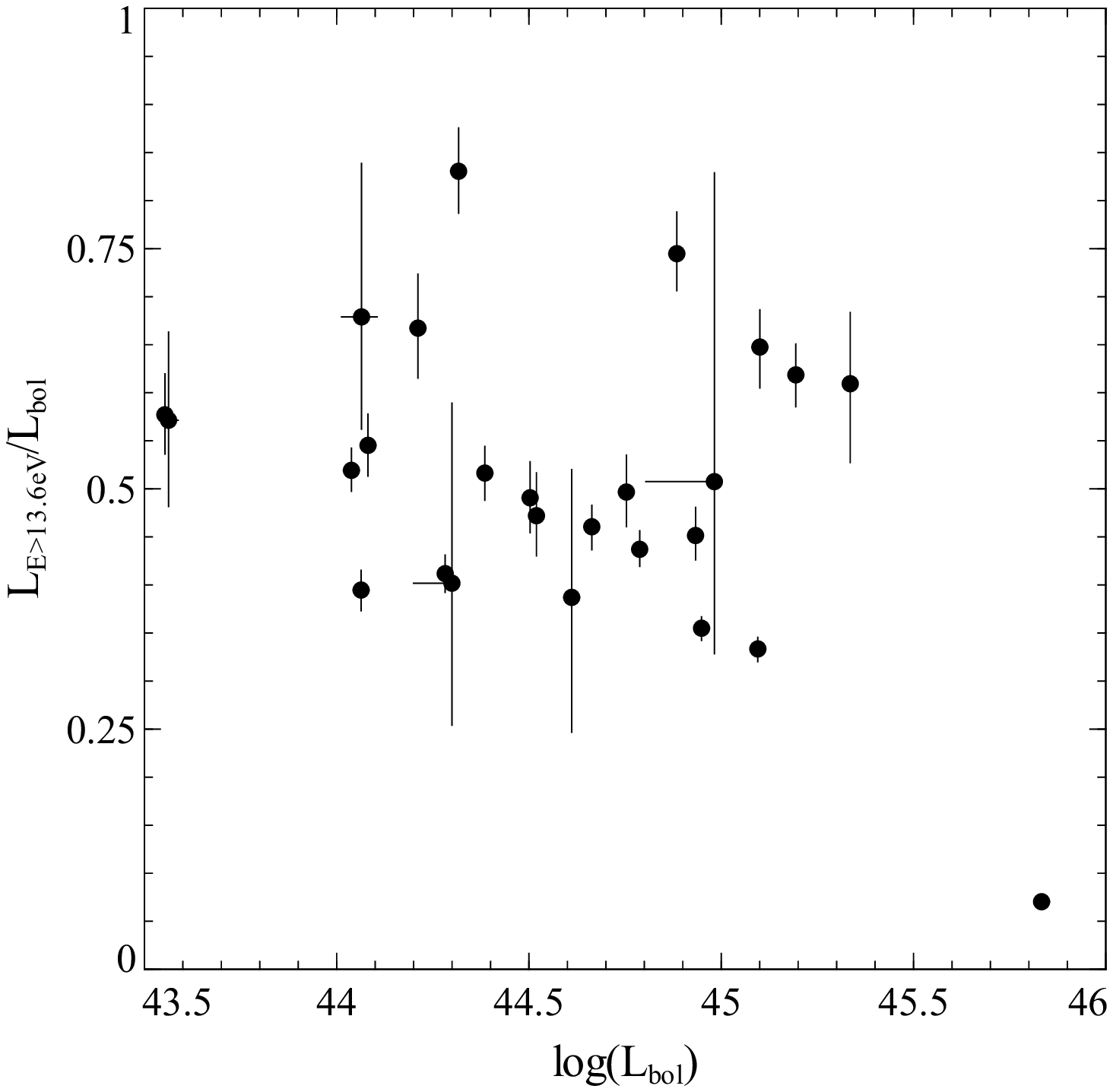}
\includegraphics[width=5cm]{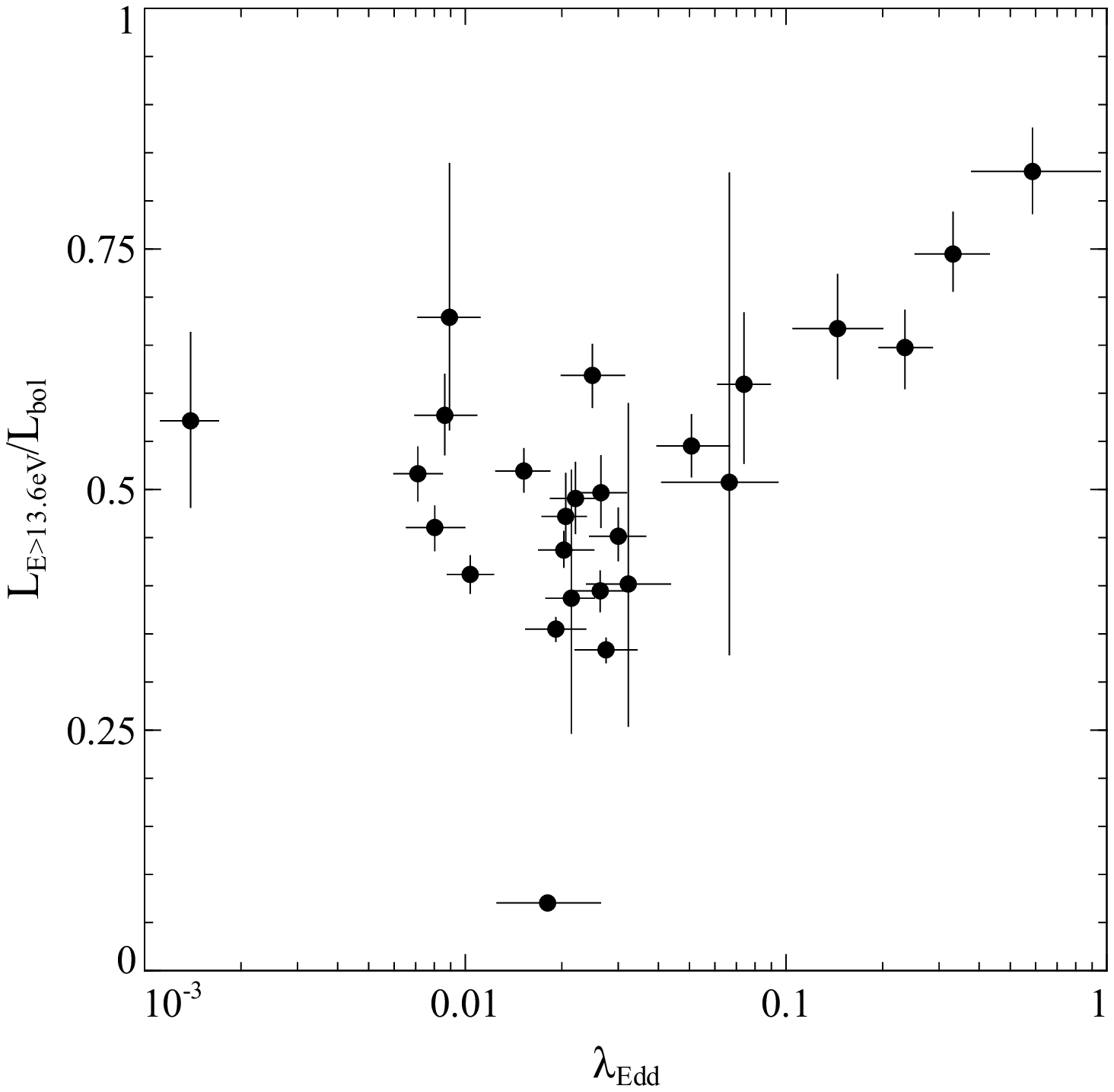}
    \caption{Ionizing fraction $L_{E>13.6\rm eV}/L_{\rm bol}$. \emph{Left panel:} $L_{E>13.6\rm eV}/L_{\rm bol}$ against X-ray luminosity. \emph{Middle panel:} $L_{E>13.6\rm eV}/L_{\rm bol}$ against bolometric luminosity. \emph{Right panel:} $L_{E>13.6\rm eV}/L_{\rm bol}$ against Eddington ratio.}
\label{ionizing_frac}
\end{figure*}

The ionizing continuum in AGN is responsible for the production of the optical and UV emission lines which are often an integral part of selection criteria in optical surveys of AGN. Variations in the ionizing fraction can therefore lead to pronounced changes in the detection of these lines (VF07).  We plot the ionizing fraction ($L_{E>13.6\rm eV}/L_{\rm bol}$) against X-ray luminosity, bolometric luminosity and Eddington ratio in Fig.~\ref{ionizing_frac} to identify any possible trends.  It appears that there is little correlation between $L_{E>13.6\rm eV}/L_{\rm bol}$ and luminosity (either X-ray or bolometric), but there may be indication of a more substantial correlation with Eddington ratio.  This is in line with the variation in SED shape reported in VF07 and VF09, despite our sample spanning only the lower-Eddington ratio regime.  The average ionizing fraction is $\sim$0.6, consistent with that adopted for the ionising luminosity corretion $b_{\rm ion}$ in \cite{2008ApJ...678..693M}, but there is a large spread; for low Eddington ratio objects this could drop to $\sim$0.4, and rise to $\sim$0.8 for high Eddington ratios.  We note that these ionizing fractions are extrapolated from the \textsc{diskpn} big blue bump model fits, and that there may be significant components contributing to this ionizing flux in the extreme-UV and soft X-ray regime, which our approach cannot quantify; the ionizing luminosity fractions presented here may therefore represent lower limits to the `true' ionizing fractions.  However, such a trend for increasing ionizing fraction with Eddington ratio is also expected from combined accretion disk and corona models such as the one by \cite{1997MNRAS.286..848W}.  The distrubution of values seen for the ionizing luminosity fraction may also need to be taken into account in the radiation pressure corrections to virial masses as discussed by \cite{2009arXiv0905.0539M}, in a refinement over their 2008 work.

\subsection{Optical--X-ray connections}

\begin{figure*}
\includegraphics[width=6cm]{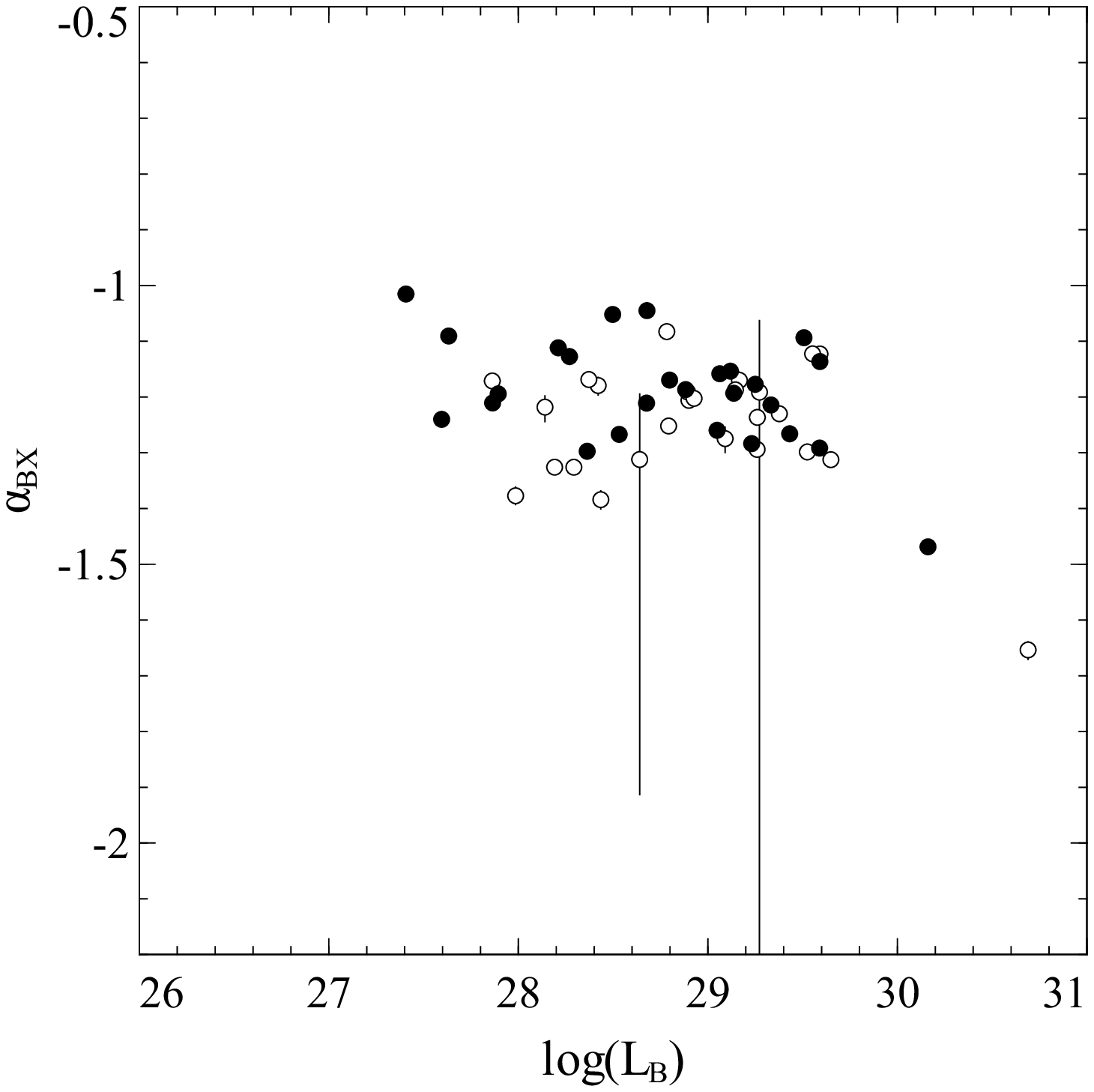}
\includegraphics[width=6cm]{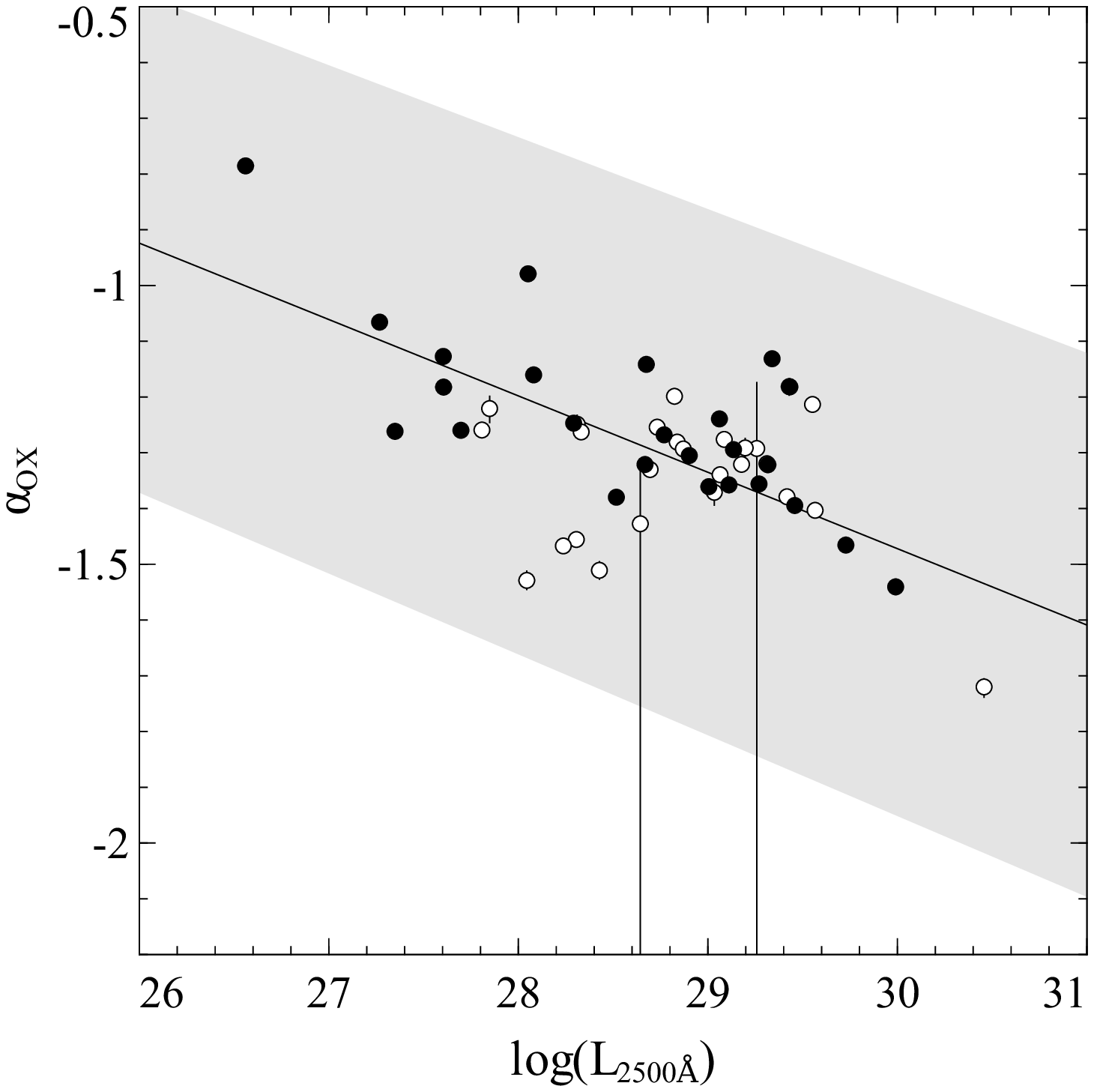}
\includegraphics[width=6cm]{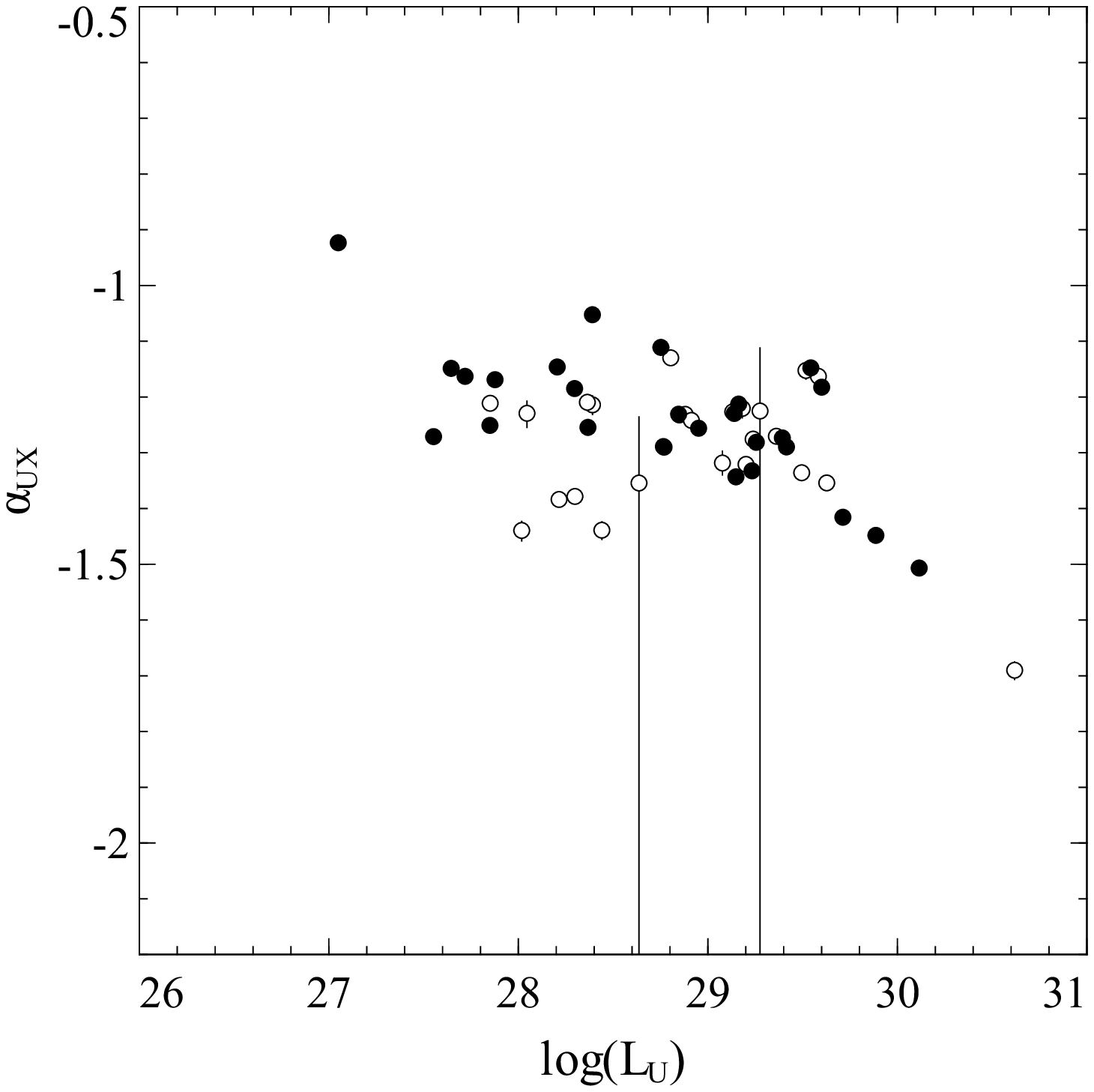}
\includegraphics[width=6cm]{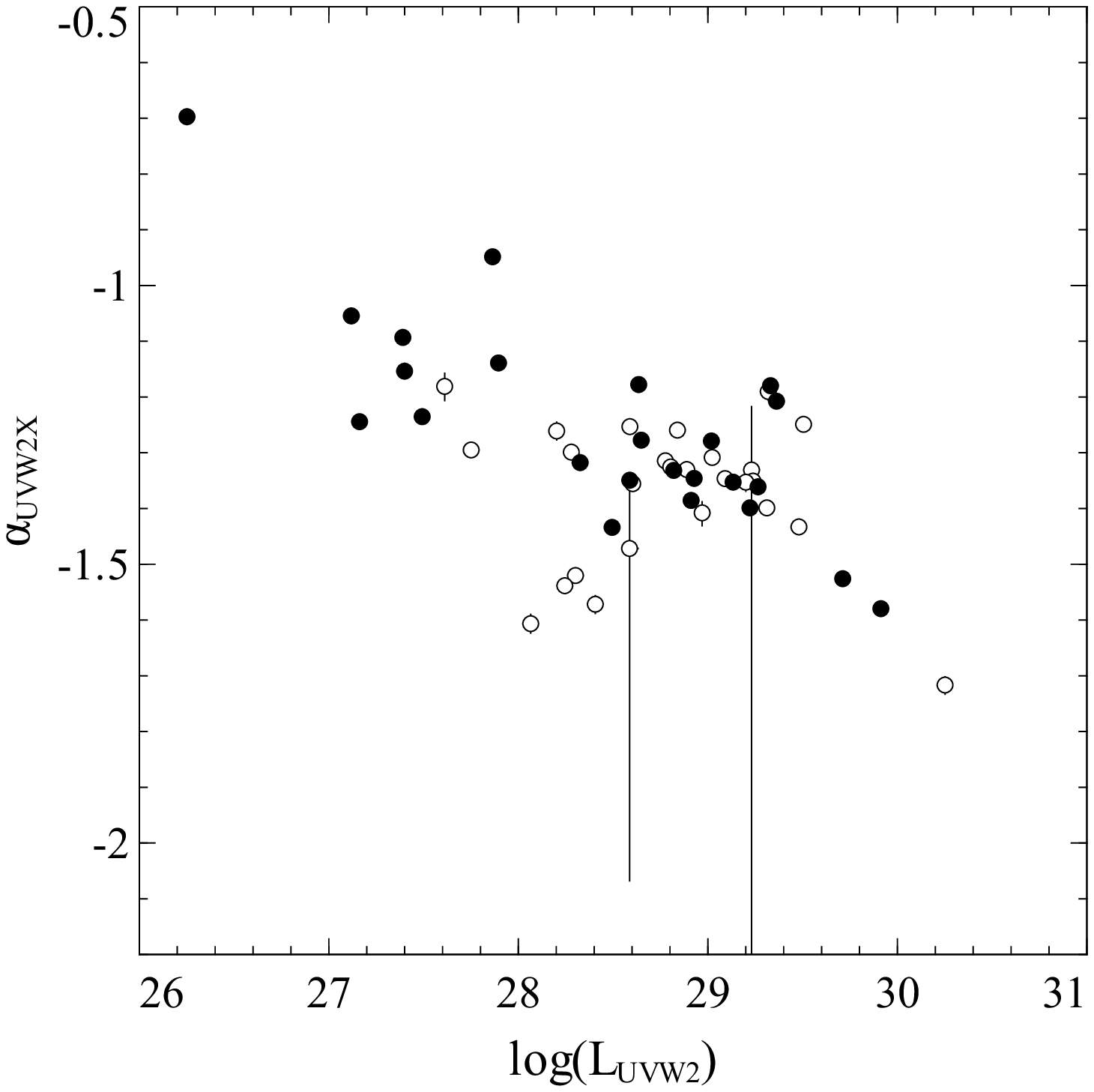}
    \caption{From left to right, top to bottom: Spectral indices connecting SED at the rest frame B-band, U-band, 2500$\rm\AA$ ($\alpha_{\rm OX}$) and UVW2-band with the SED at rest frame 2keV.  Each spectral index is plotted against the corresponding rest-frame reference luminosity in the optical/UV band (e.g. $L_{\rm B}$ for $\alpha_{\rm BX}$, $L_{\rm U}$ for $\alpha_{\rm UX}$etc.).  The black filled circles represent the values obtained from interpolating at rest-frame energies from the data points (not corrected for intrinsic reddening) and the white filled circles show the indices calculated from the de-reddened SED model fit. The solid line and shaded area in the top right panel shows the best fit and spread obtained by \protect\cite{2006AJ....131.2826S} for a sample of 333 optically selected AGN. The correlation coefficients of de-reddened, intrinsic spectral index against luminosity are as follows: $\alpha_{\rm BX}-L_{\rm B}$:-0.27,  $\alpha_{\rm UX}-L_{\rm U}$:-0.24,  $\alpha_{\rm OX}-L_{\rm 2500\AA}$:-0.19,  $\alpha_{\rm UVW2X}-L_{\rm UVW2}$:-0.16.}
\label{optical_xray_connections_lum}
\end{figure*}

The $\alpha_{\rm OX}$--$\rm L_{\nu}(2500\AA)$ correlation reported in the literature (\citealt{2005AJ....130..387S}, \citealt{2006AJ....131.2826S}) has been widely used to encapsulate a relationship between the UV accretion disk emission and the coronal X-ray emission, and the relation between optical and X-ray emission in AGN has been extensively studied observationally (e.g. the \emph{RIXOS} catalogue, \citealt{2000MNRAS.311..456M}).  The spectral index $\alpha_{\rm OX}$ is defined as $-\rm{log(L_{\nu}(2500\AA)/L_{\nu}(2keV))}/\rm{log(\nu(2500\AA)/\nu(2keV))}$ and therefore the optical flux used in its calculation is measured at an energy significantly lower than the peak of the big blue bump (in the vicinity of $\sim$1000$\rm \AA$ for typical AGN black hole masses of $\sim$$10^{7-8}M_{\odot}$, accreting at $\sim$10 per cent of the Eddington limit).  We assess how changing the optical reference point in the calculation of this optical--to--X-ray spectral index alters the correlations seen, and denote the new spectral indices as $\alpha_{BX}$, $\alpha_{UX}$ and $\alpha_{UVW2X}$ (which use the rest-frame B, U and UVW2 bands as the optical reference point instead of $2500\rm \AA$).  It is evident from the center-left panel of Fig.~\ref{optical_xray_connections_lum} that this sample follows the established correlation from \cite{2006AJ....131.2826S}, within the expected scatter.  We present spectral indices calculated both from the raw data (by interpolating to the desired optical wavelength, black filled circles) and from the de-reddened model fit (white filled circles).  In the case of $\alpha_{\rm BX}$, there seems to be little correlation as the B-band is at the low-energy end of the disk component, where variations in the disk emission have relatively little effect.  However the correlations linking $\alpha_{\rm UX}$ and $\alpha_{\rm UVW2X}$ to their respective reference luminosities seem more substantial for the raw, interpolated values: the correlation coefficients decrease monotonically from $-0.46$ to $-0.79$ as the reference wavelength moves from the B to the UVW2 band, demonstrating the stronger anti-correlations seen using a shorter reference wavelengths.   However, the intrinsic spectral indices do not show such trends: the correlation coefficients actually increase from $-0.27$ to $-0.16$ going from B to UVW2, which on visual inspection of these plots reveals a large degree of scatter between the reddening-corrected spectral indices and their reference luminosities.

This demonstrates the importance of accounting for reddening.  Previous studies on $\alpha_{\rm OX}$ have attempted to minimize the effects of reddening by carefully selecting the samples used; for example \cite{2005AJ....130..387S} require that their AGN are required to be bluer than a template AGN SED with fixed reddening; redshifting this template then provides a redshift-dependent colour threshold (see \citealt{2003AJ....126.1131R}) which they use to exclude `red' AGN.  This should statistically exclude heavily reddened AGN, but we find from \emph{individually} de-reddening our optical--UV AGN SEDs, that reddening effects could still be significant.  The anti-correlations between the raw spectral indicies (uncorrected for reddening) and their respective reference luminosities appear strong because reddening in an object produces a flatter optical--X-ray spectral index and reduces the UV luminosity; this will tend to populate the top-left regions of each plot with reddened objects while the lower-right regions will contain unreddened objects, producing an artifically robust-looking relationship.  This effect may contribute to the strength of the $\alpha_{\rm OX}-\rm L_{\rm 2500 \AA}$ correlation in the literature.  From the perspective of understanding the accretion process, however, it is the \emph{intrinsic}, de-reddened optical--X-ray spectral indices which are of interest.  We only use a small sample here, but if individually de-reddened optical SEDs for AGN can be used to construct the $\alpha_{\rm OX}-\rm L_{\rm 2500 \AA}$ relation anew with larger samples, it will clarify whether the versions of this relation seen in the literature can be used to robustly constrain the accretion process in AGN.

\begin{figure*}
\includegraphics[width=6cm]{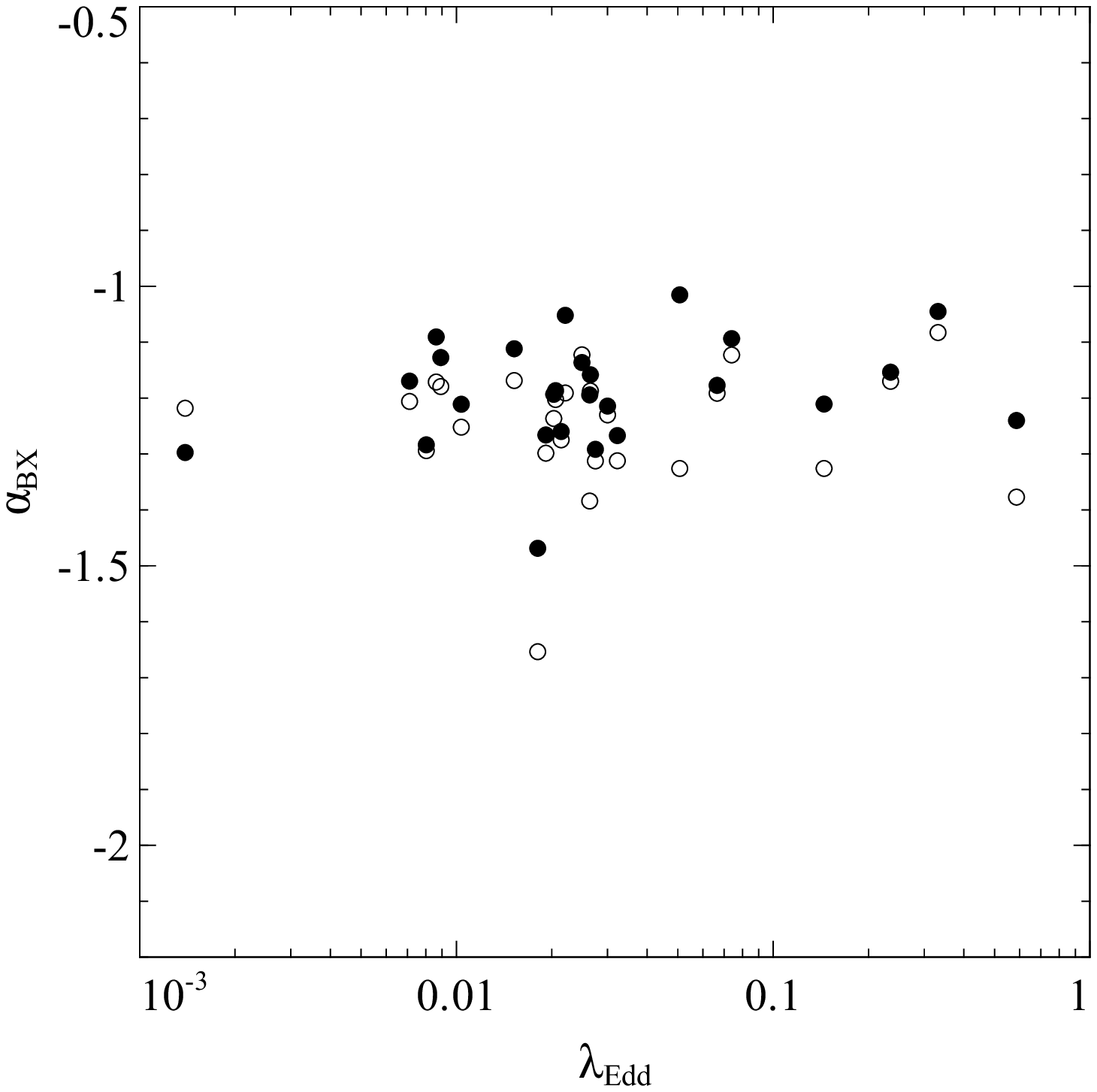}
\includegraphics[width=6cm]{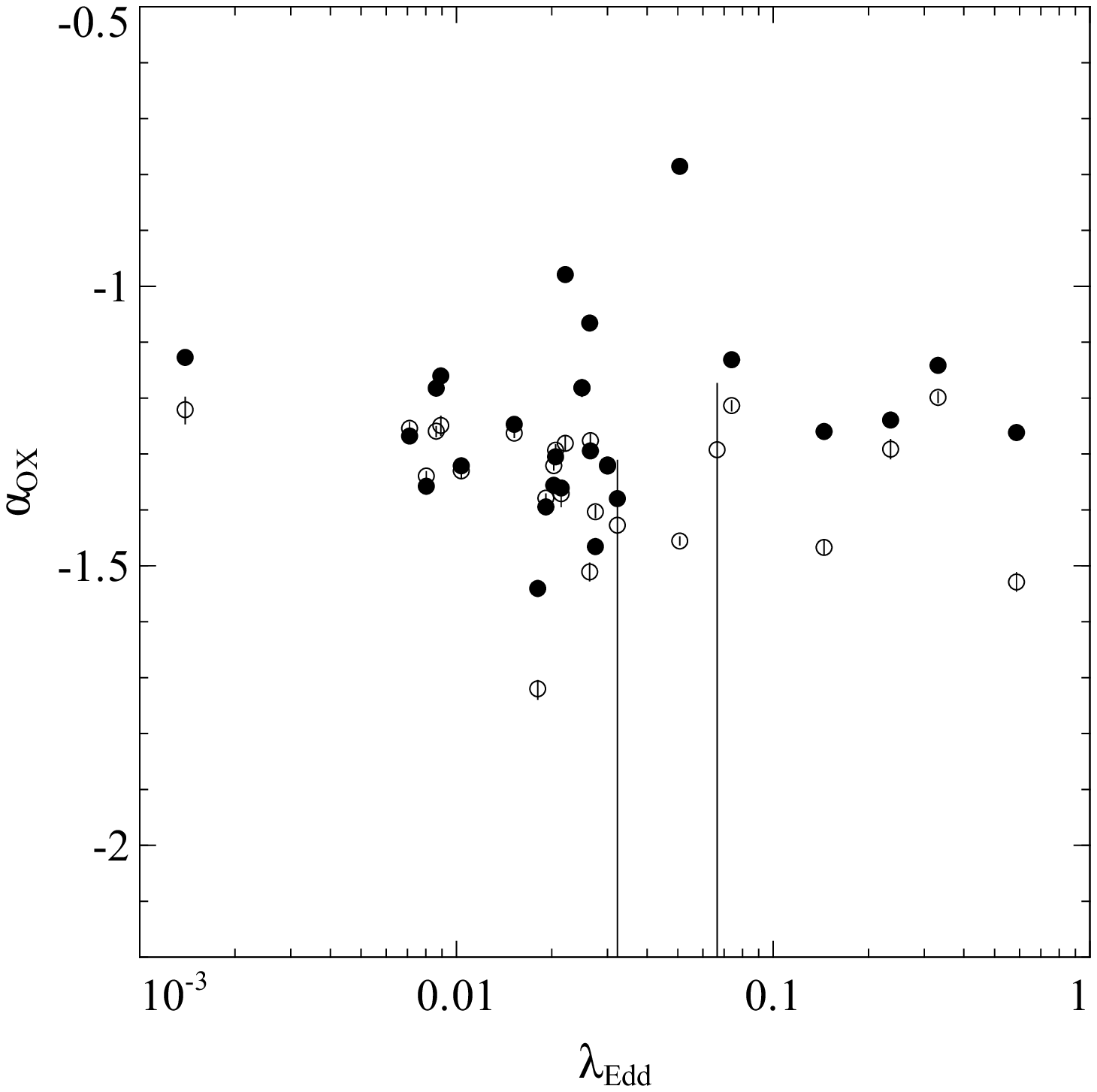}
\includegraphics[width=6cm]{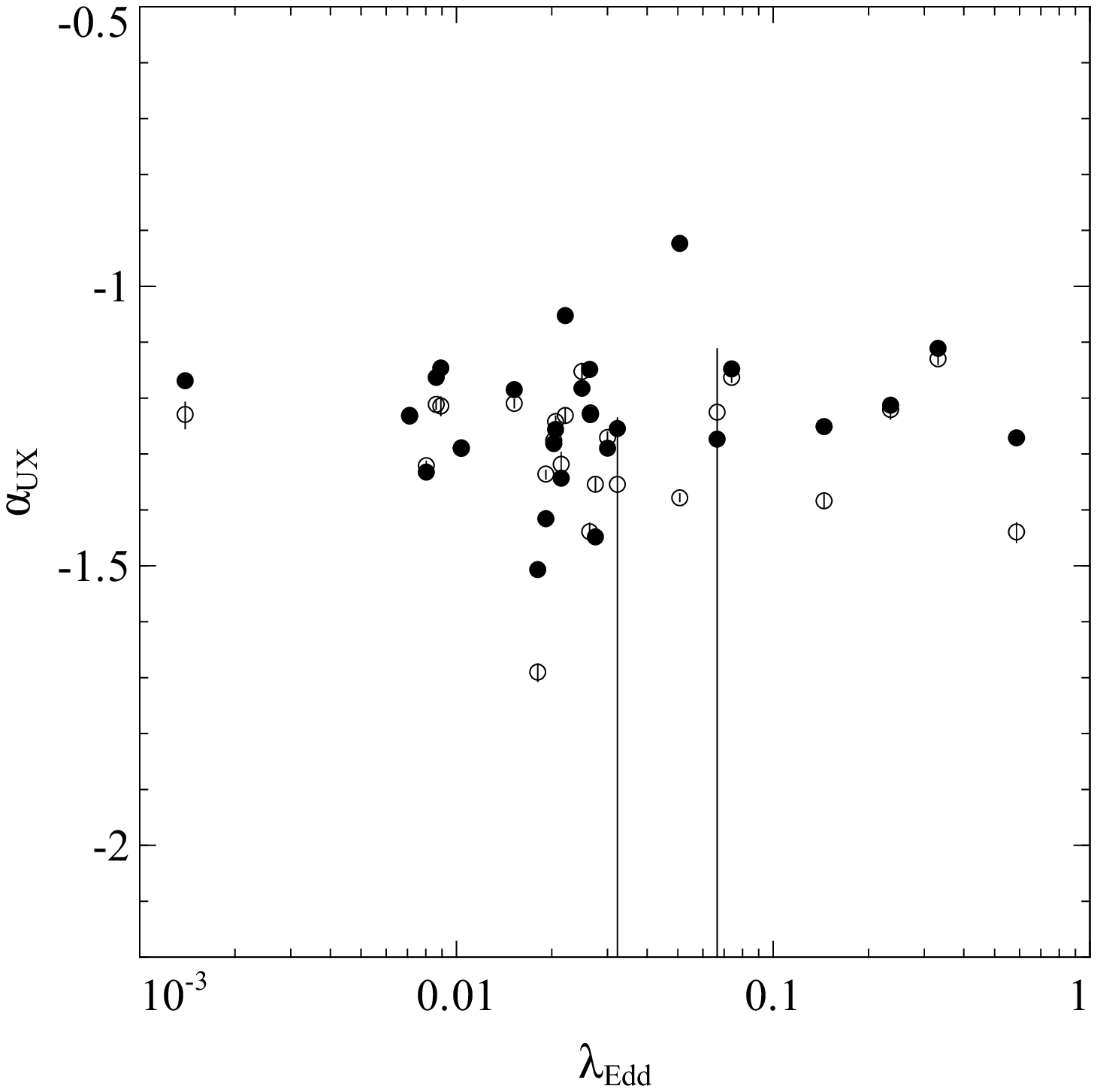}
\includegraphics[width=6cm]{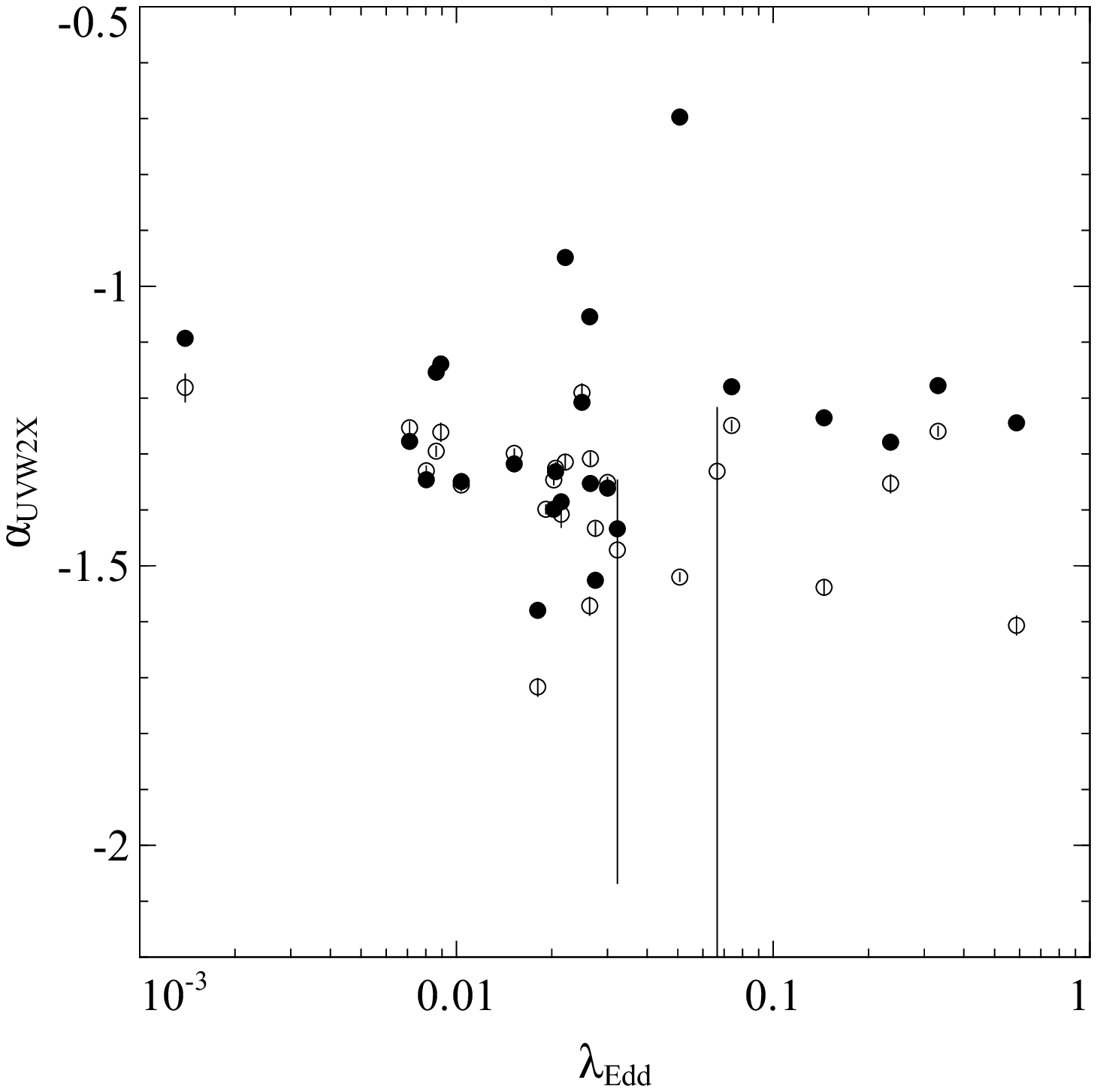}
    \caption{From left to right, top to bottom: Spectral indices connecting SED at the rest frame B-band, 2500$\rm\AA$ ($\alpha_{\rm OX}$), U-band and UVW2-band with the SED at rest frame 2keV.  Each spectral index is plotted against Eddington ratio.  All conventions are as in Fig.~\ref{optical_xray_connections_lum}.}
\label{optical_xray_connections_edd}
\end{figure*}

When plotted against Eddington ratio instead of optical/UV reference luminosity, even the anti-correlations between the raw, uncorrected indices and optical/UV reference luminosities disappear (see Fig.~\ref{optical_xray_connections_edd}).  This phenomenon was also observed by VF07 and \cite{2008ApJ...682...81S}. The latter study suggests that the lack of significant correlation between optical--X-ray spectral index and Eddington ratio could be because of inaccuracies in mass determination methods.  They also discuss how the use of an X-ray selected sample (such as the hard X-ray selected BAT catalogue used here) should alleviate some selection effects which artificially could blur any correlation with Eddington ratio, but this study seems to reinforce the lack of connection between optical--X-ray spectral index and Eddington ratio.  This seems to be the case for both the `raw' values of $\alpha_{\rm OX}$ etc. from the data and the `intrinsic' values calculated from de-reddened SED fits.  This may imply that the optical threshold needs to be moved further into the far--UV before an Eddington ratio correlation is reported, if present.

\begin{figure}
\includegraphics[width=6cm]{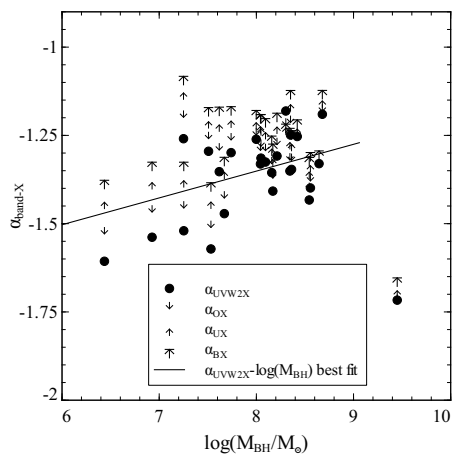}
    \caption{$\alpha_{\rm O,B,U,UVW2-X}$ against black hole mass.  The key provides details of the symbols used.  The best fit (solid line) excludes outlier IRAS 09149-6206.}
\label{alphabandxvsmass}
\end{figure}

\begin{figure}
\includegraphics[width=6cm]{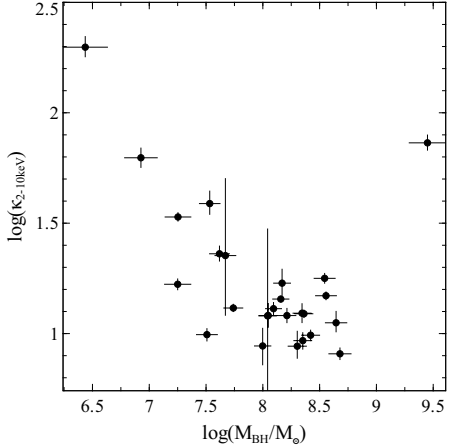}
    \caption{X-ray (2--10keV) bolometric correction against black hole mass.  The best fit (excluding the outlier IRAS 09149-6206) has the form $\rm log(\kappa_{\rm 2-10keV}) = 2.86 - 0.199 \rm log(M_{\rm BH})/M_{\odot}$ (the spread in bolometric corrections is a factor of $\sim2$, by visual inspection).}
\label{bolcorvsmass}
\end{figure}

We also briefly discuss possible correlations of these spectral indices with mass.  We remind the reader of the potential uncertainties with our K-band mass estimates discussed in \S\ref{bhmassesfrom2mass}, but we have found evidence that such estimates do trace more direct black hole mass estimators.  We find only weak indications of a correlation between $\alpha_{\rm band,X}$ and the black hole mass, but find that the correlation coefficient increases with decreasing reference wavelength, increasing from 0.30 (B-band) to 0.61 (UVW2 band; see Fig.~\ref{alphabandxvsmass}), if the outlier IRAS 09149-6206 (the AGN with the highest black hole mass in this sample) is excluded. This relation implies that the spectrum flattens for higher mass objects, presumably indicating that the fraction of UV emission directly from the accretion disc is less.  A plot of bolometric correction against black hole mass (Fig.~\ref{bolcorvsmass}) displays this in a more pronounced fashion, confirming that the fraction of the luminosity appearing in X-rays, $1/{\kappa_{\rm 2-10keV}}$ (i.e. from the corona rather than the disc) increases with mass. This lends credibility to a scenario in which local, high mass black holes are radiating predominantly at low Eddington ratios: not only does the high mass lead to a low Eddington ratio, but lower bolometric correction reduces the bolometric luminosity further.  Conversely, it is then expected that lower mass black holes exhibit higher Eddington ratios by virtue of their large bolometric corrections, in line with `cosmic downsizing' scenarios in which the bulk of accretion activity shifts to lower mass objects at low redshift \citep{2004ApJ...613..109H}.  Since we are using the bulge luminosity to calculate the black hole mass, the possibility of a correlation of bolometric correction with bulge luminosity is also interesting to consider.  If real, it suggests that brighter bulges galaxies contain AGN accreting in states with a greater fraction of X-ray emission.  This may indicate a link between the stellar populations of the host galaxy, the accretion state of the central black hole and the fuelling process; these themes are discussed in more detail by \cite{2008arXiv0812.1224K}.

The data from VF09 on the reverberation mapped sample do not exhibit any similar trends between either $\alpha_{\rm OX}$ or bolometric correction and the black hole mass, despite the better quality mass estimates used.  However, the reverberation sample is not well-selected and the quantities in VF09 do not include reddening and host galaxy corrections in the optical--UV, so do not necessarily provide a useful comparison for our purposes.

\subsection{Host galaxy properties}

The host galaxy properties of the 9-month BAT catalogue are discussed in some detail in \cite{2009ApJ...690.1322W}.  Specific conclusions from their study include the lack of a convincing relation between host galaxy inclination and X-ray absorbing column $N_{\rm H}$; the large fraction of interacting systems in the whole catalogue ($\sim$54 per cent) and suggestions of evolution away from the elliptical, red hosts seen around AGN at z$\sim$1 to systems with bluer colours.  The accretion rate estimates in their paper do not suggest, however, that interactions/mergers are triggers for AGN activity, as the interacting systems do not exhibit a different distribution of accretion rates to the whole population.

Our UVOT photometry results may provide interesting clues in understanding the host galaxy population of the catalogue. Here we present a discussion of the host-galaxy colours for our sample using the UVOT host galaxy magnitudes, and compare our findings with the literature (the more detailed work using KPNO data, Koss et al. in prep, will provide colours for the entire 9-month catalogue). 

There has been much previous work suggesting that AGN hosts occupy the so-called `green valley' (SDSS $g-r$ colour in the range $\sim0.5-0.9$) in the colour-magnitude diagram, mid-way between the late-type `red sequence' ($g-r\sim0.8-1.0$) and star-forming `blue cloud' ($g-r\sim0.2-0.5$) galaxies (see for example \citealt{2007ApJ...660L..11N}, \citealt{2009arXiv0901.4121H}).  Recently, \cite{2009ApJ...692L..19S} present the colours for a small sample of 16 AGN from the BAT catalogue which have SDSS data and find that their hosts also lie in this intermediate region.  The studies of \cite{2008ApJ...675.1025S} and \cite{2007MNRAS.382.1541W} are also of interest as they probe high and low redshift populations (from the Extended \emph{Chandra} Deep Field South and the Sloan Digital Sky Survey, respectively). \cite{2008ApJ...675.1025S} find that, accounting for biases caused by large-scale structure, AGN hosts at $z\gtrsim0.8$ have bluer hosts ($U-V<0.7$) but the lower redshift counterparts ($z\lesssim0.6$) tend to exhibit redder colours.  \cite{2007MNRAS.382.1541W} find a broad distribution of colours for their composite AGN sample, towards the red end of the colour distribution for normal galaxies, and covering the green valley.  We note that, in making our comparison with the results of \cite{2008ApJ...675.1025S}, we do not have any biases with respect to large-scale structure due to the nature of the \emph{Swift}-BAT catalogue, and additionally the only bias with respect to host galaxy properties would be the preferential selection of face-on galaxies (by virtue of our low $N_{\rm H}$ selection). 

\begin{figure*}
    \includegraphics[width=7cm]{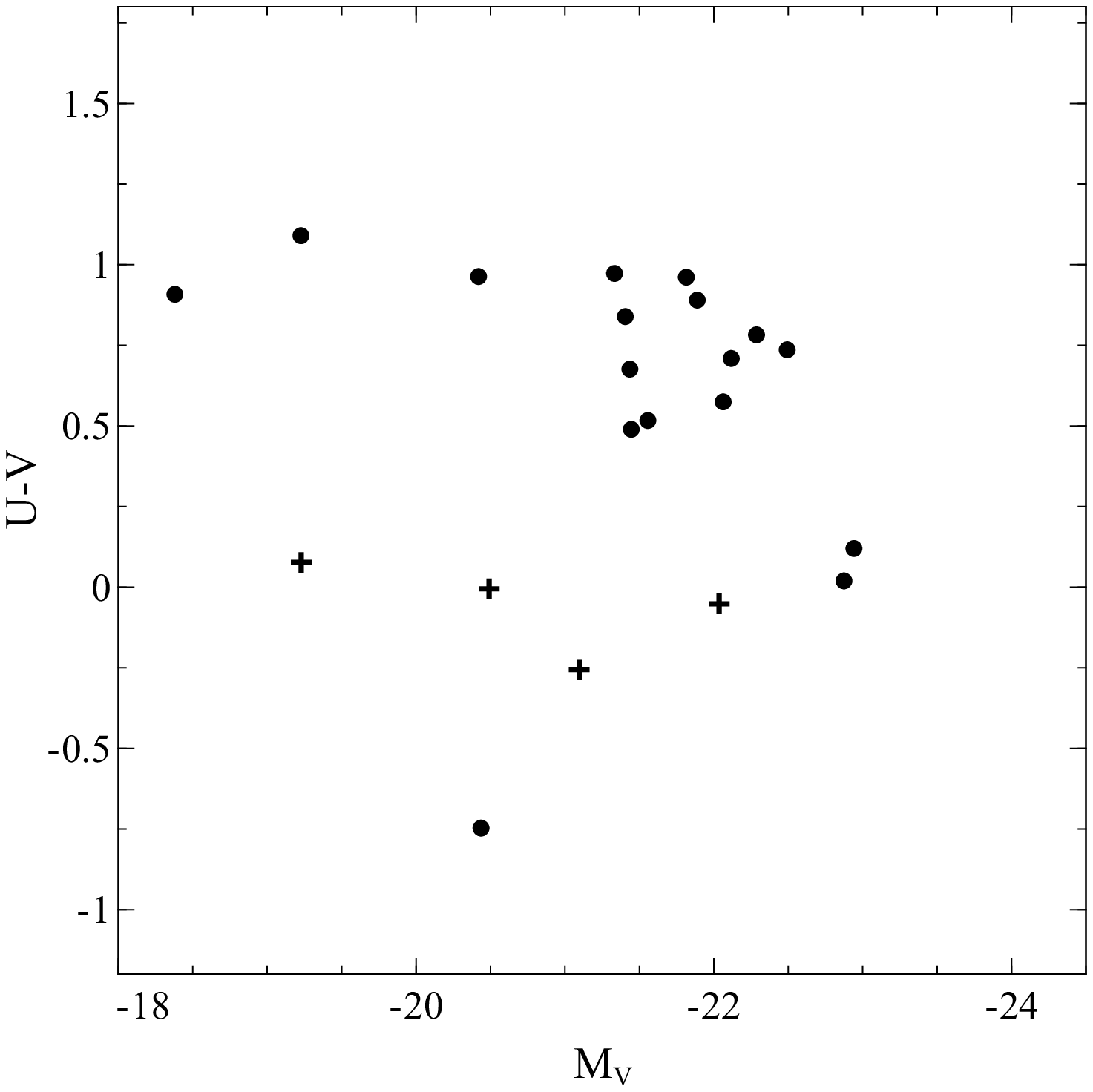}
\includegraphics[width=7cm]{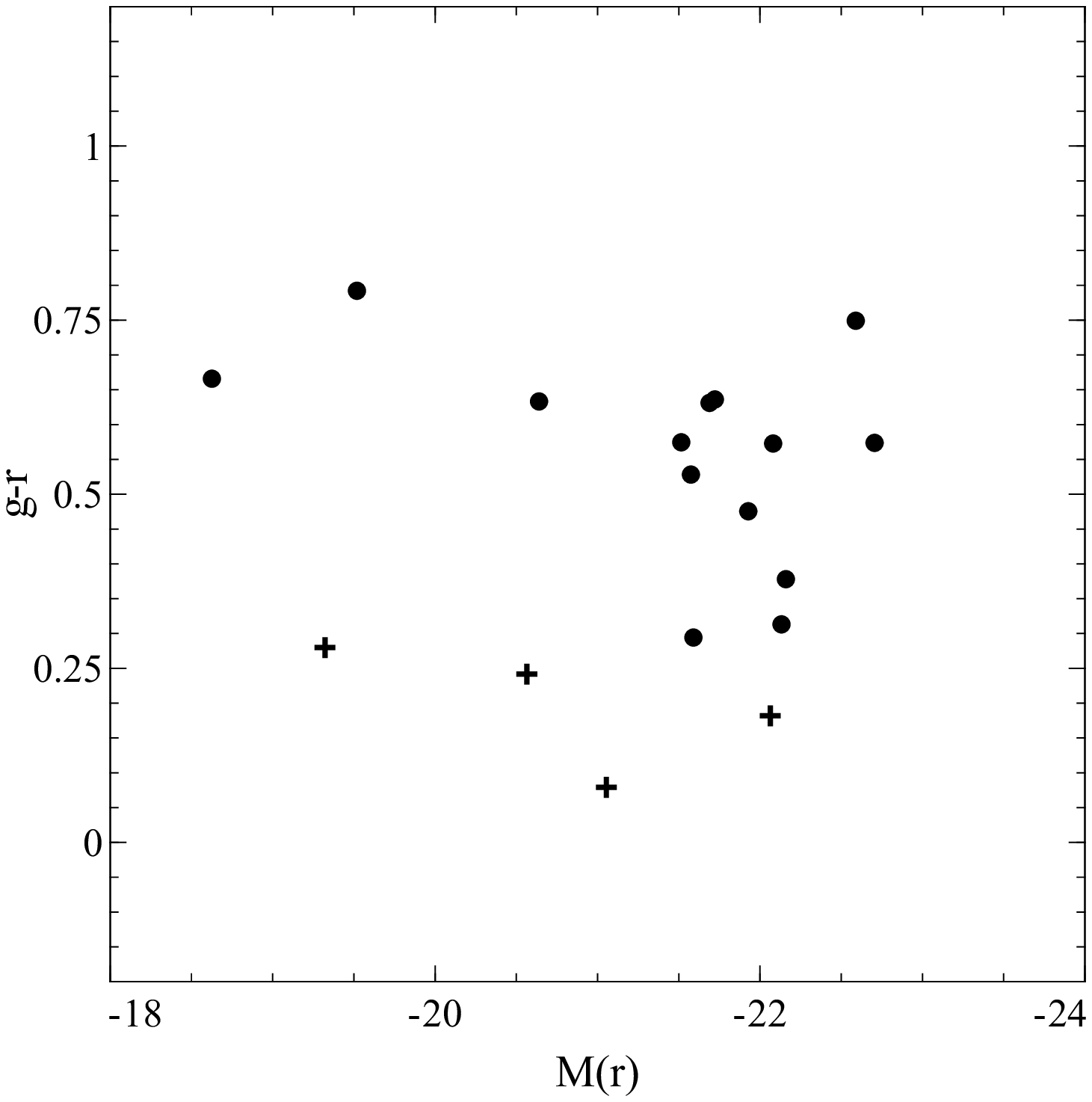}
    \caption{\emph{Left panel:} Host galaxy colour $U-V$ against absolute V-band magnitude for the host galaxies in the sample. \emph{Right panel:} SDSS $g-r$ colour against absolute r-band magnitude for the sample.  Crosses represent objects with point-like morphologies, for which the host galaxy colours should be treated with caution.  GALFIT returns errors of $\pm \sim0.01$ mag on magnitudes, but systematic errors due to the resolution of UVOT are likely to dominate (see Fig.~\ref{galaxycolours_hstcorr})}
\label{galaxycolours}
\end{figure*}

Using the prescription in the UVOT CALDB\footnote{http://heasarc.nasa.gov/docs/heasarc/caldb/swift/docs/uvot/} for converting the UVOT magnitudes into standard Johnson magnitudes, along with the prescription for calculating SDSS colours from Johnson colours, we present our colour-magnitude diagrams in Fig.~\ref{galaxycolours}.  We present both the Johnson $U-V$ colour against absolute V-band magnitude $M_V$ and SDSS $g-r$ colour against absolute SDSS r-band magnitude $M(r)$; the $g-r$ colour is calculated using the transformation $g-r=1.023(B-V)+0.16(U-B)-0.187$ \citep{2005PASA...22...24K} and the $r$--band magnitude is interpolated using the UVOT points.  We provide $U-V$ colours for 21 out of our 26 objects for which both U and V host galaxy magnitudes were available from our GALFIT analysis; 3 of these objects did not have a B-band magnitude for the transformation to $g-r$, reducing our sample of objects to 18 for the latter plot.  

 We see from Fig.~\ref{galaxycolours} that the colours of our local sample show a similar distribution to the X-ray selected AGN in Fig. 3 of \cite{2008ApJ...675.1025S}, but without a significant noticeable red population.  In comparison with \cite{2007MNRAS.382.1541W}, we find a distribution similar to their composite AGN sample, with a handful of very blue AGN.  We therefore seem to corroborate the suggestions in \cite{2009ApJ...690.1322W} of an evolution away from red type host galaxies to bluer host galaxies, with the majority of objects in the `green valley'. We also find somewhat bluer colours than those found for the 16 BAT AGN hosts in \cite{2009ApJ...692L..19S} although there are only five objects overlapping between the two samples (NGC 7469, NGC 5548, Mrk 766, MCG +04-22-042 and Mrk 1018; this list was obtained by private communication with the author).  We caution that four of the bluest objects in Fig.~\ref{galaxycolours} (Ark 120, Mrk 509, Mrk 841 and Mrk 352) have point-like morphologies in the UVOT images, and despite our efforts to minimize PSF mismatch, if such mismatch is present the wings of the PSF could be mistaken for the host galaxy giving rise to artificial `hosts' which are influenced by the bluer colours of the AGN.  These objects are depicted using crosses in Fig.~\ref{galaxycolours} for easy identification. This may indicate that the blue-ward shift is less pronounced than it may initially seem, but the colours for the rest of our sample do lie in the blue end of the green valley.  The more accurate and complete analysis Koss et al. (in prep) will clarify these issues.

We also caution that the PSF size of the UVOT may impose limitations on the accuracy of our host galaxy magnitudes.  In \S\ref{hstcomparison_nucmags}, we estimate the degree by which the nuclear flux is overestimated with UVOT by comparison with HST ACS images for one source, NGC 4593.  Based on the assumption that the flux spuriously attributed to the nucleus should actually be attributed to the host galaxy, we can make a na\"ive correction for this effect as follows.  If the nuclear flux reported by UVOT is $F_{\rm nuc}^{UVOT}=F_{\rm nuc}^{HST}+F_{\rm spurious}$ (for nuclear flux from HST $F_{\rm nuc}^{HST}$ and spurious flux $F_{\rm spurious}$), we can solve for $F_{\rm spurious}$ using the estimate of the error in the UVOT nuclear flux from \S\ref{hstcomparison_nucmags}.  We can then obtain a better estimate of the host galaxy flux seen by HST, using $F_{\rm galaxy}^{HST}=F_{\rm galaxy}^{UVOT}+F_{\rm spurious}$.  The correction to the galaxy flux emerges as $F_{\rm spurious}=F_{\rm nuc}^{UVOT}[1 - (F_{\rm nuc}^{HST}/F_{\rm nuc}^{UVOT})]$.  We employ the fractions $F_{nuc}^{HST}/F_{nuc}^{UVOT}\approx2.4/5.0$ for the V-band and $\approx1.2/1.8$ for the U-band based on the analysis of NGC 4593, and calculate the corrections to the host galaxy magnitudes as follows:

\begin{equation}
\label{mag_psfcorr}
\Delta M = -2.5 \rm log (1 + \frac{F_{nuc}^{UVOT}}{F_{galaxy}^{UVOT}}(1-\frac{F_{nuc}^{HST}}{F_{nuc}^{UVOT}}))). 
\\
\end{equation}

\begin{figure}
    \includegraphics[width=7cm]{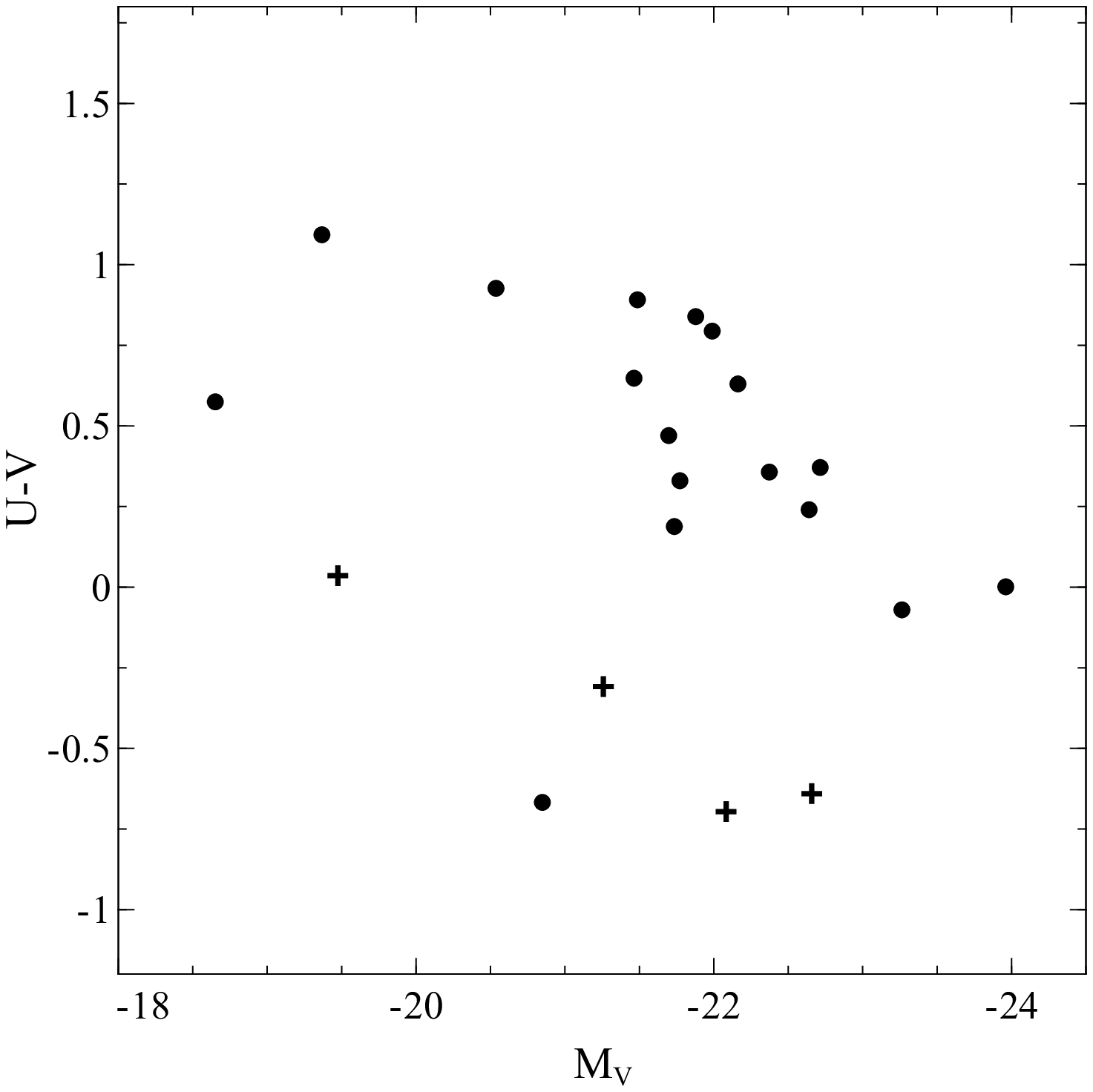}
    \caption{Host galaxy colour U-V against absolute V-band magnitude, partially corrected for large PSF of UVOT.  Key as in Fig.~\ref{galaxycolours}}
\label{galaxycolours_hstcorr}
\end{figure}

It is not straightforward to predict whether the colours should generally increase or decrease under this correction, since the correction is dependent on two things: the ratio of host galaxy flux to nuclear flux, and the degree of error in the nuclear flux intrinsic to UVOT, in a particular band.  We plot the corrected U-V colours in Fig.~\ref{galaxycolours_hstcorr}.  If the corrective factors assumed here are reasonable, we find that the $U-V$ colours generally become bluer, in some cases substantially so.  As expected, the corrected host galaxy V-band magnitudes are brighter due to the extra contribution from the nucleus.  We do not make an attempt to correct the SDSS $g-r$ magnitudes as these would require an estimate of the nuclear flux overestimate in the B-band as well, and the r-band magnitudes require interpolation to wavelengths between the central UVOT filter wavelengths.  It is therefore possible that better quality images may reveal even bluer hosts than our results indicate, but the limitations of the UVOT highlighted by this analysis suggest that the upcoming KPNO/SDSS work of Koss et al. (in prep) will provide more definitive estimates of the host galaxy colours.

\section{Summary and Conclusions}
\label{summaryconclusions}

We have presented SEDs for a well-selected subsample of local AGN from the \emph{Swift} BAT 9-month catalogue of AGN.  This study makes use of the unbiased and representative nature of the catalogue and offers numerous advancements over previous studies of the AGN SED.  The simultaneous optical--to--X-ray observations from \emph{Swift} used here provide a more accurate snapshot of the accretion energy budget in AGN, as pioneered by \cite{2006MNRAS.366..953B} and adopted by VF09.  We are also able to increase confidence in measurements of the accretion luminosity by employing a subsample of AGN with minimal spectral complexity (no strong signs of partial covering or strong reflection) and lack of heavy absorption [log($N_{\rm H}/\rm cm^{-2}$)$<$22] which could complicate the picture: the X-ray spectra for these objects can all be fit with absorbed power-laws according to previous studies \citep{2009ApJ...690.1322W}, allowing straightforward calculation of absorption-corrected X-ray luminosities.  The data from \emph{Swift}-UVOT typically cover at least four, if not all six available optical--UV filters on the instrument, and allow for a detailed reconstruction of the optical--UV SED.  This is useful on two counts: firstly, in understanding the degree of host galaxy contamination in the observed SED for these objects; and secondly, in estimating how much dust extinction is responsible for the shape of the host-galaxy corrected nuclear SED.  The inclusion of black hole mass estimates from K-band bulge luminosities allows accretion rates to be calculated using a uniform method for the sample.

We consistently remove the host-galaxy contribution from the UVOT data points by fitting a `PSF+disk' profile to the UVOT images using GALFIT, and obtain a corrected nuclear flux in all available filters.  The host galaxy is often found to be a significant contaminant in the optical (V, B and U) bands, but far less significant in the UV bands.  We also find that optical/UV emission lines are unlikely to contaminate the nuclear continuum significantly given the relatively large bandwidths of the UVOT filters (500--1000$\rm \AA$).  When fitted along with the simultaneous X-ray data with a simple `multicolour accretion disc + power-law' model, including both Galactic and intrinsic absorption, we are able to recover many important parameters of the AGN SED, and produce an average model SED for lower-luminosity Seyferts with low accretion rates.  The GALFIT fitting on the UVOT images also provides potential for understanding the host galaxy population of the unobscured BAT catalogue AGN.

\begin{itemize}
\item  In this representative sample, the 2--10 keV bolometric corrections cluster around values of 10--20.  These bolometric corrections are essential for calculations of bolometric luminosities and accretion rates from X-ray observations, and therefore are of importance in the wider picture of scaling up the total X-ray energy density due to SMBH accretion in the X-ray background.
\item The Eddington ratios for the sample are typically below 0.1.  Low bolometric corrections are expected for these low Eddington ratios (VF07, VF09), and furthermore, this may indicate that local, unobscured AGN are generally in an accretion state analogous to the `low/hard' state in Galactic black holes.  This reinforces previous findings that local Seyferts have low accretion rates, such as in early work by \cite{1989ApJ...346...68S}, who fit accretion disc models to optical--to--IR SEDs for a sample of 60 bright quasars and Seyferts.  We arrive at the same conclusion using a different approach and employing a different sample: here we also use X-ray data in the calculation of accretion rates on a lower-luminosity sample selected for minimal absorption.
\item Some objects possess significant optical--UV dust reddening, even though these objects were selected for low absorbing gas column. This is broadly consistent with the observations of \cite{2000ApJ...535...53K}, \cite{2003MNRAS.339..757B} and \cite{2005A&A...444...79M} amongst others, and adds to the body of work suggesting that optical and X-ray obscuration classifications can show significant discrepancies (\citealt{2001A&A...365...28M}, \citealt{2001MNRAS.327..781C} and \citealt{2005ApJ...618..123S} are examples of studies which find evidence for the opposite effect in part or all of their samples; namely AGN with X-ray absorption but negligible optical--UV obscuration signatures).  If we adopt a threshold of $E(B-V)\sim0.1$ to divide heavily dust reddened and unreddened objects, $\sim5-6$ out of our 26 AGN ($\sim20$ per cent) are heavily dust reddened, consistent with the proportions of `anomalous' AGN with mismatching optical and X-ray classifications found in some of the above studies, but this obviously depends on the threshold adopted.
\item The reddening-corrected ionizing luminosity fraction increases with Eddington ratio, as expected from combined accretion disc and corona models.
\item We confirm that spectral indicies linking the optical to the X-ray ($\alpha_{\rm OX}$ etc.) show evidence for correlation with luminosity but not with Eddington ratio, corroborating the findings of previous studies. We find that extinction effects may need to be more carefully considered when considering the correllation with luminosity, than has been done in previous works.  De-reddening objects \emph{individually} in constructing the $\alpha_{\rm OX}-L_{\rm 2500 \AA}$ relation yields a weaker correlation, which may need to be investigated in larger samples to check the robustness of this correlation and its use in constraining the accretion process.  The presence of a possible correlation of $\alpha_{\rm UVW2X}$ with mass, coupled with the attendant anti-correlation of bolometric correction with mass, may reinforce cosmic downsizing scenarios where lower mass black holes dominate AGN activity in the local universe by virtue of their larger accretion rates.
\item The host galaxies for this subsample generally lie in the green valley or blue-wards of it, in the standard colour-magnitude diagram.   However, the UVOT images are of limited resolution and a better quality analysis with KPNO/SDSS (Koss et al. in prep) will shed more light on the host galaxy properties.
\item Adding the hard X-ray BAT data (14--195 keV) to these observations can help with ascertaining the presence of a reflection hump in the spectrum.  We selected four sources with the lowest hard X-ray variability, and find that three of these four showed a significant excess peaking at around $\sim$30 keV which can plausibly be fit with a reflection hump.
\end{itemize}

We have also discussed at length, the calibration of the black hole mass estimation method used to obtain these conclusions.  The Eddington ratios and bolometric corrections could be significantly affected by these uncertainties.  We estimate the degree of this effect by comparison with results using reverberation mass estimates.  This analysis reveals somewhat higher bolometric corrections and Eddington ratios, but the bulk of the sample lies consistently at lower Eddington ratios and exhibits low bolometric corrections.  The bulge luminosity-based and reverberation mapping methods require more accurate calibration before more definitive statements on the magnitudes of these uncertainties can be constrained.

The hard X-ray selection criteria used to specify the BAT catalogue eliminate potential biases.  As an illustration of this, we notice the presence of substantially more low accretion rate objects in this sample than the samples of VF07 and VF09 (the former was UV selected by \emph{FUSE}, the latter selected for being optically bright enough for reverberation mapping).  The low accretion rates found here broadly corroborate the results of \cite{2009ApJ...690.1322W} using X-ray luminosities (with no bolometric correction applied).  Although these objects have low-absorption and are situated at low redshift, they are similar in power to the obscured, higher redshift sources thought to be responsible for the bulk of the X-ray background.  As suggested in \cite{2004cbhg.symp..446F}, we confirm that a lower bolometric correction is appropriate for these objects, based on our multi-pronged approach to recovering the true bolometric luminosities in these sources. Bolometric corrections of the magnitude seen here are appropriate for reconciling the energy density from the X-ray background with the local black hole density.  It remains to be seen if such properties are displayed in large samples of obscured AGN, the dominant class of AGN contributing to the X-ray background. 

Useful extensions to this work would be to acquire simultaneous \emph{Swift} observations for the remaining objects in the BAT catalogue, which would potentially confirm the trends we identify here, or highlight unknown biases in our chosen sample.  What is of particular importance next is to understand the workings of the central engine of obscured AGN.  In a companion paper (Vasudevan et al. \emph{in prep}) we hope to address this issue for both obscured and unobscured AGN in the Swift/BAT catalogue, using the reprocessed IR emission  along with the hard X-ray BAT observations to estimate their bolometric properties, analogous to the work done by \cite{2007A&A...468..603P} for eight higher redshift quasars.  The well-studied 9-month catalogue again provides an excellent starting point for such work, and follow-up studies on the 22-month catalogue \citep{2009arXiv0903.3037T} will offer new scope for constraining the accretion physics of a larger, representative sample of AGN.

\section{Acknowledgements}

RVV acknowledges support fom the Science and Technology Facilities Council (STFC) and ACF thanks the Royal Society for Support.  We thank the \emph{Swift}/BAT team for the 9-month AGN catalogue data.  We thank Stephen Holland for help with UVOT data analysis, and Alice Breeveld for kindly providing and customizing her PSF generation code for UVOT images.  We thank Jack Tueller for the use of results derived from the 8-channel BAT data, Kevin Schawinksi for providing host galaxy colours and magnitudes from his paper for comparison with our study. We thank Richard McMahon for help in understanding the correct use of the data in the 2MASS catalogues.  We also thank the anonymous referee for useful comments and suggestions which improved this work.  This research has made use of the NASA Extragalactic Database (NED) and the NASA/IPAC Infrared Science Archive, which are operated by the Jet Propulsion Laboratory, California Institute of Technology, under contract with the National Aeronautics and Space Administration.

\bibliographystyle{mnras} 
\bibliography{swiftuvotseds}

\newpage
\newpage

\begin{appendix}
\appendix

\section{Results from SED fitting}
\label{appendix1}
\begin{table*}
\begin{tabular}{|l|l|l|l|l|l|l|l|l|l}
\hline
AGN&$M_{\rm BH}/M_{\odot}$&$\Gamma$&$\rm{L_{2-10keV}}$&$\alpha_{\rm ox,A}$&$E(B-V)$&$\rm{L_{bol}}$&$\alpha_{\rm ox,B}$&$\lambda_{\rm Edd}$&$\kappa_{\rm 2-10keV}$\\
&   (1)&   (2)&   (3)&  (4)&   (5)&   (6)&   (7)&   (8)&   (9)\\\hline
1RXS J045205.0+493248&$8.04$&$1.86$&$43.8$&(none)&$0.0$&$45.0$&$-1.29$&$0.067^{+0.028}_{-0.026}$&$12^{+18}_{-6}$\\
2MASX J21140128+8204483&$8.68$&$1.85$&$44.3$&$-1.18$&$0.000056^{+0.035000}_{-0.000056}$&$45.2$&$-1.18$&$0.0249^{+0.0064}_{-0.0050}$&$8.10^{+0.51}_{-0.47}$\\
3C 120&$8.35$&$1.78$&$43.8$&$-1.32$&$0.0$&$44.9$&$-1.32$&$0.0300^{+0.0065}_{-0.0056}$&$12.4^{+1.3}_{-1.2}$\\
3C 390.3&$8.35$&$1.75$&$44.4$&$-1.13$&$0.061^{+0.019}_{-0.021}$&$45.3$&$-1.21$&$0.074^{+0.015}_{-0.013}$&$9.29^{+0.80}_{-0.79}$\\
Ark 120&$8.54$&$1.90$&$43.8$&$-1.47$&$0.0^{+0.0}_{-0.0}$&$45.1$&$-1.40$&$0.0275^{+0.0068}_{-0.0055}$&$17.79^{+0.93}_{-0.88}$\\
ESO 490-G026&$8.05$&$1.91$&$43.4$&$-0.979$&$0.266^{+0.019}_{-0.023}$&$44.5$&$-1.28$&$0.0221^{+0.0044}_{-0.0036}$&$12.1^{+1.6}_{-1.4}$\\
ESO 548-G081&$7.74$&$2.03$&$42.9$&$-1.25$&$0.0$&$44.0$&$-1.26$&$0.0152^{+0.0031}_{-0.0028}$&$13.06^{+0.52}_{-0.47}$\\
IRAS 05589+2828&$8.64$&$1.61$&$43.6$&$-1.36$&$0.0^{+0.0}_{-0.0}$&$44.7$&$-1.34$&$0.0080^{+0.0019}_{-0.0015}$&$11.2^{+1.4}_{-1.0}$\\
IRAS 09149-6206&$9.45$&$1.74$&$44.0$&$-1.54$&$0.28^{+-0.23}_{-0.25}$&$45.8$&$-1.72$&$0.0181^{+0.0083}_{-0.0055}$&$73.1^{+6.2}_{-5.3}$\\
MCG +04-22-042&$8.09$&$1.94$&$43.4$&$-1.30$&$0.0$&$44.5$&$-1.29$&$0.0206^{+0.0033}_{-0.0032}$&$12.97^{+0.86}_{-0.84}$\\
MCG -06-30-015&$7.25$&$1.62$&$42.6$&$-0.785$&$0.598^{+0.025}_{-0.026}$&$44.1$&$-1.46$&$0.051^{+0.016}_{-0.011}$&$33.7^{+1.5}_{-1.4}$\\
Mrk 1018&$8.21$&$1.95$&$43.7$&$-1.29$&$0.0$&$44.8$&$-1.28$&$0.0265^{+0.0053}_{-0.0042}$&$12.07^{+0.94}_{-0.86}$\\
Mrk 279&$8.42$&$1.88$&$43.4$&$-1.27$&$0.0$&$44.4$&$-1.25$&$0.0071^{+0.0014}_{-0.0011}$&$9.83^{+0.53}_{-0.54}$\\
Mrk 352&$6.93$&$1.68$&$42.4$&$-1.26$&$0.190^{+0.016}_{-0.015}$&$44.2$&$-1.47$&$0.145^{+0.055}_{-0.040}$&$62.5^{+6.6}_{-6.0}$\\
Mrk 509&$8.56$&$1.83$&$43.8$&$-1.39$&$0.0^{+0.0}_{-0.0}$&$44.9$&$-1.38$&$0.0192^{+0.0046}_{-0.0037}$&$14.84^{+0.60}_{-0.60}$\\
Mrk 590&$8.30$&$1.88$&$42.6$&$-1.13$&$0.096^{+0.084}_{-0.008}$&$43.6$&$-1.22$&$0.00139^{+0.00031}_{-0.00027}$&$8.8^{+1.5}_{-1.0}$\\
Mrk 766&$7.53$&$1.76$&$42.5$&$-1.07$&$0.376^{+0.021}_{-0.024}$&$44.1$&$-1.51$&$0.0263^{+0.0061}_{-0.0050}$&$38.8^{+5.4}_{-4.1}$\\
Mrk 841&$8.17$&$1.89$&$43.4$&$-1.36$&$0.016^{+0.020}_{-0.016}$&$44.6$&$-1.37$&$0.0214^{+0.0038}_{-0.0036}$&$16.9^{+2.6}_{-2.0}$\\
NGC 4593&$7.51$&$1.62$&$42.6$&$-1.18$&$0.087^{+0.018}_{-0.019}$&$43.6$&$-1.26$&$0.0086^{+0.0022}_{-0.0017}$&$9.89^{+0.64}_{-0.63}$\\
NGC 5548&$8.00$&$1.51$&$43.1$&$-1.16$&$0.087^{+0.019}_{-0.027}$&$44.1$&$-1.25$&$0.0089^{+0.0022}_{-0.0018}$&$8.8^{+1.8}_{-1.6}$\\
NGC 7469&$8.16$&$1.98$&$43.1$&$-1.32$&$0.017^{+0.010}_{-0.017}$&$44.3$&$-1.33$&$0.0104^{+0.0019}_{-0.0016}$&$14.33^{+0.42}_{-0.42}$\\
NGC 985&$8.36$&$1.80$&$43.7$&$-1.36$&$0.0^{+0.0}_{-0.0}$&$44.8$&$-1.32$&$0.0203^{+0.0049}_{-0.0033}$&$12.30^{+0.48}_{-0.48}$\\
SBS 1136+594&$7.62$&$1.94$&$43.7$&$-1.24$&$0.074^{+0.015}_{-0.016}$&$45.1$&$-1.29$&$0.235^{+0.051}_{-0.040}$&$23.0^{+1.9}_{-1.7}$\\
SBS 1301+540&$7.25$&$1.81$&$43.7$&$-1.14$&$0.054^{+0.016}_{-0.016}$&$44.9$&$-1.20$&$0.332^{+0.098}_{-0.079}$&$16.73^{+0.95}_{-0.93}$\\
UGC 06728&$6.44$&$1.82$&$42.0$&$-1.26$&$0.242^{+0.022}_{-0.021}$&$44.3$&$-1.53$&$0.59^{+0.37}_{-0.21}$&$198^{+23}_{-19}$\\
WKK 1263&$7.67$&$1.68$&$42.9$&$-1.38$&$0.079^{+0.150}_{-0.079}$&$44.3$&$-1.43$&$0.032^{+0.011}_{-0.008}$&$23^{+28}_{-10}$\\
\hline
\end{tabular}
\caption{\label{uvotxrtresults}Results from SED fitting. \protect\\ (1) - Log of central BH mass from 2MASS K-band magnitude. Errors from K-band magnitudes translate into errors of $\pm0.1$ in values for $ \rm{log} (M_{\rm BH}/M_{\odot})$, but we refer the reader to the systematics discussed in the text. \protect\\ (2) - Photon index from fit to 0.3--10keV regime. \protect\\ (3) - Log of 2--10keV luminosity from power-law fit to 0.3--10keV regime. \protect\\ (4) - Spectral index $\alpha_{\rm ox}$ calculated by interpolation between available UVOT data points to determine the 2500$\rm \AA$ luminosity.  \protect\\ (5) - Intrinsic extinction $E(B-V)$ from fitting \textsc{zdust(diskpn)} model combination to UVOT data. \protect\\ (6) - Log of bolometric (0.001--100keV) luminosity, corrected for X-ray absorption and optical--UV dust reddening. \protect\\ (7) - Spectral index $\alpha_{\rm ox}$ calculated from the full optical--to--X-ray model fit, corrected for optical--UV dust reddening. \protect\\ (8) - Eddington ratio. \protect\\ (9) - Hard X-ray 2--10keV bolometric correction $L_{\rm bol}/L_{\rm 2-10keV}$. \protect\\Random error estimates are provided on values for intrinsic extinction, Eddington ratio and bolometric correction; random errors on all other quantities are omitted as systematic errors generally dominate.}
\end{table*}

\end{appendix}

\clearpage

\end{document}